\documentclass[twocolumn,english,aps,superscriptaddress,floatfix,longbibliography,nobibnotes,nofootinbib]{revtex4-1}
\usepackage{mathptmx}

\usepackage[T1]{fontenc}
\usepackage[latin9]{inputenc}
\usepackage{color}
\usepackage{babel}
\usepackage{dsfont}
\usepackage{amsmath}
\usepackage{amssymb}
\usepackage{graphicx}
\usepackage{wasysym}
\usepackage[unicode=true,pdfusetitle,
 bookmarks=true,bookmarksnumbered=false,bookmarksopen=false,
 breaklinks=true,pdfborder={0 0 0},pdfborderstyle={},backref=false,colorlinks=true]
 {hyperref}
\hypersetup{
 linkcolor=blue,citecolor=blue,urlcolor=blue}

\makeatletter
\usepackage{dsfont}
\usepackage{orcidlink}

\makeatother

\begin{document}
\title{Random free-fermion quantum spin chain with multispin interactions}
\author{Francisco C. Alcaraz\orcidlink{0000-0002-5014-3590}}
\affiliation{Instituto de F\'isica de S\~ao Carlos, Universidade de S\~ao Paulo,
C.P. 369, S\~ao Carlos, SP, 13560-970, Brazil}
\author{José A. Hoyos\orcidlink{0000-0003-2752-2194}}
\affiliation{Instituto de F\'isica de S\~ao Carlos, Universidade de S\~ao Paulo,
C.P. 369, S\~ao Carlos, SP, 13560-970, Brazil}
\affiliation{Max Planck Institute for the Physics of Complex Systems, N\"othnitzer
Str. 38, 01187 Dresden, Germany}
\author{Rodrigo A. Pimenta\orcidlink{0000-0003-2971-0455}}
\affiliation{Department of Physics and Astronomy, University of Manitoba, Winnipeg
R3T 2N2, Canada}
\begin{abstract}
We study the effects of quenched disorder in a class of quantum chains
with ($p+1$)-multispin interactions exhibiting a free fermionic spectrum,
paying special attention to the case $p=2$. Depending if disorder
couples to (i) all the couplings or just to (ii) some of them, we
have two distinct physical scenarios. In case (i), we find that the
transitions of the model are governed by a universal infinite-randomness
critical point surrounded by quantum Griffiths phases similarly as
happens to the random transverse-field Ising chain. In case (ii),
we find that quenched disorder becomes an irrelevant perturbation:
the clean critical behavior is stable and Griffiths phases are absent.
Beyond the perturbative regime, disorder stabilizes a line of finite-randomness
critical points (with nonuniversal critical exponents), that ends
in a multicritical point of infinite-randomness type. In that case,
quantum Griffiths phases also appear surrounding the finite-disorder
transition point. We have characterized the correlation functions
and the low-temperature thermodynamics of these chains. Our results
are derived from a strong-disorder renormalization-group technique
and from finite-size scaling analysis of the spectral gap computed
exactly (up to sizes $\sim10^{7}$) via an efficient new numerical
method recently introduced in the literature {[}Phys. Rev. B \textbf{104},
174206 (2021){]}.\\
\\
Published in \href{https://journals.aps.org/prb/abstract/10.1103/PhysRevB.108.214413}{Phys. Rev. B {\bf 108}, 214413 (2023)};
DOI: \href{https://doi.org/10.1103/PhysRevB.108.214413}{10.1103/PhysRevB.108.214413}
\end{abstract}
\maketitle

\section{Introduction}

The importance of studying one-dimensional quantum models is invaluable.
Due to the peculiarities of the phase space in $d=1$, many models
can be precisely described and, thus, serve as important test beds
for physical insights~\citep{giamarchi-book}. Among those models,
there is an important class that goes under the general name of free
systems. These are systems (of volume $V$) whose the exponentially
large Hilbert space of dimension $\sim e^{V}$ can be expressed as
a combination of $\sim V$ quasienergies of a free-particle system.
Despite the name, they exhibit nontrivial phenomena, such as phase
transitions and zero-energy fractional edge modes, and are, thus,
good starting points to understand many complex behaviors of more
general systems~\citep{lieb-schultz-mattis,pfeuty-ap70,kitaev-pu-2001}.

Initially, the known free systems models were linked to a Jordan-Wigner
transformation which maps interacting spin models into free fermionic
particles, i.e., into models which are bilinear in the fermionic operators
or fields. Later, it was realized that the existence of this transformation
is not a necessary condition~\citep{baxter-pla89}. It is possible
to construct the eigen-spectrum of the spin system from the free-fermion
pseudo-energies obtained from the roots of a characteristic polynomial.
This polynomial is constructed thanks to the infinite number of conserved
charges of the model~\citep{fendley-jpa19,alcaraz-pimenta-prb20a,alcaraz-pimenta-prb20b}.
Interestingly, these latter free systems, \emph{a priori}, cannot
be written in term of local bilinears of fermion operators. 

The first models in this class are the $\text{Z}_{N}$-symmetric free
parafermionic models~\citep{baxter-pla89,baxter-jsp89,baxter-jsp04,fendley-jpa14}
and the three-spin interacting Fendley model~\citep{fendley-jpa19}.
 Actually, it has been found that Fendley model belongs to a large
family of spin chains with multispin interactions and $\text{Z}_{N}$-symmetry~\citep{alcaraz-pimenta-prb20a,alcaraz-pimenta-prb20b}.
For $N>2$, the spectrum is complex and has a free parafermionic form.
Furthermore, it was shown that the free-particle pseudo-energies of
some of the above models are also the pseudo-energies of a multispin
U(1) symmetric XY model~\citep{alcaraz-etal-prb21,alcaraz-pimenta-pre21}.
Additional developments related to Fendley model can be found in Refs.~\citep{elman-champan-flammia-cmp21,miao-sp22,alcaraz-pimenta-sirker-prb23,chapman2023unified}.

Recently, it was shown that, in general, the roots of the characteristic
polynomial yielding the free fermionic spectrum can be efficiently
obtained numerically~\citep{alcaraz-etal-prb21}. As a consequence,
the finite-size gaps of the spin system can be computed straightforwardly
for quite large system sizes. Furthermore, the numerical cost for
exactly computing the finite-size gap with machine precision (and,
hence, the associated dynamical critical exponent) is minimum at criticality
increasing only linearly with the chain size. While this may seems
innocuous for translation-invariant systems where analytical results
can often be obtained, it is of great applicability to quenched disordered
systems where analytical results are scarce and exactly numerical
results are plagued by numerical instabilities inherent to the extreme
slow critical dynamics of these systems~\citep{getelina-hoyos-ejpb20}.
This brings us to the topic of our research. What are the effects
of quenched disorder on these generalized free fermionic systems?

It is well known that the effects of quenched disorder (i.e., static
random inhomogeneities) in strongly interacting systems can lead to
interesting new phenomena. For instance, even the small amount of
inhomogeneities can change the singular behavior of a critical system~\citep{harris-jpc74,vojta-hoyos-prl14},
change the sharp character of a transition by smearing~\citep{vojta-prl03,hoyos-vojta-prl08,hoyos-vojta-prb12},
or even destroy the phase transition itself~\citep{imry-ma-prl75,imry-wortis-prb79}.
For reviews, see, e.g., Refs.~\citealp{vojta-review06,vojta-jltp-10}.

In the (free fermionic) transverse-field Ising chain, quenched disorder
completely changes the critical behavior of the clean (homogeneous)
system as expected from the Harris criterion~\citep{harris-jpc74}.
The conventional critical dynamical scaling of the clean system $\tau\sim\xi^{z}$
(with $\tau$ and $\xi$ being, respectively, time and length scales,
and $z=1$ being the dynamical exponent) is replaced by an activated
dynamical scaling $\ln\tau\sim\xi^{\psi}$ (with universal, i.e.,
disorder-independent, tunneling exponent $\psi=1/2$), which is now
recognized as a hallmark of the so-called infinite-randomness quantum
criticality~\citep{fisher92,fisher95}. Technically, this means that
the disorder-induced statistical fluctuations among the relevant energy
scales increases without bounds under the renormalization-group coarse-grain
procedure. Additionally, it was realized that this exotic phenomena
appears in all dimensions~\citep{senthil-sachdev-prl96,pich-etal-prl98,kovacs-igloi-prb11}
and in many other contexts, such as in Heisenberg-like quantum chains
and ladders~\citep{fisher94-xxz,boechat-etal-ssc96,monthus-s1-PRL,refael-fisher-s32,damle-huse-multicritical,yusuf-zigzag,hoyos-ladders,hoyos-sun,bonesteel-yang-prl07,hrahsheh-hoyos-vojta-prb12,quito-hoyos-miranda-prl15,quito-hoyos-miranda-prb16,shu-etal-prb16,quito-etal-prb19,quito-etal-epjb20},
in the Hubbard chain~\citep{melin-igloi-prb06,yu-arxiv22}, in aperiodic
quantum chains~\citep{vieira-aperiodic-PRL,vieira-aperiodic-PRB,fleury-etal-jsm12,casa-grande-etal-prb14},
in open quantum rotors systems~\citep{hoyos-kotabage-vojta-prl07,vojta-kotabage-hoyos-prb09},
in reaction-diffusion models such as the contact process~\citep{hooyberghs-prl,hooyberghs-pre,vojta-dickison-nonequ,hoyos-pre08,vojta-pre12,vojta-hoyos-epl15,fiore-etal-pre18},
in unbiased random walkers in random media~\citep{fisher-etal-prl98},
and in Floquet criticality~\citep{monthus-jsm-17,berdanier-etal-pnas18}.
For a review, see, e.g., Refs.~\citealp{igloi-review,igloi-monthus-review2}.
Despite these overwhelming situations in which infinite-randomness
is theoretically found, experimental evidence is slight~\citep{guo-etal-prl08,ubaidkassis-vojta-schroeder-prl10,demko-etal-prl12,xing-etal-science15,lewellyn-etal-prb19}.

Despite all the progress on characterizing the infinite-randomness
criticality, we currently do not know what are the necessary conditions
ensuring its appearance.\footnote{Our best educated guess comes from the classification of the Griffiths
singularities near the transition point~\citep{vojta-schmalian-prb05,vojta-hoyos-prl14}.
Due to statistical fluctuations inherent of quenched disorder systems,
there may be arbitrarily large regions in space which are locally
ordered and virtually disconnected from the bulk (the so-called rare
regions). If there rare regions are in their lower critical dimension
and that $\min\left\{ d,d_{c}^{+}\right\} \nu<2$, where $d$ is the
dimension of the system, $d_{c}^{+}$ is the upper critical dimension
of the problem, and $\nu$ is the correlation-length critical exponent
of the clean theory, then, very likely, infinite-randomness is expected.
However, this criterion can only be regarded as a sufficient one,
not a necessary one because infinite-randomness occurs in aperiodic
systems which do not have rare regions or Griffiths singularities.
Thus, it is desirable to further study systems exhibiting infinite-randomness
criticality to better understand the fundamental ingredients yielding
it.} It is then desirable to further study other aspects which were not
considered in the previous studies. Here, we consider interactions
involving more than the conventional two-body interactions. We pay
special attention to the $\text{Z}_{2}$-symmetric free fermionic
case in which the interactions involve three consecutive spins~\citep{fendley-jpa19}.
The clean phase diagram has three critical lines separating three
gapful phases which are related to each other by triality.\footnote{This is a generalization of the duality as happens in the $\text{Z}_{2}$-symmetric
transverse-field Ising chain. The model is self-dual if under the
duality transformation the ferromagnetic and the paramagnetic phase
are interchanged.} The universality classes of the associated transitions are that of
the transverse-field Ising chain, but that of the multicritical point
(where these three critical lines meet) is yet to be determined. Currently,
only its dynamical critical exponent $z=\frac{3}{2}$~\citep{fendley-jpa19}
and the specific-heat exponent $\alpha=0$~\citep{alcaraz-pimenta-prb20b}
are known. An inhomogeneous quantum Ising chain sharing the same energy
spectrum of this multicritical point was introduced in Ref.~\citealp{alcaraz-pimenta-sirker-prb23}.
In this related Ising chaing the order-parameter exponent is $\beta=\frac{1}{8}$
like the standard Ising chain.

Similarly to the random transverse-field Ising chain and to the spin-1/2
XX chain with random couplings, we show in this paper that generic
quenched disorder stabilizes quantum Griffiths phases surrounding
the three transition lines. In addition, the corresponding universality
classes of these transitions and that of the multicritical point are
the same and are of infinite-randomness type (with disorder-independent
tunneling exponent $\psi=\frac{1}{2}$). Our results are based on
a generalization of the strong-disorder renormalization-group (SDRG)
method~\citep{MDH-PRL,MDH-PRB,bhatt-lee} and on exact numerical
calculations of the spectral gap using the aforementioned method based
on the characteristic polynomial that allows us to obtain exact numerical
results for very large lattice sizes~\citep{alcaraz-etal-prb21}.
After unveiling the structure of the SDRG flow, we generalize our
results to the case of ($p+1$)-multispin interactions ($p=1,2,3,\dots$)
and arrive basically at the same conclusions. All the transitions
are in the same universality class of infinite-randomness type with
the same tunneling exponent $\psi=\frac{1}{2}$. 

In addition, we have also studied the case in which disorder couples
only to one type of coupling constants (the others remaining homogeneous).
In that case, some of the clean transition lines remain stable for
weak disorder strength while a line of finite-disorder fixed points
(with nonuniversal dynamical critical exponents $z)$ appear for large
disorder strengths, and terminates in an infinite-randomness multicritical
point.

This paper is organized as follows. In Sec.~\ref{sec:The-model}
we define the model studied and review some key results important
to our purposes. In Sec.~\ref{sec:Overview} we overview the expected
effects of quenched disorder in a heuristic way. Our arguments are
mostly based on the effects caused by rare-regions in the near-critical
Griffiths phases. In Sec.~\ref{sec:SDRG} we review the SDRG method
for the standard 2-spin interacting case and generalize it to the
3-spin interacting case. In Sec.~\ref{sec:FS-gap} we report our
numerical study of the finite-size gap statistics of the model which
are in agreement with the SDRG results. We present further discussions
and concluding remarks in Sec.~\ref{sec:Conclusions}. Finally, the
details of the renormalization-group flow are presented in Appendices
\ref{sec:blockSDRG} and \ref{sec:SDRG-flow-Eq}. 

\section{The model and review of key results\label{sec:The-model}}

We consider the $\left(p+1\right)$-multispin interacting quantum
chains $\left(p=1,2,3,\dots\right)$, whose Hamiltonian, introduced
in Refs.~\citealp{alcaraz-pimenta-prb20a,alcaraz-pimenta-prb20b},
is given by 
\begin{eqnarray}
{\cal H}_{p} & = & -\sum_{i=1}^{L-p}\lambda_{i}\sigma_{i}^{x}\prod_{j=i+1}^{i+p}\sigma_{j}^{z}-\sum_{i=L-p+1}^{L}\lambda_{i}\sigma_{i}^{x}\prod_{j=i+1}^{L}\sigma_{j}^{z}\label{eq:Hp}\\
 & = & -\sum_{i=1}^{L}\lambda_{i}h_{i},\mbox{ where }h_{i}=-\sigma_{i}^{x}\prod_{j=i+1}^{\min\{i+p,L\}}\sigma_{j}^{z},
\end{eqnarray}
 $\sigma_{i}^{x,y}$ are Pauli matrices associated with spin-$\frac{1}{2}$
degrees of freedom at site $i$, and $L$ is the total number of $\left\{ h_{i}\right\} $
energy-density operators (which is also the total number of spins
in the chain). The case $p=1$ is equivalent (up to global degeneracies)
to the inhomogeneous transverse-field Ising quantum chain. The local
interaction operator $h_{i}$ involves $\min\left\{ p+1,i\right\} $
spins and satisfy the algebra (for $i\neq j$)
\begin{equation}
\left\{ h_{i},h_{j}\right\} =0,\mbox{ if }\left|i-j\right|\le p,\mbox{ and }\left[h_{i},h_{j}\right]=0\mbox{ otherwise}.\label{eq:algebrap}
\end{equation}
 In other words, $h_{i}$ and $h_{j}$ commute if they are farther
than $p$ lattice units ($\left|i-j\right|>p$), and anticommute otherwise.
Evidently, from Eq.~(\ref{eq:Hp}), $h_{i}^{2}=\mathds{1}$. Finally,
$\lambda_{i}$ is the local multispin energy coupling. In this work,
we introduce quenched disorder by considering $\left\{ \lambda_{i}\right\} $
as independent random variables. Their precise distribution will be
defined later.

Interestingly, it was shown~\citep{alcaraz-pimenta-prb20a,alcaraz-pimenta-prb20b}
that the spectrum of \eqref{eq:Hp} has the free fermionic form 
\begin{equation}
E^{\left\{ s_{k}\right\} }=-\sum_{k=1}^{\bar{L}}s_{k}\epsilon_{k},
\end{equation}
 where $s_{k}=\pm1$, and the free fermionic pseudo-energies $\epsilon_{k}=1/\sqrt{x_{k}}$,
with $\left\{ x_{k}\right\} $ being the roots of the polynomial of
degree $\bar{L}=\left\lfloor \frac{L+p}{p+1}\right\rfloor $ (with
$\left\lfloor x\right\rfloor $ being the integer part of $x$): 
\begin{equation}
P_{L}(x)=\sum_{\ell=0}^{\bar{L}}C_{L}\left(\ell\right)x^{\ell},\label{eq:polynomial}
\end{equation}
 whose coefficients $C_{L}\left(\ell\right)$ are obtained from the
recurrence relation
\begin{equation}
P_{L}(x)=P_{L-1}(x)-x\lambda_{L}^{2}P_{L-p-1}(x),\label{eq:recurrence-P}
\end{equation}
 with the initial condition $P_{j}(x)=1$ for $j\leq0$.

It is important to notice that the free fermionic character is guaranteed
when open boundary conditions are applied. For other boundary conditions,
very likely this is not true~\citep{alcaraz-pimenta-prb20a,alcaraz-pimenta-prb20b},
and the solution of the model remains an open problem.

The first gap in the spectrum energy is $\Delta=2/\sqrt{x_{\text{max}}}$,
where $x_{\text{max}}=\max\left\{ x_{k}\right\} $ is the largest
root of the polynomial \eqref{eq:polynomial}. At and near criticality,
it was recently shown~\citep{alcaraz-etal-prb21} that $x_{\text{max}}$
can be efficiently computed for very large chains even though $C_{L}\left(\ell\right)$
grows factorially with $\ell$. (Indeed, there is no need to compute
all $C_{L}\left(\ell\right)$.) This is accomplished when one uses
the Laguerre's upper bound (LB) for the roots of a polynomial 
\begin{equation}
x_{\text{LB}}=-\frac{\alpha_{1}}{\bar{L}}+\frac{\bar{L}-1}{\bar{L}}\sqrt{\alpha_{1}^{2}-2\left(\frac{\bar{L}}{\bar{L}-1}\right)\alpha_{2}},\label{eq:LB}
\end{equation}
 where 
\begin{equation}
\alpha_{1}=\frac{C_{L}(\bar{L}-1)}{C_{L}(\bar{L})},\mbox{ and }\alpha_{2}=\frac{C_{L}(\bar{L}-2)}{C_{L}(\bar{L})},\label{eq:y1y2}
\end{equation}
 as the starting initial guess for the $x_{\text{max}}$.

In the (critical) homogeneous case $\lambda_{i}=\lambda$, it was
shown~\citep{alcaraz-etal-prb21} that, for any $p$, the quantity
$\Delta_{\text{LB}}\equiv2/\sqrt{x_{\text{LB}}}=\left(1-\epsilon\right)\Delta$
as $L\rightarrow\infty$, with $0<\epsilon<1$ being a constant. Thus,
$\Delta_{\text{LB}}$ has the same finite-size scaling properties
of the finite-size gap $\Delta$ and, thus, can be used to obtain
the dynamical critical exponent $z$, i.e., 
\begin{equation}
\Delta_{\text{LB}}\sim L^{-z_{\text{LB}}},\label{eq:clean-scaling}
\end{equation}
 with $z_{\text{LB}}=z=\frac{p+1}{2}$~\citep{alcaraz-pimenta-prb20a,alcaraz-pimenta-prb20b}.
Numerically, this is convenient since $\Delta_{\text{LB}}$ can be
efficiently computed. 

For $p=1$, it was shown~\citep{alcaraz-etal-prb21} that, in the
critical quenched disordered case ($\left\{ \lambda_{i}\right\} $
being independent and identically distributed random variables), $\Delta_{\text{LB}}=\left(1-\epsilon_{L}\right)\Delta$
with $\epsilon_{L}$ vanishing slowly as $L\rightarrow\infty$. This
provides a convenient tool to obtain the dynamical scaling relation.
In this case, 
\begin{equation}
\ln\Delta_{\text{LB}}\sim-L^{\psi},\label{eq:dirty-scaling}
\end{equation}
 with universal tunneling exponent $\psi=1/2$. By universal we mean
that $\psi$ does not depend on the particular distribution of the
disorder variables.\footnote{Provided that the distribution is not pathological or extremely broad~\citep{fisher94-xxz,karevski-etal-epjb01,krishna-bhatt-prb20,krishna-bhatt-ap21},
which is the case for most physically relevant distributions.} This result is expected since the model Hamiltonian \eqref{eq:Hp}
for $p=1$, apart from global degeneracies, has the same spectrum
as the transverse-field Ising chain. It is worthy noting that the
activated dynamical scaling \eqref{eq:dirty-scaling} has been confirmed
by many analytical and numerical studies~\citep{igloi-review}. In
addition, $\Delta_{\text{LB}}$ can also be used to study the finite-size
gap in the near critical Griffiths phase. In this phase, the system
is gapless even though it is short-range correlated (finite spin-spin
correlation length)~\citep{fisher95}. The finite-size gap $\Delta$
and the associated LB estimate, $\Delta_{\text{LB}}$, obeys the power-law
scaling \eqref{eq:clean-scaling} with $z$ being the off-critical
(Griffiths) dynamical exponent which depends on the distance from
criticality. As criticality is approached, the off-critical dynamical
exponent increases and becomes infinite at the critical point, in
accordance to the activated critical dynamical scaling \eqref{eq:dirty-scaling}.

\begin{figure}[tb]
\begin{centering}
\includegraphics[clip,width=0.6\columnwidth]{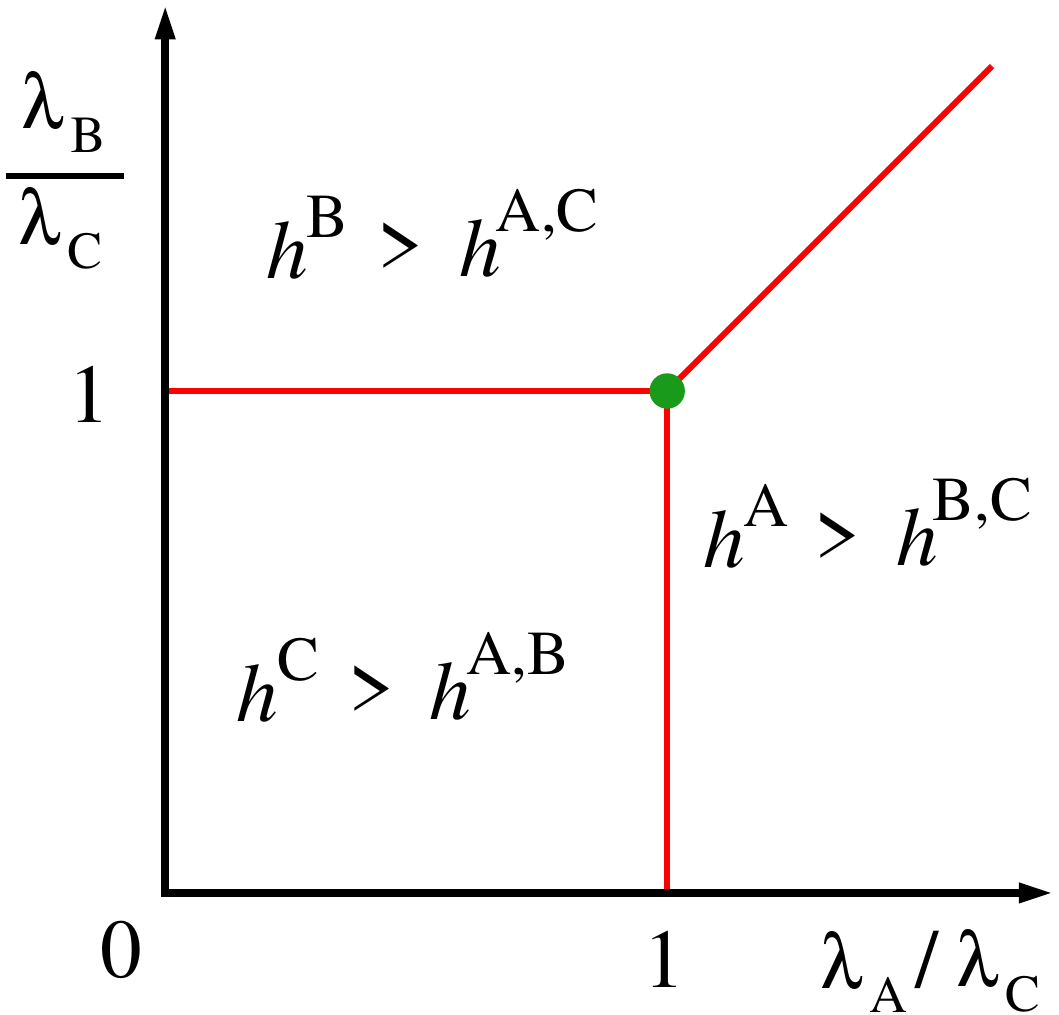}
\par\end{centering}
\caption{The phase diagram of the clean Hamiltonian \eqref{eq:Hp} for $p=2$.
Here, $\lambda_{3i-2}=\lambda_{A}$, $\lambda_{3i-1}=\lambda_{B}$,
and $\lambda_{3i}=\lambda_{C}$. The solid red dashed lines are transitions
in the Ising universality class ($z=1$). The multicritical point
$\lambda_{A}=\lambda_{B}=\lambda_{C}$ is in a different universality
class where $z=\frac{3}{2}$ and $\alpha=0$.\label{fig:PD-clean}
The three phases are gapful and are similar to each other (see text).}
\end{figure}

The model \eqref{eq:Hp} for $p=2$ and nondisordered couplings was
studied in Ref.~\citealp{fendley-jpa19} for couplings of period
$3$ {[}which is the natural period given by the algebra \eqref{eq:algebrap}{]},
i.e., 
\begin{equation}
\lambda_{3i-2}=\lambda_{A},\ \lambda_{3i-1}=\lambda_{B},\mbox{ and }\lambda_{3i}=\lambda_{C}.
\end{equation}
 Exploring the triality of the model, the phase diagram was determined
(see Fig.~\ref{fig:PD-clean}).

We now want to discuss what we mean by triality in this model. In
the bulk limit, the lattice translations $i\rightarrow i+1$ (equivalent
to $A\rightarrow B\rightarrow C\rightarrow A$) and $i\rightarrow i+2$
(equivalent to $A\rightarrow C\rightarrow B\rightarrow A$), and the
lattice reflection $\left(i\rightarrow L+1-i\right)$ do not change
the algebra (\ref{eq:algebrap}) and, therefore, the spectrum. Thus,
\begin{eqnarray}
{\cal H}\left(\lambda_{A},\lambda_{B},\lambda_{C}\right) & = & {\cal H}\left(\lambda_{B},\lambda_{C},\lambda_{A}\right)={\cal H}\left(\lambda_{C},\lambda_{A},\lambda_{B}\right)\nonumber \\
 & = & {\cal H}\left(\lambda_{C},\lambda_{B},\lambda_{A}\right).\label{eq:triality}
\end{eqnarray}

Now, consider the case $\alpha=\lambda_{A}/\lambda_{C}<1$ fixed.
If the transition happens at $\beta=\lambda_{B}/\lambda_{C}\neq1$,
the relation (\ref{eq:triality}) implies the existence of another
transition at $\beta^{-1}$ for the same $\alpha$. Since there is
only a global $\text{Z}_{2}$ symmetry to be broken, then there is
only a single phase transition. Therefore, the transition must happen
at $\beta_{c}=\beta_{c}^{-1}=1$. Successive applications of the relation
(\ref{eq:triality}) imply that the self-triality curves (red lines
in Fig.~\ref{fig:PD-clean}) are the transition lines. Notice that
this argument can be generalized to higher values of $p$ where the
self-$\left(p+1\right)$-ality hyperplanes are the transition manyfolds.
For $p=3$, see Ref.~\citealp{elman-champan-flammia-cmp21}. 

The three different phases are characterized by the following expectation
values: $h^{A}=\left|\sum_{i}\left\langle h_{3i-2}\right\rangle \right|$,
$h^{B}=\left|\sum_{i}\left\langle h_{3i-1}\right\rangle \right|$,
and $h^{C}=\left|\sum_{i}\left\langle h_{3i}\right\rangle \right|$.
For $\lambda_{A,B}<\lambda_{C}$, the system is in a phase where $h^{C}>h^{A,B}$.
By symmetry or, more precisely, by triality, there are other two phases
in which $h^{A}>h^{B,C}$ and $h^{B}>h^{A,C}$ which happen when $\lambda_{C,B}<\lambda_{A}$
and $\lambda_{A,C}<\lambda_{B}$, respectively. There are three phase
transition boundaries: $\lambda_{A}<\lambda_{B}=\lambda_{C}$, $\lambda_{C}<\lambda_{A}=\lambda_{B}$,
and $\lambda_{B}<\lambda_{C}=\lambda_{A}$. All of them are in the
2D Ising universality class, and, thus, the dynamical exponent is
$z=1$. The multicritical point $\lambda_{A}=\lambda_{B}=\lambda_{C}$
is, on the other hand, in a different universality class where $z=\frac{3}{2}$~\citep{fendley-jpa19}
and the specific-heat exponent $\alpha=0$~\citealp{alcaraz-pimenta-prb20b}. 

The purpose of present work is the study of the quenched disorder
effects in the phase transitions of the model Hamiltonian \eqref{eq:Hp}
for $p\ge2$.

As a final remark motivating this work, we recall that multibody interactions
are not an uncommon feature in condensed-matter physics. It naturally
arises in Mott insulators and, specifically, is a key ingredient for
stabilizing chiral spin liquids~\citep{wen-wilczek-zee-prb89,motrunich-prb06}.
In addition to that, transitional metal oxides naturally have four-body
interactions as described by the Kugel\textendash Khomskii model~\citep{kugel-khomskii-review82}.
Multispin interactions also appear in quantum spin chains due to another
reason: the mapping between different models. For instance, there
is an equivalence of the Ising model with two- and three-spin interactions
with the four-state Potts model~\citep{blote-jpa87}. In addition,
it is well known a mapping between the quantum Ashkin-Teller model
in 1D, which has four-spin interactions, and the XXZ spin-1/2 chain,
which has only conventional two-spin interactions~\citep{abb}. Finally,
the multispin interactions should not be seen as the main ingredient
for novel physics. The main ingredient is the algebra of the local
Hamiltonian operators (\ref{eq:algebrap}) that solely determines
the free fermionic spectrum. The representation of this algebra in
our work involves multispin interactions. But this is not a necessary
condition.

\section{Overview of the effects of quenched disorder\label{sec:Overview}}

In this work, we study the effects of quenched disorder on the Hamiltonian
\eqref{eq:Hp}, paying special attention to the first nontrivial case
$p=2$. We inquire how disorder on the coupling constants changes
the clean phase diagram Fig.~\ref{fig:PD-clean} as well as the universality
classes of the transitions. 

We build our arguments taking as the starting point the physical behavior
of the clean system~\citep{fendley-jpa19} revised in Sec.~\ref{sec:The-model}.
For this sake, we assume that the set of couplings $\left\{ \lambda_{A,i}\right\} \equiv\left\{ \lambda_{3i-2}\right\} $,
$\left\{ \lambda_{B,i}\right\} \equiv\left\{ \lambda_{3i-1}\right\} $,
and $\left\{ \lambda_{C,i}\right\} \equiv\left\{ \lambda_{3i}\right\} $
are random variables respectively distributed according to the probability
distributions ${\cal P}_{A}(\lambda)$, ${\cal P}_{B}(\lambda)$ and
${\cal P}_{C}(\lambda)$.

\subsection{The simpler case of a vanishing coupling $\lambda_{A,i}$}

When one of the couplings is vanishing, say $\lambda_{A,i}=0$, the
effective algebra (\ref{eq:algebrap}) is that of the model with $p=1$
{[}see, also, Eq.~(\ref{eq:recurrence-P}){]}, which corresponds
to the standard algebra of the transverse-field Ising chain. In that
case, the phase diagram is that of the random transverse-field Ising
chain which is well known~\citep{fisher95,hoyos-etal-epl11}. The
transition takes place when the typical values (geometric mean) of
the remaining couplings equal each other~\citep{pfeuty-pla79}, i.e.,
the system is critical when $\overline{\delta}=0$, where $\overline{\delta}=\overline{\delta_{i}}$,
with $\overline{\cdots}$ denoting the disorder average and 
\begin{equation}
\delta_{i}\equiv\ln\lambda_{B,i}-\ln\lambda_{C,i}.\label{eq:deltai}
\end{equation}
 For $\overline{\delta}>0$ ($\overline{\delta}<0$), the system is
in the $B$- ($C$-)phase.

\subsubsection{Uncorrelated disorder}

According to the Harris' criterion~\citep{harris-jpc74}, uncorrelated
quenched disorder is a relevant perturbation at $\overline{\delta}=0$
(since $d\nu=1<2$, with $d=1$ being the number of spatial dimensions
in which disorder is uncorrelated and $\nu=1$ being the correlation
length exponent of the clean theory) and, thus, the universality class
of the transition must change. As shown by Fisher~\citep{fisher92,fisher95},
the universality class is of infinite-randomness type with activated
dynamical scaling \eqref{eq:dirty-scaling}. 

In addition, surrounding this exotic quantum critical point, there
are Griffiths phases whose spectral gap vanishes and the spin-spin
correlations are short-ranged. The off-critical dynamical scaling
is a power-law (\ref{eq:clean-scaling}) with effective dynamical
exponent $z\propto\overline{\delta}^{-1}$ diverging as the system
approaches criticality.

\subsubsection{Appearance of locally correlated disorder\label{subsec:correlated-disorder}}

On the other hand for $\lambda_{B,i}=e^{\epsilon}\lambda_{C,i}$ with
$\epsilon$ being a constant, i.e., for perfectly correlation between
the local random variables in lattices B and C, the Harris criterion
has to be applied with caution. This is because the local distance
from criticality $\delta_{i}=\epsilon$ is uniform throughout the
chain. It turns out that disorder is an irrelevant perturbation~\citep{hoyos-etal-epl11}.
The clean critical behavior is stable up to some critical disorder
strength, beyond which it changes to a finite-randomness critical
behavior, where the dynamical critical scaling is a conventional power-law
\eqref{eq:clean-scaling} but with a nonuniversal critical dynamical
exponent $z$, i.e., it depends on the disorder strength~\citep{getelina-etal-prb16,getelina-hoyos-ejpb20}.
In addition, no Griffiths effects exist in this case.

In this work, we do not consider explicitly the case where $\lambda_{B,i}$
and $\lambda_{C,i}$ are locally correlated. However, we do consider
the case in which both couplings are uniform ($\lambda_{B,i}=\lambda_{B}$
and $\lambda_{C,i}=\lambda_{C}$) and the third one ($\lambda_{A,i}$)
is random. If the typical value of $\lambda_{A,i}$ is sufficiently
small, we show that quenched disorder is perturbatively irrelevant
in the renormalization-group (RG) sense. Thus, disorder can be simply
ignored, and the transition at $\lambda_{B}=\lambda_{C}$ is in the
universality class of the clean system. As we increase the values
of $\lambda_{A,i}$'s beyond the perturbative regime, disorder induces
randomness in the renormalized couplings $\tilde{\lambda}_{B,C}$.
Interestingly, the induced disorder has a perfect correlation, namely,
$\tilde{\lambda}_{B,i}=\tilde{\lambda}_{C,i}$. Thus, the long-distance
physics of locally correlated random couplings naturally appears in
this model. Consequently, a finite-randomness critical point governs
the transition for sufficiently large $\lambda_{A,i}$'s.

\subsection{The boundary phases: the case of small $\lambda_{A,i}$ couplings}

\begin{figure}[tb]
\begin{centering}
\includegraphics[clip,width=0.6\columnwidth]{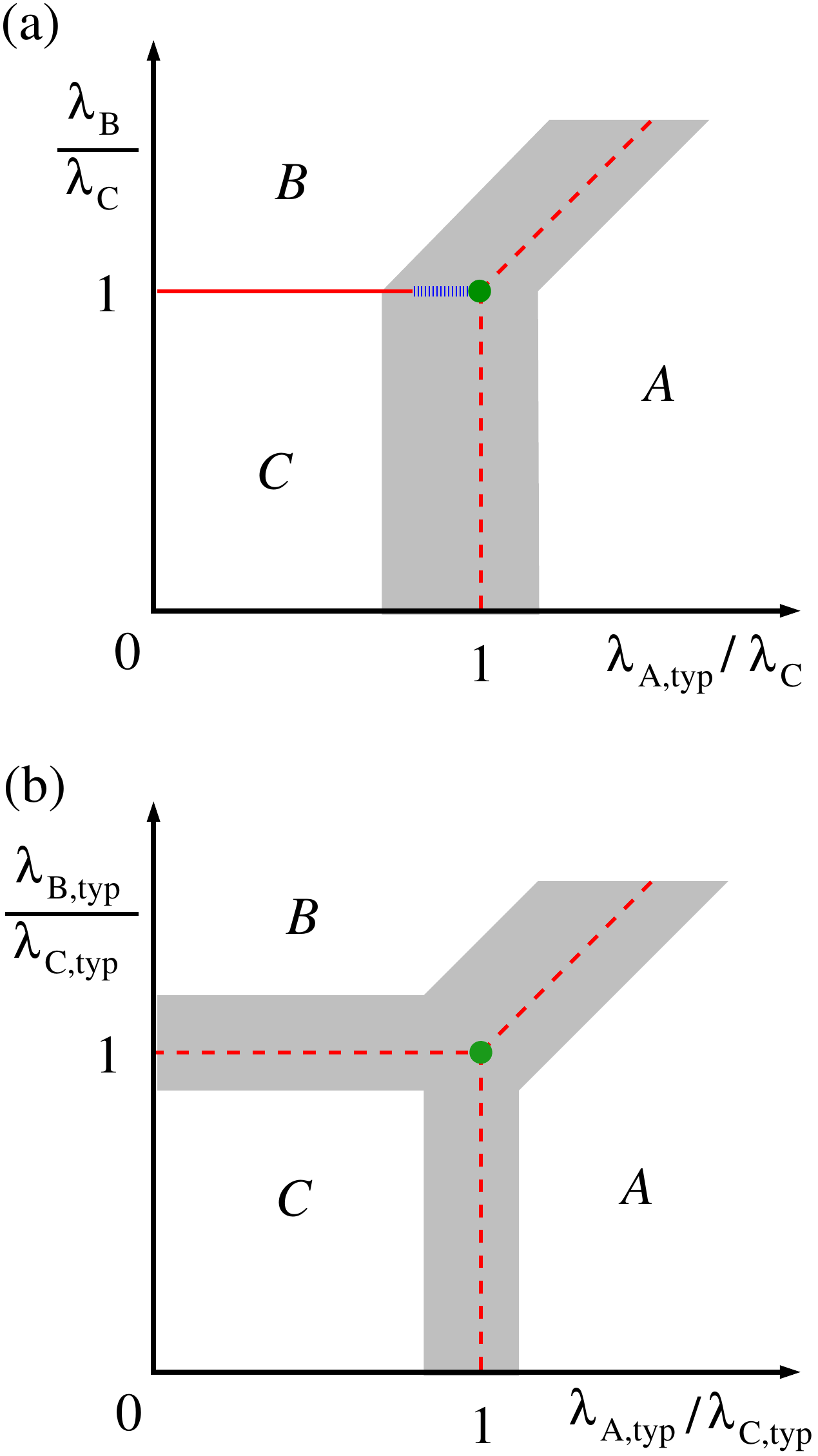}
\par\end{centering}
\caption{The phase diagram of the Hamiltonian \eqref{eq:Hp} for $p=2$. In
panel (a), $\left\{ \lambda_{3i-2}\right\} =\left\{ \lambda_{A,i}\right\} $
is a set of independent random variables while the remaining couplings
$\lambda_{3i-1}=\lambda_{B}$ and $\lambda_{3i}=\lambda_{C}$ are
homogeneous. In panel (b), at least two of the coupling constants
are independent random variables (see text). The solid red line is
a transition line in the Ising universality class of the clean system
{[}$z=1$, see Eq.~(\ref{eq:clean-scaling}){]}. The red dashed lines
are transitions in the infinite-randomness universality class {[}$\psi=1/2$,
see Eq.~(\ref{eq:dirty-scaling}){]}. The multicritical point (in
both cases) is also in the same infinite-randomness universality class.
The blue dotted line in panel (a) is a transition line where the universality
class is of finite-randomness type ($1<z<\infty)$. The phases have
the same nature as the homogeneous case (see Fig.~\ref{fig:PD-clean})
and the shaded regions delimits the associated Griffiths phases where
the spectrum gap vanishes.\label{fig:PD-dirty}}
\end{figure}

What are the effects of small coupling $\lambda_{A,i}$? In the clean
case, due to triality, small $\lambda_{A}$ cannot shift the location
of the critical point $\lambda_{B}=\lambda_{C}$ (see Fig.~\ref{fig:PD-clean})
(see Ref.~\citealp{fendley-jpa19} for another argument). In the
disordered case, we numerically show (see Sec.~\ref{sec:FS-gap})
that this remains true, i.e., the critical point remains at $\overline{\delta}=\overline{\ln\lambda_{B,i}}-\overline{\ln\lambda_{C,i}}=0$
provided that $\overline{\ln\lambda_{A,i}}<\overline{\ln\lambda_{B,i}}$.
Thus, the phase transition lines of the phase diagram, in the random
system, are equal to those of the clean system with $\lambda_{A}$,
$\lambda_{B}$, and $\lambda_{C}$ replaced by their typical values
$\lambda_{A,\text{typ}}$, $\lambda_{B,\text{typ}}$, and $\lambda_{C,\text{typ}}$
as sketched in Fig.~\ref{fig:PD-dirty}.

How about the universality classes of the transitions? Here, we consider
two cases: two competing (strongest) couplings (a) not generating
random mass, and (b) do generating random mass $\delta_{i}$, see
Eq.~(\ref{eq:deltai}).

In case (a) the two strongest couplings are homogeneous ($\lambda_{B,i}=\lambda_{B}$
and $\lambda_{C,i}=\lambda_{C}$) and $\lambda_{A,i}$ is typically
much smaller than $\lambda_{B,C}$, the transition is in the universality
class of the clean system (Ising) as shown by the solid red line in
Fig.~\hyperref[fig:PD-dirty]{\ref{fig:PD-dirty}(a)}. When approaching
the multicritical point, however, the weak disordered couplings $\left\{ \lambda_{A,i}\right\} $
become nonperturbative and a line of finite-randomness fixed points
emerges (as previously mentioned). The resulting universality class
of the transition {[}blue dotted line in Fig.~\hyperref[fig:PD-dirty]{\ref{fig:PD-dirty}(a)}{]}
has critical dynamical exponent $z$ larger than the unity and diverges
as the infinite-randomness multicritical point is reached.

In case (b) the two competing couplings generate a random mass ($\overline{\delta^{2}}-\overline{\delta}^{2}\neq0$).
This happens whenever either one or both couplings are independent
random variables. The clean (Ising) universality class in unstable
since the Harris criterion is violated. The resulting universality
class is the one of the random transverse-field Ising chain with activated
dynamical scaling \eqref{eq:dirty-scaling}, and the associated phase
boundaries are the dashed lines in the phase diagram of Fig.~\ref{fig:PD-dirty}.

\subsection{The multicritical point: the case of strong $\lambda_{A,i}$ couplings}

What is the change of the clean multicritical point in the presence
of disorder? We cannot apply the Harris criterion since the clean
correlation length exponent $\nu$ is not known. To answer this question,
we develop an appropriate strong-disorder renormalization-group technique
(see Sec.~\ref{sec:SDRG}). Our results strongly indicate that quenched
disorder is a relevant perturbation. In addition, we show that the
resulting universality class is of infinite-randomness type with activated
dynamics \eqref{eq:dirty-scaling} and universal tunneling exponent
$\psi=\frac{1}{2}$. We have also confirmed these results by numerically
studying the finite-size gap of the system (see Sec.~\ref{sec:FS-gap}).

\subsection{The presence of quantum Griffiths phases}

We now inquire about the off-critical properties. Are they affected
by quenched disorder? The nature of the phases do not change since
disorder on the coupling constants does not break neither a symmetry
of the Hamiltonian nor a symmetry of the order-parameter field, i.e.,
disorder does not couple directly to the order-parameter field in
the associated underlying field theory. Thus, the phase diagram has
the same phases as the clean one. However, near the transition lines,
random mass induces Griffiths phases. In those regions of the phase
diagram (shaded areas in Fig.~\ref{fig:PD-dirty}), the spectral
gap vanishes and the correlations remain short-ranged. The finite-size
gap scaling is of power-law type \eqref{eq:clean-scaling} with nonuniversal
(i.e., disorder-dependent) effective dynamical exponent $z$.

Here, the Griffiths phases can be understood through the lenses of
the so-called rare regions (RRs): large and rare spatial regions in
a phase locally different from the bulk. For definiteness, consider
the case in which $\lambda_{C,i}=\lambda_{C}$ (i.e., uniform) and
$\lambda_{A,i}$'s are distributed between $\lambda_{A,\text{min}}$
and $\lambda_{A,\text{max}}$, with $0<\lambda_{A,\text{min}}<\lambda_{A,\text{max}}$.
Evidently the typical value $\lambda_{A,\text{typ}}=\exp(\overline{\ln\lambda_{A}})$
is between $\lambda_{A,\text{min}}$ and $\lambda_{A,\text{max}}$.
To start, let us analyze the phase transition between the $A$- ($h^{A}>h^{B,C}$)
and $C$-phases ($h^{C}>h^{A,B}$) where we can disregard the weak
coupling $\lambda_{B,i}$ (say, for simplicity, that $\max\{\lambda_{B,i}\}<\min\{\lambda_{A,\text{min}},\lambda_{C}\}$).
As previously discussed, the transition takes place when $\lambda_{C}=\lambda_{A,\text{typ}}$.
When $\lambda_{C}\gg\lambda_{A,\text{max}}$, the system is deep in
the homogeneous $C$-phase. Its properties are just the one of the
clean systems with the random couplings $\lambda_{A(B),i}$ replaced
by their typical value $\lambda_{A(B),\text{typ}}$. Importantly,
the spectral gap is finite. 

When $\lambda_{A,\text{typ}}<\lambda_{C}<\lambda_{A,\text{max}}$,
on the other hand, there are RRs where the local couplings $\lambda_{A,i}$
are typically greater than $\lambda_{C}$. Being locally in the $A$
phase, they endow the system a high $A$-phase susceptibility. The
spin in the domain walls between the $A$- and $C$-phases can be
arbitrarily weakly coupled and are responsible for the low-lying excitations
closing the spectral gap. As neither the bulk nor the RRs are critical,
the corresponding correlation length is finite. By duality, an analogous
Griffiths phase appears when $\lambda_{A,\text{min}}<\lambda_{C}<\lambda_{A,\text{typ}}$. 

Interestingly, there is a simple quantitative argument providing the
closing of the spectral gap in the Griffiths phase. Consider a RR
of size $L_{\text{RR}}$. The effective interaction between the domain
wall spins are thus of order $J_{\text{DW}}\sim e^{-L_{\text{RR}}/\xi_{\text{RR}}}$,
where $\xi_{\text{RR}}$ is the corresponding correlation length in
that particular RR. For simplicity, we will consider $\xi_{\text{RR}}=\xi$
to be RR-independent (a more precise treatment can be found in Ref.~\citealp{vojta-hoyos-prl14}).
The reason for $J_{\text{DW}}$ being exponentially small is because
the RR itself does not harbor Goldstone modes since the symmetry of
the Hamiltonian interactions is discrete. The system low-energy density
of states $\rho_{\text{DOS}}$ is dominated by the excitations of
the weakly coupled domain walls. Ignoring the even weaker coupling
to other domain wall spins belonging to other RRs, then $\rho_{\text{DOS}}\left(\omega\right)\sim\int dL_{\text{RR}}w_{\text{RR}}\left(L_{\text{RR}}\right)\delta\left(\omega-J_{\text{DW}}\right)$.
Here, we simply sum of all possible RRs weighting their contribution
by their existence probability $w_{\text{RR}}\sim e^{-L_{\text{RR}}/\ell}$,
with $\ell\propto-1/\ln p$, and $p$ being the probability of $\lambda_{A,i}$
being greater than $\lambda_{C}$. Notice that the probability of
finding a RR decreases exponentially with its volume, and $\ell$
is a constant that depends on the distribution's details of the coupling
constants. Consequently, one finds that $\rho_{\text{DOS}}\sim\omega^{-1+1/z}$,
with dynamical Griffiths exponent $z=\ell/\xi$. Notice the absence
of a gap or a pseudogap in $\rho_{\text{DOS}}$. Actually, there is
a divergence in the low-energy density of states when $z>1$ and $\omega\rightarrow0$.
We see that $z$ diverges $\sim1/\overline{\delta}$ when approaching
the transition.

By triality, the resulting Griffiths phases are those sketched in
Fig.~\hyperref[fig:PD-dirty]{\ref{fig:PD-dirty}(b)} if the $B$-couplings
are also randomly distributed between $\lambda_{B,\text{min}}$ and
$\lambda_{B,\text{max}}$ ($0<\lambda_{B,\text{min}}<\lambda_{B,\text{max}}$). 

Finally, near the multicritical point there are Griffiths phases with
RRs locally belonging to, say, either $A$-phase or $B$-phase, while
the bulk is in the $C$-phase. A similar feature also appeared in
the Griffiths phase of quantum Ashikin-Teller chain~\citep{hrahsheh-etal-prb14,barghathi-etal-PS15}.
In those cases, the effective dynamical exponent $z=\text{max}\{z_{A},z_{B}\}$,
where $z_{A(B)}$ is the dynamical exponent provided by the Griffiths
singularities of the $A$- ($B$-)RRs.

\subsection{The absence of Griffiths phases}

If, on the other hand, the $B$-couplings are also homogeneous ($\lambda_{B,i}=\lambda_{B}$),
the resulting transition between the $B$- and $C$-phases is the
one of the clean transverse-field Ising chain for sufficiently weak
$\lambda_{A,\text{max}}$ (as previously discussed). In addition,
there is no associated Griffiths phase since there are no RRs ($\lambda_{A,\text{max}}<\lambda_{B}=\lambda_{C}$)
as sketched in Fig.~\hyperref[fig:PD-dirty]{\ref{fig:PD-dirty}(a)}.
However, when approaching the multicritical point, RRs in the $A$-phase
appear and enhance the low-energy density of states. As a result,
the gap closes around the transition. At criticality, those $A$-RRs
can even provide a larger dynamical exponent $z$. In that case, the
clean critical point is replaced by a line of finite-disorder critical
points {[}dotted boundary line in Fig.~\hyperref[fig:PD-dirty]{\ref{fig:PD-dirty}(a)}{]}.
Finally, at the multicritical point, the approximation of weak $A$-couplings
completely breaks down and the most general theory contains a random-mass
term. Therefore, this critical point is of infinite-randomness type.

\section{The strong-disorder renormalization-group method\label{sec:SDRG}}

In this section, we develop a strong-disorder renormalization-group
(SDRG) method suitable for studying the long-distance physics of the
Hamiltonian \eqref{eq:Hp} for $p=1$ and $2$ and with random coupling
constants. It is an energy-based RG method where strongly coupled
degrees of freedom are locally decimated out hierarchically. Namely,
we search for the strongest coupled local degrees of freedom and freeze
them in their local ground state. The couplings between the remaining
degrees of freedom are renormalized perturbatively. This procedure
becomes more and more accurate if the local energy scales become more
and more disordered (broadly distributed). In that case, the perturbative
renormalization procedure becomes more accurate after each RG decimation
step. This method was originally devised to conventional spin-1/2
models~\citep{MDH-PRL,MDH-PRB,bhatt-lee} and later on generalized
to many other models. For a review, see Refs.~\citealp{igloi-review,igloi-monthus-review2}.

\subsection{Case $p=1$\label{subsec:p1}}

\subsubsection{The decimation procedure}

To start, let us consider the case $p=1$ in which the model Hamiltonian
\eqref{eq:Hp} simplifies to 
\begin{equation}
H=-\sum_{j}\lambda_{j}h_{j}=-\sum_{j}\lambda_{j}\sigma_{j}^{x}\sigma_{j+1}^{z}.\label{eq:Hp1}
\end{equation}
 For simplicity, we disregard the boundary conditions. Although (\ref{eq:Hp1})
and the random transverse-field Ising chain are distinct, they share
(apart of global degeneracies) the same eigenenergies. 

Following the SDRG philosophy, we search for the largest local energy
scale $\Omega=\max\left\{ \left|\lambda_{j}\right|\right\} $, say
$\left|\lambda_{2}\right|$. We then project the Hamiltonian onto
the low-energy sector of $H_{0}=-\lambda_{2}h_{2}=-\lambda_{2}\sigma_{2}^{x}\sigma_{3}^{z}$.
Denoting $\sigma_{i}^{z}\left|\uparrow_{i}\right\rangle =\left|\uparrow_{i}\right\rangle $,
$\sigma_{i}^{z}\left|\downarrow_{i}\right\rangle =-\left|\downarrow_{i}\right\rangle $,
$\sigma_{i}^{x}\left|\rightarrow_{i}\right\rangle =\left|\rightarrow_{i}\right\rangle $,
and $\sigma_{i}^{x}\left|\leftarrow_{i}\right\rangle =-\left|\leftarrow_{i}\right\rangle $,
then the ground-state sector of $H_{0}$ is spanned by $\left|\pm\right\rangle =(\left|\rightarrow_{2},\uparrow_{3}\right\rangle \pm\left|\leftarrow_{2},\downarrow_{3}\right\rangle )/\sqrt{2}$,
if $\lambda_{2}>0$, and $\left|\pm\right\rangle =(\left|\rightarrow_{2},\downarrow_{3}\right\rangle \pm\left|\leftarrow_{2},\uparrow_{3}\right\rangle )/\sqrt{2}$,
otherwise. As a result, the projection can be interpreted as a replacement
of spins $\boldsymbol{\sigma}_{2}$ and $\boldsymbol{\sigma}_{3}$
by an effective spin-1/2 degree of freedom $\tilde{\boldsymbol{\sigma}}$
which is defined by $\tilde{\sigma}^{z}\left|\tilde{\pm}\right\rangle =\pm\left|\tilde{\pm}\right\rangle $.
Notice in addition that $\left\langle \pm\left|h_{2}\right|\pm\right\rangle =\left\langle \pm\left|\sigma_{2}^{x}\sigma_{3}^{z}\right|\pm\right\rangle =\text{sign}\left(\lambda_{2}\right)\ne0$. 

The effective system Hamiltonian is obtained by treating $H_{1}=-\lambda_{1}h_{1}-\lambda_{3}h_{3}$
as a perturbation to $H_{0}$. In second order of perturbation theory,
we find that 
\begin{equation}
\tilde{H}_{1}=-\tilde{\lambda}\tilde{h}+\text{const},\mbox{ with }\tilde{\lambda}=\frac{\lambda_{1}\lambda_{3}}{\Omega}\label{eq:H1tilde}
\end{equation}
 and $\tilde{h}=\sigma_{1}^{x}\tilde{\sigma}^{z}\sigma_{4}^{z}$.
For our purposes, the constant term is harmless and can be disregarded.

Notice that $\tilde{h}$ is now a three-spin interaction. However,
since $\tilde{\boldsymbol{\sigma}}$ appears only in $\tilde{h}$,
it is a local gauge variable whose role is simply to double the degeneracy
of the spectrum. The SDRG decimation procedure \eqref{eq:H1tilde}
can be straightforwardly generalized to operators involving an arbitrary
number of ``internal'' degrees of freedom since the algebra \eqref{eq:algebrap}
is preserved.

Alternatively, the additional degeneracy induced by the effective
internal spin $\tilde{\boldsymbol{\sigma}}$ can be interpreted as
if the renormalized chain is, actually, two decoupled new chains.
In the first one, $\tilde{\boldsymbol{\sigma}}$ is fixed in state
$\left|\tilde{+}\right\rangle $ (the original spins $2$ and $3$
fixed in the ground state $\left|+\right\rangle $ of $H_{0}$) and
the corresponding renormalized Hamiltonian is simply $\tilde{H}_{1}=-\tilde{\lambda}\sigma_{1}^{x}\sigma_{4}^{z}$,
while in the second one $\tilde{\boldsymbol{\sigma}}$ is fixed at
$\left|\tilde{-}\right\rangle $ (the original spins frozen in the
state $\left|-\right\rangle $) and $\tilde{H}_{1}=+\tilde{\lambda}\sigma_{1}^{x}\sigma_{4}^{z}$.
These two chains have the same spectrum and the subsequent SDRG decimations
are identical (apart from the signs of the renormalized couplings
which, for our purposes, are not important).

Thus, $\tilde{h}$ can be simplified back to a two-spin interaction
at the expense of dealing with two ``twins'' renormalized chains.
The only difference between them being the sign of the renormalized
coupling constant $\tilde{\lambda}$.

To use this simplification and keep track of the degeneracies, we
are then required to introduce the quantity $g_{\Omega}$. It measures
the total number of gauge (extra) spin-1/2 degrees of freedom at the
energy scale $\Omega$. Clearly, after each decimation it renormalizes
to 
\begin{equation}
g_{\Omega}\rightarrow g_{\Omega}+1,\label{eq:g-tilde}
\end{equation}
 with the initial condition $g_{\Omega_{0}}=0$. The total number
of effective degrees of freedom in the chain $N_{\Omega}$ renormalizes
to 
\begin{equation}
N_{\Omega}\rightarrow N_{\Omega}-2,\label{eq:N-tilde}
\end{equation}
 with the initial condition $N_{\Omega_{0}}=L$. Notice that $N_{\Omega}+2g_{\Omega}=L$
is a constant throughout the RG flow. 

\begin{figure}[t]
\begin{centering}
\includegraphics[clip,width=1\columnwidth]{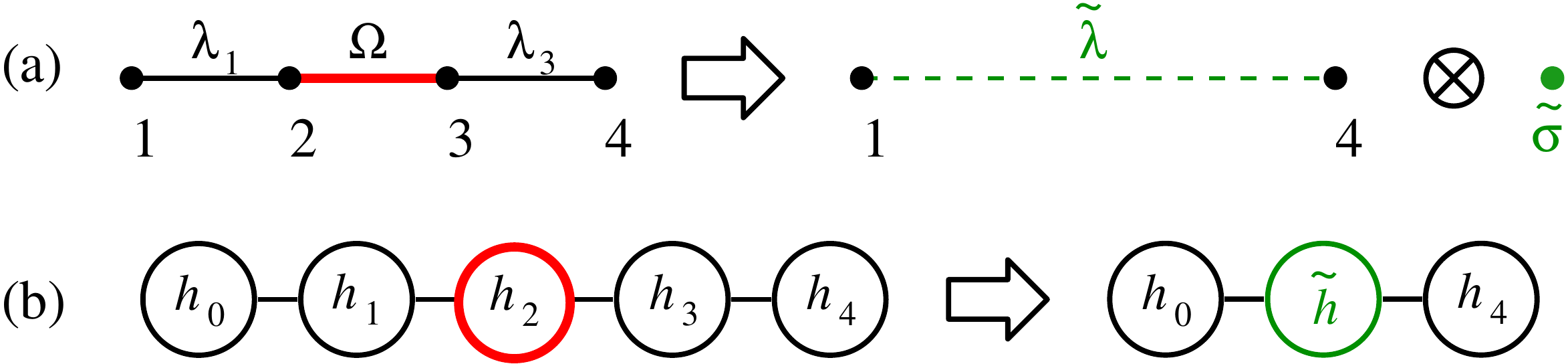}
\par\end{centering}
\caption{Decimation scheme for the Hamiltonian \eqref{eq:Hp} in the $p=1$
case. In (a), the decimation is sketched in the real space with points
and lines representing spin sites and coupling constants, respectively.
In (b), the decimation is sketched in the Hamiltonian space with circles
representing the local energy operators and the lines connecting anticommuting
operators.\label{fig:Decimation-p1}}
\end{figure}

In Fig.~\hyperref[fig:Decimation-p1]{\ref{fig:Decimation-p1}(a)}
we sketch the decimation procedure \eqref{eq:H1tilde}\textendash \eqref{eq:N-tilde}.
Regarding the local energy scales, the decimation procedure \eqref{eq:H1tilde}
is identical to that of the random spin-1/2 XX chain~\citep{fisher94-xxz,hyman-dimer}
and that of the random transverse-field Ising chain~\citep{fisher95}.
This is not a surprise since the free-particle spectra of all these
models are the same.

Finally, it is instructive to recast the decimation procedure in the
Hamiltonian space as shown in Fig.~\hyperref[fig:Decimation-p1]{\ref{fig:Decimation-p1}(b)}.
The $j$th circle represents the local energy operator $h_{j}$. A
line connecting different circles means that the sharing operators
anticommute with each other. Disconnected operators act on different
Hilbert spaces and, thus, trivially commute with each other. In the
decimation procedure, $h_{2}$ and the ``neighboring'' operators are
replaced by $\tilde{h}$ which anticommutes with the neighboring operators
$h_{0}$ and $h_{4}$. The algebra structure is, thus, preserved along
the SDRG flow.

\subsubsection{The SDRG flow}

Since the SDRG decimation rule \eqref{eq:H1tilde} is the same as
that for the spin-1/2 XX chain, the renormalization-group flow of
the coupling constants is already known~\citep{fisher94-xxz,fisher95,hyman-dimer}.
Let 

\begin{equation}
\delta\equiv\frac{\overline{\ln\lambda_{\text{odd}}}-\overline{\ln\lambda_{\text{even}}}}{\sigma_{\ln\lambda_{\text{odd}}}^{2}+\sigma_{\ln\lambda_{\text{even}}}^{2}},\label{eq:delta}
\end{equation}
 with $\sigma_{x}^{2}=\overline{x^{2}}-\overline{x}^{2}$ being the
variance of $x$. For $\delta\gg1$, the SDRG flow is towards a stable
fixed point in which only the odd couplings are decimated. This implies
that only the even couplings are renormalized and, thus, are much
smaller than the odd ones. This corresponds to a phase in which $\left|\left\langle h_{2i-1}\right\rangle \right|>\left|\left\langle h_{2i}\right\rangle \right|$.
In the spin-1/2 XX chain, this corresponds to the odd-dimer phase
where spin singlets are formed over the odd bonds, i.e., $\left|\left\langle \mathbf{S}_{2i-1}\cdot\mathbf{S}_{2i}\right\rangle \right|>\left|\left\langle \mathbf{S}_{2i}\cdot\mathbf{S}_{2i+1}\right\rangle \right|$.
The correspondence of this phase in the transverse-field Ising chain
is not so straightforward since the XX spin-1/2 chains maps into two
independent random transverse-field Ising chains~\citep{fisher94-xxz}.
In the first one, the odd couplings of our model play the role of
the transverse fields of the Ising chain. In the second, these roles
are exchanged. Thus, the phase $\left|\left\langle h_{2i-1}\right\rangle \right|>\left|\left\langle h_{2i}\right\rangle \right|$
corresponds to the paramagnetic (ferromagnetic) phase in the first
(second) quantum Ising chain.

If $0<\delta\ll1$, the system is in the associated Griffiths phase.
Typically, $\left|\left\langle h_{\text{odd}}\right\rangle \right|>\left|\left\langle h_{\text{even}}\right\rangle \right|$,
but there are some ``defects'' inside which $\left|\left\langle h_{\text{odd}}\right\rangle \right|<\left|\left\langle h_{\text{even}}\right\rangle \right|$.
These defects form the rare regions discussed in Sec~\ref{sec:Overview}.
Surrounding a rare region, there are two (domain-wall) spins weakly
coupled. As a result of their weak coupling, the typical and mean
values of the finite-size gap vanish $\sim L^{-z}$ {[}Eq.~\eqref{eq:clean-scaling}{]},
which defines an off-critical dynamical (Griffiths) exponent $z$.
As the critical point is approached, this exponent diverges as $z\sim\left|\delta\right|^{-1}$.

We now further discuss on the effects of the RRs in the Griffiths
phase through the lenses of the SDRG method. Recall that, in the transverse-field
Ising chain, the origin of the gapless modes in, say, the paramagnetic
phase is due the RRs which are locally in the ferromagnetic phase
and fluctuate between the two ferromagnetic states. As this is a coherent
tunneling process involving many spins, the associated relaxation
time increasing exponentially with the RR's volume. In the XX spin-1/2
chain, these RRs correspond to patches which are locally in the even-dimer
phase while the bulk is in the odd-dimer phase. The two domain walls
delimiting a RR are, in zeroth order of approximation, simply free
(unpaired) spins. To lowest nonvanishing order in perturbation theory,
these spins (say, at sites $1$ and $\ell$) actually interact via
an effective coupling constant equal to $\tilde{\lambda}_{\ell}=\lambda_{1}\lambda_{3}\dots\lambda_{\ell-1}/\lambda_{2}\lambda_{4}\dots\lambda_{\ell-2}$
{[}which could also be obtained by a successive iteration of Eq.~\eqref{eq:H1tilde}{]}.
Thus, the gap of a finite chain is simply the excitation energy of
the weakest coupled spins in these domain walls: $\min\{|\tilde{\lambda}_{\ell}|\}$.
Notice that, as expected, $\tilde{\lambda}_{\ell}$ vanishes exponentially
with the RR size implying an exponentially large relaxation time.
An analogous physical picture appears in our model. For simplicity,
consider a compact RR in which the local even couplings $\lambda_{2},\,\lambda_{4},\,\dots,\,\lambda_{\ell-2}$
($\ell$ even) are greater than the local odd couplings $\lambda_{1},\,\lambda_{3},\,\dots,\,\lambda_{\ell-1}$.
In that case, after decimating the even operators $h_{2i}$ in that
RR, an effective operator linking spins $1$ and $\ell$ appear $\tilde{h}_{1}=-\tilde{\lambda}_{\ell}\sigma_{1}^{x}\sigma_{\ell}^{z}$
with $\tilde{\lambda}_{\ell}=\lambda_{1}\lambda_{3}\dots\lambda_{\ell-1}/\lambda_{2}\lambda_{4}\dots\lambda_{\ell-2}$
(disregarding an unimportant sign). Thus, $\left\langle h_{2}\right\rangle =\left\langle h_{4}\right\rangle =\dots=\left\langle h_{2\ell-2}\right\rangle =\pm1$
and a longer correlation between spins $1$ and $\ell$ develop, $\left\langle \sigma_{1}^{x}\sigma_{\ell}^{z}\right\rangle \neq0$.
Consequently, a low-energy mode arises with excitation energy of order
$\tilde{\lambda}$. Evidently, by duality, there are analogous conventional
and Griffiths phases for $\delta<0$.

At criticality $\delta=0$, the fixed point is universal in the sense
that critical exponents do not depend on the details of the disorder
distributions. For instance, the finite-size gap distribution for
sufficiently large systems is~\citep{fisher-young-RTFIM} 
\begin{eqnarray}
{\cal P}_{\text{SDRG}}\left(\eta\right) & = & \frac{4}{\sqrt{\pi}}\sum_{k=0}^{\infty}\left(-1\right)^{k}\left(k+\frac{1}{2}\right)e^{-\eta^{2}\left(k+\frac{1}{2}\right)^{2}},\nonumber \\
 & = & \frac{4\pi}{\eta^{3}}\sum_{k=0}^{\infty}\left(-1\right)^{k}\left(k+\frac{1}{2}\right)e^{-\pi^{2}\left(k+\frac{1}{2}\right)^{2}/\eta^{2}},\quad\label{eq:PSDRG}
\end{eqnarray}
 where 
\begin{equation}
\eta=\frac{\ln\left(2\Omega_{0}/\Delta\right)}{\sigma_{0}\left(L/2\right)^{\psi}},\label{eq:eta}
\end{equation}
 $\sigma_{0}=\sqrt{\frac{1}{2}\sigma_{\ln\lambda_{\text{odd}}}^{2}+\frac{1}{2}\sigma_{\ln\lambda_{\text{even}}}^{2}}$,
$\Omega_{0}$ is the maximum value of $\lambda$ in the bare system,
$\Delta$ is the finite-size gap, and $\psi=1/2$ is the universal
tunneling exponent. The distribution ${\cal P}_{\text{SDRG}}$ is
$L$ independent for $L\gg\gamma_{D}^{-1}=\pi/8\sigma_{\ln\lambda}^{2}$,
the inverse of the Lyapunov exponent~\citep{mard-etal-prb14} which
plays the role of a clean-dirty crossover length~\citep{laflorencie-correlacao-PRB,wada-hoyos-prb22}.
The relation between length and energy scales follow from the scaling
variable in Eq.~\eqref{eq:eta}, from which follows the activated
dynamical scaling $\overline{\ln\Delta}\sim-L^{\psi}$ in Eq.~\eqref{eq:dirty-scaling}.

\subsubsection{Thermodynamics\label{subsec:Thermodynamics}}

The thermodynamic observables follow straightforwardly. For instance,
the low-temperature entropy is simply $S\sim\frac{1}{L}\left(N_{T}+g_{T}\right)\ln2$,
which simply counts the total number of active (undecimated) spins
at the energy scale $\Omega=T$. The reasoning is the following~\citep{bhatt-lee}.
At low temperatures $T\ll\Omega_{0}$, the distribution of effective
coupling constants is singular and, thus, the majority of the couplings
are much smaller than the maximum energy scale $\Omega=T$. In sum,
the active spins are essentially free. At criticality ($\delta=0$),
$N_{\Omega}\approx L/\left(1+\ln\left(\Omega_{0}/\Omega\right)/\sigma_{\ln\lambda}^{2}\right)^{1/\psi}$~\citep{igloi-juhasz-lajko-prl01,hoyosvieiralaflorenciemiranda}
and $g_{\Omega}=\frac{1}{2}\left(L-N_{\Omega}\right)$. Then, 
\begin{equation}
S=\frac{1}{2}\left(1+\left(\frac{\sigma_{\ln\lambda}^{2}}{\ln\left(\frac{\Omega_{0}}{T}\right)}\right)^{1/\psi}\right)\ln2.\label{eq:S}
\end{equation}
 Notice the residual zero-temperature entropy coming from the exponentially
large ground-state degeneracy. 

The specific heat follows from $C=T\partial S/\partial T$. In the
low-temperature limit, 
\begin{equation}
C\sim\ln^{-\left(1+1/\psi\right)}\left(\Omega_{0}/T\right),\label{eq:SpecificHeat}
\end{equation}
 which is similar to that of the random spin-1/2 XX chain.

\subsubsection{Ground-state degeneracy and the spectrum of other models}

It is interesting to further explore the connection between the $p=1$
model, the spin-1/2 XX chain, and the transverse-field Ising chain
in view of the SDRG decimation procedure.

Within the SDRG framework, one can obtain the whole spectrum of the
transverse-field Ising chain in the following way~\citep{pekker-etal-prx14}.
When performing the decimating procedure, one can either search for
the ground state (and then project onto ground state of the local
Hamiltonian) or, alternatively, search for the excited states (and
then project onto the excited state of the local Hamiltonian). If
one projects onto the excited states, the effective coupling of field
pics up a different sign but the magnitude is the same {[}see Eq.~\eqref{eq:H1tilde}{]}.
However, for decimation purposes, all remains the same as the sign
is irrelevant. Performing the decimation for all possibilities of
low and excited states, one constructs the entire spectrum of the
Ising chain: $2^{L/2}$ states in total. (Recall that the associated
Ising chain has half of the sites, but the same number of operators
in the Hamiltonian.) Notice that this is equivalent to consider two
``twins'' renormalized chains as previously outlined. Thus, all ground
states of the $p=1$ model are equivalent to the entire spectrum of
the transverse-field Ising chain. 

The connection to the XX model is alike. Here, the local Hamiltonian
has $3$ energy levels: one corresponds to the spin-$0$ singlet state,
another to the zero-magnetization spin-$1$ triplet state, and the
remaining one is doubly degenerate corresponding to the $\pm1$-magnetization
spin-$1$ states. If one disregards the doubly degenerate $\pm1$-magnetization
states, the decimation procedure recovers that of the Ising chain.
Thus, the many ground-states of the $p=1$ model corresponds to a
small fraction of the states in the spectrum of the XX spin-1/2 chain.

We now inquire about the spectrum of the $p=1$ chain. When projecting
the system onto to local excited states, the only difference with
respect to projecting onto the local ground state is a sign picked
up by the effective energy scales and a flip of one of the spins.
Thus, the entire spectrum can be easily related to the states in the
ground-state manifold. There will be $2^{\frac{L}{2}}$ states each
of which is $2^{\frac{L}{2}}$ degenerate. Precisely, all states can
be represented by 
\begin{equation}
\otimes_{j=1}^{L/2}\left|\phi_{j}\right\rangle ,\label{eq:states}
\end{equation}
 where $\left|\phi_{j}\right\rangle $ is either $\left|\phi_{j,+}\right\rangle =(\left|\rightarrow_{j_{1}},\uparrow_{j_{2}}\right\rangle \pm\left|\leftarrow_{j_{1}},\downarrow_{j_{2}}\right\rangle )/\sqrt{2}$
or $\left|\phi_{j,-}\right\rangle =(\left|\rightarrow_{j_{1}},\downarrow_{j_{2}}\right\rangle \pm\left|\leftarrow_{j_{1}},\uparrow_{j_{2}}\right\rangle )/\sqrt{2}$.
Here, $\left\{ j_{1},j_{2}\right\} $ is the pair of spin sites decimated
together in the $j$th decimation which is the same pair for all states.
The precise state $\left|\phi_{j}\right\rangle $ (if $\left|\phi_{j,+}\right\rangle $
or $\left|\phi_{j,-}\right\rangle $ and with the $+$ or $-$ sign)
depends on the sign of coupling constant and on whether the projection
was made into the local ground or excited states. All of these, are
determined by the history of decimation procedure.

\subsubsection{Spin-spin correlations}

In the SDRG approach, any of the $2^{L/2}$ ground states of $H_{0}$
is a simply product state as specified in \eqref{eq:states}. In that
case, the spins in the $j$th pair are strongly correlated and dominates
the average value of the spin-spin correlation. The probability that
a pair of length $r=j_{2}-j_{1}$ is formed along the SDRG flow is
proportional to $r^{-2}$ for sufficiently large $r$~\citep{fisher94-xxz,hoyosvieiralaflorenciemiranda}.
Hence, the mean correlation function decays only algebraically, 
\begin{equation}
\overline{\left|\left\langle \sigma_{i}^{x}\sigma_{i+r}^{z}\right\rangle \right|}\sim r^{-\eta},\label{eq:meanC-p1}
\end{equation}
 with universal exponent $\eta=2$. (Here, $\overline{\cdots}$ denotes
the disorder average.) The typical value of the correlation, on the
other hand, decays stretched exponentially fast~\citep{fisher94-xxz},
\begin{equation}
\overline{\ln\left|\left\langle \sigma_{i}^{x}\sigma_{i+r}^{z}\right\rangle \right|}\sim-r^{\psi}.\label{eq:typC-p1}
\end{equation}
 The distribution of the values of the correlation function is also
known. For that, we refer the reader to Refs.~\onlinecite{getelina-hoyos-ejpb20,wada-hoyos-prb22}.

The off-critical ($\delta\neq0$) correlations are also known~\citep{fisher94-xxz,fisher95,hoyos-etal-epl11}.
They decay exponentially faster $\sim e^{-r/\xi}$ with a diverging
correlation length 
\begin{equation}
\xi\sim\delta^{-\nu},\label{eq:corr-length}
\end{equation}
 where $\nu=2$ for the mean correlations and $\nu=1$ for the typical
correlations.

Interestingly, all those results apply for any state as well provided
that only the magnitude of the correlations are concerned.

\subsection{Case $p=2$\label{subsec:SDRGp2}}

\subsubsection{The usual SDRG method\label{subsec:usual-SDRG}}

We now derive the SDRG decimation rules for the Hamiltonian \eqref{eq:Hp}
with $p=2$.

Following the usual SDRG receipt, we treat $H_{1}=-\lambda_{3}h_{3}-\lambda_{4}h_{4}-\lambda_{6}h_{6}-\lambda_{7}h_{7}$
(where $h_{j}=\sigma_{j}^{x}\sigma_{j+1}^{z}\sigma_{j+2}^{z}$) as
a perturbation to $H_{0}=-\lambda_{5}h_{5}$. (We are assuming that
the energy cutoff $\Omega=\left|\lambda_{5}\right|$.) Up to second
order in perturbation theory, the renormalized Hamiltonian is $\tilde{H}_{1}=P_{0}H_{1}^{2}P_{0}/\left(-2\left|\lambda_{5}\right|\right)$,
where $P_{0}$ is the projector onto the ground state of $H_{0}$.
The direct terms in the square of $H_{1}$ are unimportant constants
since $h_{i}^{2}=1$. The cross terms are proportional to the anticommutators
$\left\{ h_{i},h_{j}\right\} $. Thus, many of those vanish because
of the algebra \eqref{eq:algebrap}. The surviving ones are those
which commute with each other. Thus, $\tilde{H}_{1}$ simplifies to
$\tilde{H}_{1}=-\tilde{\lambda}_{C}\tilde{h}_{C}-\tilde{\lambda}_{AC}\tilde{h}_{AC}-\tilde{\lambda}_{A}\tilde{h}_{A}$
where the operators are 
\begin{eqnarray}
\tilde{h}_{C} & = & P_{0}h_{3}h_{6}P_{0}=\sigma_{3}^{x}\sigma_{4}^{z}\tilde{\sigma}^{x}\tilde{\tau}^{z}\sigma_{8}^{z},\nonumber \\
\tilde{h}_{AC} & = & P_{0}h_{3}h_{7}P_{0}=\sigma_{3}^{x}\sigma_{4}^{z}\tilde{\sigma}^{x}\tilde{\tau}^{x}\sigma_{8}^{z}\sigma_{9}^{z},\label{eq:hab}\\
\tilde{h}_{A} & = & P_{0}h_{4}h_{7}P_{0}=-\sigma_{4}^{x}\tilde{\sigma}^{y}\tilde{\tau}^{y}\sigma_{8}^{z}\sigma_{9}^{z},\nonumber 
\end{eqnarray}
 and the renormalized couplings are 
\begin{equation}
\tilde{\lambda}_{C}=\frac{\lambda_{3}\lambda_{6}}{\Omega},\ \tilde{\lambda}_{AC}=\frac{\lambda_{3}\lambda_{7}}{\Omega},\mbox{ and }\tilde{\lambda}_{A}=\frac{\lambda_{4}\lambda_{7}}{\Omega}.\label{eq:Jtilde-p=00003D2}
\end{equation}
 Here, the effective degrees of freedom $\tilde{\boldsymbol{\sigma}}$
and $\tilde{\boldsymbol{\tau}}$ span the ground-state subspace $\{\left|\rightarrow\uparrow\uparrow\right\rangle ,$
$\left|\rightarrow\downarrow\downarrow\right\rangle ,$ $\left|\leftarrow\downarrow\uparrow\right\rangle ,$
$\left|\leftarrow\uparrow\downarrow\right\rangle \}$ for $\lambda_{5}>0,$
otherwise the set of states is $\{\left|\leftarrow\uparrow\uparrow\right\rangle ,$
$\left|\leftarrow\downarrow\downarrow\right\rangle ,$ $\left|\rightarrow\downarrow\uparrow\right\rangle ,$
$\left|\rightarrow\uparrow\downarrow\right\rangle \}$. These states
are recognized as $\{\left|\uparrow,\uparrow\right\rangle ,$ $\left|\uparrow,\downarrow\right\rangle ,$
$\left|\downarrow,\uparrow\right\rangle ,$ $\left|\uparrow,\uparrow\right\rangle \}$
in the $\tilde{\sigma}^{z}\otimes\tilde{\tau}^{z}$-basis. 

\begin{figure}[t]
\begin{centering}
\includegraphics[clip,width=1\columnwidth]{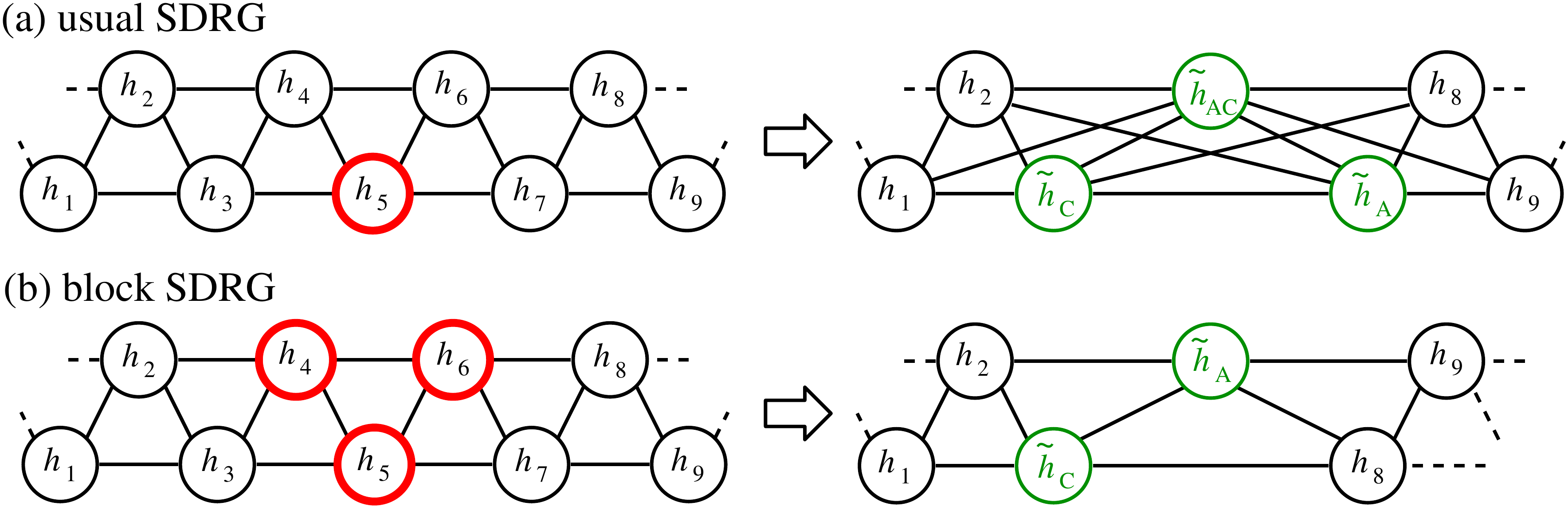}
\par\end{centering}
\caption{Strong-disorder RG decimation scheme for the Hamiltonian \eqref{eq:Hp}
in the $p=2$ case. The decimation is depicted in the Hamiltonian
space analogous to Fig.~\hyperref[fig:Decimation-p1]{\ref{fig:Decimation-p1}(b)}
where circles represent the local energy operators and lines connect
operators which anticommute. In (a), the SDRG method is implemented
in its simpler form (usual SDRG) where a newly generated operator
$\tilde{h}_{AB}$ disrupts the algebra \eqref{eq:algebrap}. In (b),
the SDRG method is implemented in a slightly more general fashion
(block SDRG) which preserves the algebra \eqref{eq:algebrap} in the
renormalized chain . \label{fig:Decimation-p2}}
\end{figure}

The decimation procedure \eqref{eq:hab} and \eqref{eq:Jtilde-p=00003D2}
is represented in the Hamiltonian space in Fig.~\hyperref[fig:Decimation-p2]{\ref{fig:Decimation-p2}(a)}.
Unlike the $p=1$ case {[}see Fig.~\hyperref[fig:Decimation-p1]{\ref{fig:Decimation-p1}(b)}{]},
the structure of the anticommuting algebra \eqref{eq:algebrap} of
the original Hamiltonian is not preserved and new operators (involving
$\sigma^{y}$) appear.

\subsubsection{The block SDRG method\label{subsec:block-SDRG}}

The arising of the new operator $\tilde{h}_{AC}$ in (\ref{eq:hab})
hinders the practical implementation of the method. First, it prevents
us from projecting the renormalized system onto the ground states
of the new effective spin operators $\tilde{\boldsymbol{\sigma}}$
and $\tilde{\boldsymbol{\tau}}$ as in the $p=1$ case. Second, and
more importantly, it requires a generalization of the SDRG procedure
to take into account these new operators. While this is possible and
cumbersome, we found, surprisingly, a quite simpler route guided by
the algebra of the energy-density operators (\ref{eq:algebrap}).
We generalize the ``usual'' SDRG approach above reported to (in the
lack of a better therminology) a ``block'' SDRG approach. In the latter,
we consider a larger unperturbed Hamiltonian (and thus, a larger Hilbert
space) when performing the decimation procedure (see details in Appendix~\ref{sec:blockSDRG}).
The size of the block is the maximum number of operators which anticommute
among themselves.

When decimating the largest local energy scale $\Omega=\left|\lambda_{5}\right|$,
instead of considering $H_{0}=-\lambda_{5}h_{5}$ (which is a $B$-type
operator), we consider a larger block involving the $A$- and $C$-type
``nearest-neighbor'' operators, i.e., $H_{0}=-\lambda_{4}h_{4}-\lambda_{5}h_{5}-\lambda_{6}h_{6}$.
This is the largest block which encompass $h_{5}$ and still have
only two energy levels (as $h_{5}$),\footnote{Actually, there are other 2 blocks: $-\lambda_{3}h_{3}-\lambda_{4}h_{4}-\lambda_{5}h_{5}$
and $-\lambda_{5}h_{5}-\lambda_{6}h_{6}-\lambda_{7}h_{7}$. However,
only the symmetric one (with respect to $\lambda_{5}$) provides the
convenient SDRG decimation rules.} i.e., the eigenenergies of $H_{0}$ are $\pm\sqrt{\lambda_{4}^{2}+\lambda_{5}^{2}+\lambda_{6}^{2}}$.
Then, we project the $H-H_{0}$ on the ground-state subspace of $H_{0}$.
The degeneracy of the ground state is $2^{4}$ and, thus, can be spanned
by four effective spin-1/2 degrees of freedom $\tilde{\boldsymbol{\sigma}}_{a}$,
$\tilde{\boldsymbol{\sigma}}_{b}$, $\tilde{\boldsymbol{\sigma}}_{c}$,
and $\tilde{\boldsymbol{\sigma}}_{d}$. In practice, we have to project
$\lambda_{2}h_{2}$, $\lambda_{3}h_{3}$, $\lambda_{7}h_{7}$ and
$\lambda_{8}h_{8}$. The result is that, in the regime $\left|\lambda_{5}\right|\gg\left|\lambda_{4,6}\right|$,
the renormalized operators are 
\begin{equation}
\tilde{h}_{2}=\sigma_{2}^{x}\sigma_{3}^{z}\tilde{\sigma}_{a}^{z}\left(\sin\theta\tilde{\sigma}_{b}^{x}\tilde{\sigma}_{d}^{z}+\cos\theta\tilde{\sigma}_{b}^{z}\right)\tilde{\sigma}_{c}^{z},\label{eq:h2}
\end{equation}
 where 
\begin{equation}
\cos\theta=\frac{-\text{sign}\left(\lambda_{5}\right)\left|\lambda_{4}\right|}{\sqrt{\lambda_{4}^{2}+\lambda_{6}^{2}}}\mbox{ and }\sin\theta=\frac{\lambda_{6}}{\sqrt{\lambda_{4}^{2}+\lambda_{6}^{2}}},\label{eq:theta}
\end{equation}
\begin{equation}
\tilde{h}_{8}=-\text{sign}\left(\lambda_{5}\right)\left(\cos\theta\tilde{\sigma}_{d}^{x}+\sin\theta\tilde{\sigma}_{b}^{y}\tilde{\sigma}_{d}^{y}\right)\sigma_{9}^{z}\sigma_{10}^{z},\label{eq:h8}
\end{equation}
\begin{equation}
\tilde{h}_{3}=\sigma_{3}^{x}\tilde{\sigma}_{a}^{y}\tilde{\sigma}_{b}^{y}\tilde{\sigma}_{c}^{z}\tilde{\sigma}_{d}^{z},\mbox{ and }\tilde{h}_{7}=\tilde{\sigma}_{c}^{x}\tilde{\sigma}_{d}^{z}\sigma_{9}^{z}.
\end{equation}
 The corresponding renormalized coupling constants are $\tilde{\lambda}_{2}=\lambda_{2}$,
$\tilde{\lambda}_{8}=\lambda_{8}$, 
\begin{equation}
\tilde{\lambda}_{3}=-\text{sign}\left(\lambda_{4}\right)\frac{\lambda_{3}\lambda_{6}}{\Omega},\mbox{ and }\tilde{\lambda}_{7}=\frac{\left|\lambda_{4}\right|\lambda_{7}}{\Omega}.
\end{equation}

Surprisingly, the renormalized operators are different in character
from those in Eq.~\eqref{eq:hab}. Interestingly, they preserve the
algebra structure \eqref{eq:algebrap} of the original system as depicted
in Fig.~\hyperref[fig:Decimation-p2]{\ref{fig:Decimation-p2}(b)}
if we identify $\tilde{h}_{2}\rightarrow h_{2}$, $\tilde{h}_{3}\rightarrow\tilde{h}_{C}$,
$\tilde{h}_{7}\rightarrow\tilde{h}_{A}$, and $\tilde{h}_{8}\rightarrow h_{8}$.
Furthermore, the ``hybrid'' operator $\tilde{h}_{AC}$ {[}generated
from the usual SDRG approach, see Fig.~\hyperref[fig:Decimation-p2]{\ref{fig:Decimation-p2}(a)}{]}
is not generated in the block SDRG approach {[}see Fig.~\hyperref[fig:Decimation-p2]{\ref{fig:Decimation-p2}(b)}{]}.
Instead, there are only ``pure'' operators except for the $BC$-type
operator in \eqref{eq:h2} and the $AB$-type operator in \eqref{eq:h8}.
This is very convenient because they can be neglected at strong-disorder
fixed points (corresponding to situations near and at the dashed transitions
in Fig.~\ref{fig:PD-dirty}). The reasoning is the following. Since
the effective disorder at and near the transition is very large (which
we show \emph{a posteriori}), very likely either $\left|\lambda_{4}\right|\gg\left|\lambda_{6}\right|$
or the $\left|\lambda_{4}\right|\ll\left|\lambda_{6}\right|$. In
the former case, $\sin\theta\approx0$ in Eq.~\eqref{eq:theta},
and, thus, the hybrid type operators can be neglected. The renormalized
$B$-type operators then simplify to $\tilde{h}_{2}=-\text{sign}\left(\lambda_{5}\right)\sigma_{2}^{x}\sigma_{3}^{z}\tilde{\sigma}_{a}^{z}\tilde{\sigma}_{b}^{z}\tilde{\sigma}_{c}^{z}$,
$\tilde{h}_{8}=\tilde{\sigma}_{d}^{x}\sigma_{9}^{z}\sigma_{10}^{z}$.
Taking $\sin\theta=1-\cos\theta=0$ and noticing that the new effective
spin-1/2 degrees of freedom $\tilde{\boldsymbol{\sigma}}_{a}$ and
$\tilde{\boldsymbol{\sigma}}_{b}$ appear only in combinations that
commute with each other ($\tilde{\sigma}_{a}^{z}\tilde{\sigma}_{b}^{z}$
in $\tilde{h}_{2}$ and $\tilde{\sigma}_{a}^{y}\tilde{\sigma}_{b}^{y}$
in $\tilde{h}_{3}$), we can project the resulting renormalized Hamiltonian
in the common eigenstates of $\tilde{\sigma}_{a}^{z}\tilde{\sigma}_{b}^{z}$
and $\tilde{\sigma}_{a}^{y}\tilde{\sigma}_{b}^{y}$. As a result,
we have four ``twins'' renormalized systems. They are 
\begin{equation}
\tilde{H}_{1}=\pm\lambda_{2}h_{2}\pm\tilde{\lambda}_{C}\tilde{h}_{C}-\tilde{\lambda}_{A}\tilde{h}_{A}-\lambda_{8}h_{8},\label{eq:H1-tilde}
\end{equation}
 where $h_{2}=\sigma_{2}^{x}\sigma_{3}^{z}\tilde{\sigma}_{c}^{z}$,
$\tilde{h}_{C}=\sigma_{3}^{x}\tilde{\sigma}_{c}^{z}\tilde{\sigma}_{d}^{z}$,
$\tilde{h}_{A}=\tilde{\sigma}_{c}^{x}\tilde{\sigma}_{d}^{z}\sigma_{9}^{z}$,
$h_{8}=\tilde{\sigma}_{d}^{x}\sigma_{9}^{z}\sigma_{10}^{z}$, 
\begin{equation}
\tilde{\lambda}_{C}=\frac{\lambda_{3}\lambda_{6}}{\Omega},\mbox{ and \ensuremath{\tilde{\lambda}_{A}}=\ensuremath{\frac{\left|\lambda_{4}\right|\lambda_{7}}{\Omega}}.}\label{eq:JAC-tilde}
\end{equation}
 The decimation procedure \eqref{eq:H1-tilde} and \eqref{eq:JAC-tilde}
is schematically depicted in Fig.~\hyperref[fig:Decimation-p2]{\ref{fig:Decimation-p2}(b)}.
By symmetry, an analogous decimation procedure is obtained in the
case $\left|\lambda_{4}\right|\ll\left|\lambda_{6}\right|$, with
the exchanges $\lambda_{2}\leftrightharpoons\lambda_{8}$ and $\tilde{\lambda}_{A}\leftrightharpoons\tilde{\lambda}_{C}$,
which is obtained after a convenient change in the definition of the
effective operators $\tilde{\boldsymbol{\sigma}}_{a}$, $\tilde{\boldsymbol{\sigma}}_{b}$,
$\tilde{\boldsymbol{\sigma}}_{c}$, and $\tilde{\boldsymbol{\sigma}}_{d}$.

The decimation procedure \eqref{eq:H1-tilde} and \eqref{eq:JAC-tilde}
{[}see Fig.~\hyperref[fig:Decimation-p2]{\ref{fig:Decimation-p2}(b)}{]}
is very convenient. It preserves the algebra structure \eqref{eq:algebrap}
and the operators of the original Hamiltonian $\tilde{h}_{i}=\sigma_{i}^{x}\sigma_{i+1}^{z}\sigma_{i+2}^{z}$.
This procedure is a straightforward generalization of the decimation
procedure of the $p=1$ case in the following sense. For $p=1$, a
decimation of an $A$-type coupling implies the renormalization of
the neighboring $B$-type couplings and vice versa {[}see Eq.~\eqref{eq:H1tilde}{]}.
For $p=2$, due to triality, the decimation of a $B$-type operator
implies the renormalization of the neighboring $A$- and $C$-type
couplings {[}see Eq.~\eqref{eq:JAC-tilde}{]}.

\subsubsection{On the equivalence between the usual and the block SDRG approaches}

In the regime $\left|\lambda_{5}\right|\gg\left|\lambda_{4,3}\right|$,
there should be no difference between the usual and block SDRG approaches
as the ground state of the block $H_{0}=-\lambda_{4}h_{4}-\lambda_{5}h_{5}-\lambda_{6}h_{6}$
is simply that of the local $H_{0}=-\lambda_{5}h_{5}$ furnished with
trivial degeneracies. It is somewhat surprising that seemingly fundamentally
different decimation procedures arise from these approaches. Thus,
we inquire whether these two decimating procedures are really different.

In Ref.~\onlinecite{hoyos-ladders} the SDRG method was carried out
in the antiferromagnetic Heisenberg model in a zigzag and two-leg
ladder geometries. In both cases, it was shown that, in the early
stages of the SDRG flow, further neighbors interactions arise, just
like what was found in the usual SDRG decimation {[}see Fig.~\hyperref[fig:Decimation-p2]{\ref{fig:Decimation-p2}(a)}{]}.
However, in the final stages of the SDRG flow, the renormalized geometry
of the system always converged to a chain geometry. (Evidently, one
had to disregard exceedingly small couplings that were generated along
the flow.) This may be a general feature of quasi-one-dimensional
critical or near-critical systems. The low-energy long-wavelength
effective theory is that of a chain. Technically, couplings of ``long''
operators that connect exceedingly distant spins arise after many
renormalizations and, thus, are typically much smaller than those
of ``shorter'' operators.

In sum, the longer range character of the new hybrid operator $\tilde{h}_{AC}$
in the usual SDRG approach may be an irrelevant ``operator'' which
vanishes in the latter stages of the SDRG flow. In the block SDRG
approach, the new hybrid operators $\tilde{h}_{AB}$ and $\tilde{h}_{BC}$
could be clearly determined as irrelevant ones near and at criticality,
where the flow is towards strong disorder, a self-consistent assumption
that we have yet to prove. Therefore, it is plausible that $\tilde{h}_{AC}$
in the usual SDRG approach can be neglected by the same reasons. In
that case, both the usual and block SDRG approaches are equivalent
near and at criticality. This is a fundamental feature of the renormalization-group
philosophy. Details on defining the coarse-grained operators should
not matter in the long-wavelength regime.

\subsubsection{SDRG flow corresponding to conventional phases}

Having derivied the SDRG decimation rules, we now analyze the flow
of the coupling constants.

Let us start by analyzing the simpler case corresponding to flow towards
the gapped phases. In this case, one of the coupling constants, say,
of $C$-type, is always greater than the others. Precisely, $\min\left\{ \lambda_{3i}\right\} >\max\left\{ \lambda_{3i-1},\lambda_{3i-2}\right\} $.
In that case, notice that only the bare $h_{3i}$ operators are decimated.
Therefore, the corresponding fixed point is noncritical and, thus,
represents a phase: the $h^{C}>h^{A,B}$ conventional phase in Fig.~\ref{fig:PD-dirty}.
Notice that both the usual and the block SDRG can be applied here
since only the original $h_{3i}$ operators are decimated. Thus, in
the SDRG framework, the spectrum gap is simply $\Delta=2\min\left\{ \lambda_{3i}\right\} $.

Evidently, by triality, there are additional two phases in which $h^{A}>h^{B,C}$
and $h^{B}>h^{A,C}$ as shown in Fig.~\ref{fig:PD-dirty}. Finally,
we call attention to the fact that, in the SDRG framework, these phases
exist because a decimation of $h_{i}$, renormalizes the neighboring
interactions $h_{j}$ with $\left|j-i\right|\le p=2$. Thus, for a
generic value of $p$ in the Hamiltonian \eqref{eq:Hp}, we expect,
at least, $p+1$ different phases. 

\subsubsection{SDRG flow corresponding to conventional Griffiths phases and phase
transitions between two phases.}

First, we analyze the case in which at least two of the three types
of couplings are random, i.e., the SDRG flow associated with the transitions
and Griffiths phases shown in Fig.~\hyperref[fig:PD-dirty]{\ref{fig:PD-dirty}(b)}
away from the multicritical point.

Let us consider the case in which $\max\left\{ \lambda_{3i-2}\right\} <\min\left\{ \lambda_{3i-1},\lambda_{3i}\right\} $.
In this case, the SDRG flow is dictated only by the competition between
$\left\{ \lambda_{3i}\right\} $ and $\left\{ \lambda_{3i-1}\right\} $
as the $A$-type couplings will never be decimated. Thus, we can simply
drop the $\lambda_{A}$'s in the decimation procedure \eqref{eq:H1-tilde}
and \eqref{eq:JAC-tilde} as the distribution $P_{A}$ renormalizes
to an extremely singular one. In that case, the zigzag-chain geometry
of the renormalized system becomes that of a single chain. Therefore,
the flow is simply that of the case $p=1$. 

To analyze it, we simply generalize the distance from criticality
Eq.~\eqref{eq:delta} to 
\begin{equation}
\delta_{BC}\equiv\frac{\overline{\ln\lambda_{3i}}-\overline{\ln\lambda_{3i-1}}}{\sigma_{\ln\lambda_{B}}^{2}+\sigma_{\ln\lambda_{C}}^{2}}.\label{eq:delta-BC}
\end{equation}
 From here, all energy-related results from the $p=1$ case follows
straightforwardly. The transition is governed by an infinite-randomness
critical point and happens for $\delta_{BC}=0$. The gap distribution
is that of Eq.~\eqref{eq:PSDRG} with the scaling variable \eqref{eq:eta}
redefined as 
\begin{equation}
\eta=\frac{\ln\left(2\Omega_{0}/\Delta\right)}{\sigma_{0}\left(L/3\right)^{\psi}}.\label{eq:etap2}
\end{equation}
 This redefinition is due to the fact that the total number of decimations
in the $p=1$ case is $L/2$ while it is $L/3$ for the $p=2$ case.
Thus, $L/2$ in the $p=1$ case translates to $L/3$ in the $p=2$
case.

In addition, there are associated Griffiths phases for $\left|\delta_{BC}\right|\ll1$.
The off-critical dynamical exponent $z$ diverges as $z\sim\left|\delta_{BC}\right|^{-1}$
when it approaches criticality. The nature of the low-energy modes
are also associated with domain wall spins surrounding a rare region,
just like for $p=1$.

By triality, there are other two boundary transitions for $\delta_{AB}=0$
and $\delta_{AC}=0$. These boundaries and the Griffiths phases are
shown in Fig.~\ref{fig:PD-dirty} as dashed lines and shaded regions,
respectively. The SDRG results here reported, thus, put the heuristic
arguments of Sec.~\ref{sec:Overview} in solid grounds.

Now we analyze the case in which the two competing (strongest) couplings
are uniform, i.e., $\lambda_{B,i}=\lambda_{B}$, $\lambda_{C,i}=\lambda_{C}$,
and $\lambda_{A,i}$ random with $\lambda_{A,\text{typ}}<\lambda_{B,C}$.
This corresponds to the region in the phase diagram of Fig.~\hyperref[fig:PD-dirty]{\ref{fig:PD-dirty}(a)}
surrounding the clean and finite-randomness transition but sufficiently
far from the multicritical point.

When $\max\left\{ \lambda_{A,i}\right\} <\lambda_{B,C}$, the SDRG
method does not need to be applied since weak $\lambda_{A}$ coupling
is irrelevant. The transition at $\lambda_{B}=\lambda_{C}$ is in
the clean Ising universality class, and the spectral gap vanishes
only at criticality.

When $\lambda_{A,\text{typ}}<\lambda_{B}<\lambda_{C}<\max\left\{ \lambda_{A,i}\right\} $,
the system in the Griffiths $C$-phase with high $A$-susceptibility
as in the generic case discussed above. (Analogously for $\lambda_{A,\text{typ}}<\lambda_{C}<\lambda_{B}<\max\left\{ \lambda_{A,i}\right\} $.)

Finally, we discuss the interesting case when $\lambda_{A,\text{typ}}<\lambda_{B}=\lambda_{C}<\max\left\{ \lambda_{A,i}\right\} $.
The system is globally critical between $B$- and $C$-phases but
there are rare regions locally in the $A$-phase. What are their effects?
Applying the block-SDRG decimation procedure, we simply decimate $A$-operators
in the first stages of the flow. After decimating all $A$-operators
with $\lambda_{B}=\lambda_{C}<\lambda_{A,i}<\max\left\{ \lambda_{A,i}\right\} $,
we can simply ignore the remaining $A$-operators since they would
be surrounded by locally stronger $B$- and $C$-operators. The effective
chain is, thus, that with $B$- and $C$-operators only. However,
the effective coupling constants are not uniform. Interestingly, notice
that the effective couplings obey the condition $\tilde{\lambda}_{B,i}=\tilde{\lambda}_{C,i}$.
The precise condition for the absence of random distances from criticality
(random mass) discussed in Sec.~\ref{subsec:correlated-disorder}.
When the $A$-rare regions are small, the effective couplings $\tilde{\lambda}_{B,C,i}$
are weakly renormalized and their effective disorder (variance of
their distribution) is weak. As a result, the clean critical behavior
is stable. However, when approaching the multicritical point we expect
the renormalization of $\tilde{\lambda}_{B,C,i}$ to become more and
more relevant. This means that their effective disorder increases
to the point where the clean critical behavior is destabilized. The
resulting phase transition is thus of finite-randomness type with
nonuniversal critical exponent $z$. This result should be valid up
to the multicritical point where $z$ reaches its maximum value. Interesting,
$z$ is formally infinity in the other transition lines meeting at
the multicritical point.

We end this section by noticing that the above results straightforwardly
generalizes to any value of $p$.

\subsubsection{The SDRG flow at the multicritical point}

Finally, we deal with the case in which all couplings compete.

At first glance, one could think in analyzing the flow equations for
the distributions $P_{A}$, $P_{B}$, and $P_{C}$ analytically. The
flow equation for $P_{A}$ is (see Appendix~\ref{sec:SDRG-flow-Eq})
\begin{equation}
-\frac{\partial P_{A}}{\partial\Omega}=P_{A}\left(\Omega\right)P_{A}-\left(P_{B}\left(\Omega\right)+P_{C}\left(\Omega\right)\right)\left(P_{A}-P_{A}\otimes P_{A}\right),\label{eq:flow-PA}
\end{equation}
 where $P_{X}=P_{X}\left(\lambda;\Omega\right)$, $P_{X}\left(\Omega\right)=P_{X}\left(\Omega;\Omega\right)$,
and 
\begin{equation}
P_{X}\otimes P_{X}=\int d\lambda_{1}d\lambda_{4}P_{X}\left(\lambda_{1};\Omega\right)P_{X}\left(\lambda_{4};\Omega\right)\delta\left(\lambda-\frac{\lambda_{1}\lambda_{4}}{\Omega}\right).
\end{equation}
 By triality, the flow equations for $P_{B}$ and $P_{C}$ follow
from exchanging the labels accordingly. The last term on the r.h.s.
of \eqref{eq:flow-PA} implements the renormalization of the $A$-type
coupling {[}$\tilde{\lambda}_{A}$ in Eq.~\eqref{eq:JAC-tilde},
and we are considering only the magnitude of the coupling constants{]}
when a $B$- or $C$-type coupling is decimated. The remaining terms
ensure that $P_{A}$ remains normalized when the cutoff energy scale
$\Omega$ is changed. As shown in App.~\ref{sec:SDRG-flow-Eq}, the
critical point (where $P_{A}=P_{B}=P_{C}$) of the flow equation \eqref{eq:flow-PA}
is of infinite-randomness type {[}see Eq.~\eqref{eq:Pa*}{]} and
is related to a different universal tunneling exponent $\psi=\frac{2}{3}$
which is greater than that $\psi=\frac{1}{2}$ of the case $p=1$.
A larger tunneling exponent means larger effective disorder as the
typical value of the finite-size gap is even smaller {[}see Eq.~\eqref{eq:etap2}{]}.
This sounds intuitive since two coupling constants are renormalized
in each decimation instead of one renormalization per decimation as
happens in the $p=1$ case.

Although this gives us a prescription to find a new universality class
beyond that of the permutation-symmetric universality class (where
$\psi=1/N$, with $N$ being an integer~\citep{damle-huse-multicritical,hoyos-sun,fidkowski-etal-prb09,quito-etal-epjb20,quito-etal-prb19}),
the analytical analysis of this problem is not correct. The flow equation
\eqref{eq:flow-PA} assumes no correlation between the three types
of couplings. However, when, say, an $B$-type coupling is decimated
the renormalized $A$- and $C$-type couplings are added as neighbors
in the renormalized system {[}see Fig.~\hyperref[fig:Decimation-p2]{\ref{fig:Decimation-p2}(b)}{]}.
Being neighbors diminishes their chance of being further renormalized
(which would increase the effective disorder strength by producing
even more singular couplings) when compared with the case described
by Eq.~\eqref{eq:flow-PA} where they are inserted in the chain in
an uncorrelated fashion.

As treating the correlated case analytically is not simple, we then
proceed our study numerically. We implement the block SDRG rules \eqref{eq:H1-tilde}
and \eqref{eq:JAC-tilde} for a system where all the coupling constants
are independent random variable and identically distributed according
to 
\[
P\left(\lambda;\Omega_{0}\right)=\frac{1}{D_{0}\Omega_{0}}\left(\frac{\Omega_{0}}{\lambda}\right)^{1-1/D_{0}},
\]
 where $\Omega_{0}$ is the initial energy cutoff and $D_{0}$ parametrizes
the disorder strength of the bare system. 

\begin{figure}
\begin{centering}
\includegraphics[clip,width=0.88\columnwidth]{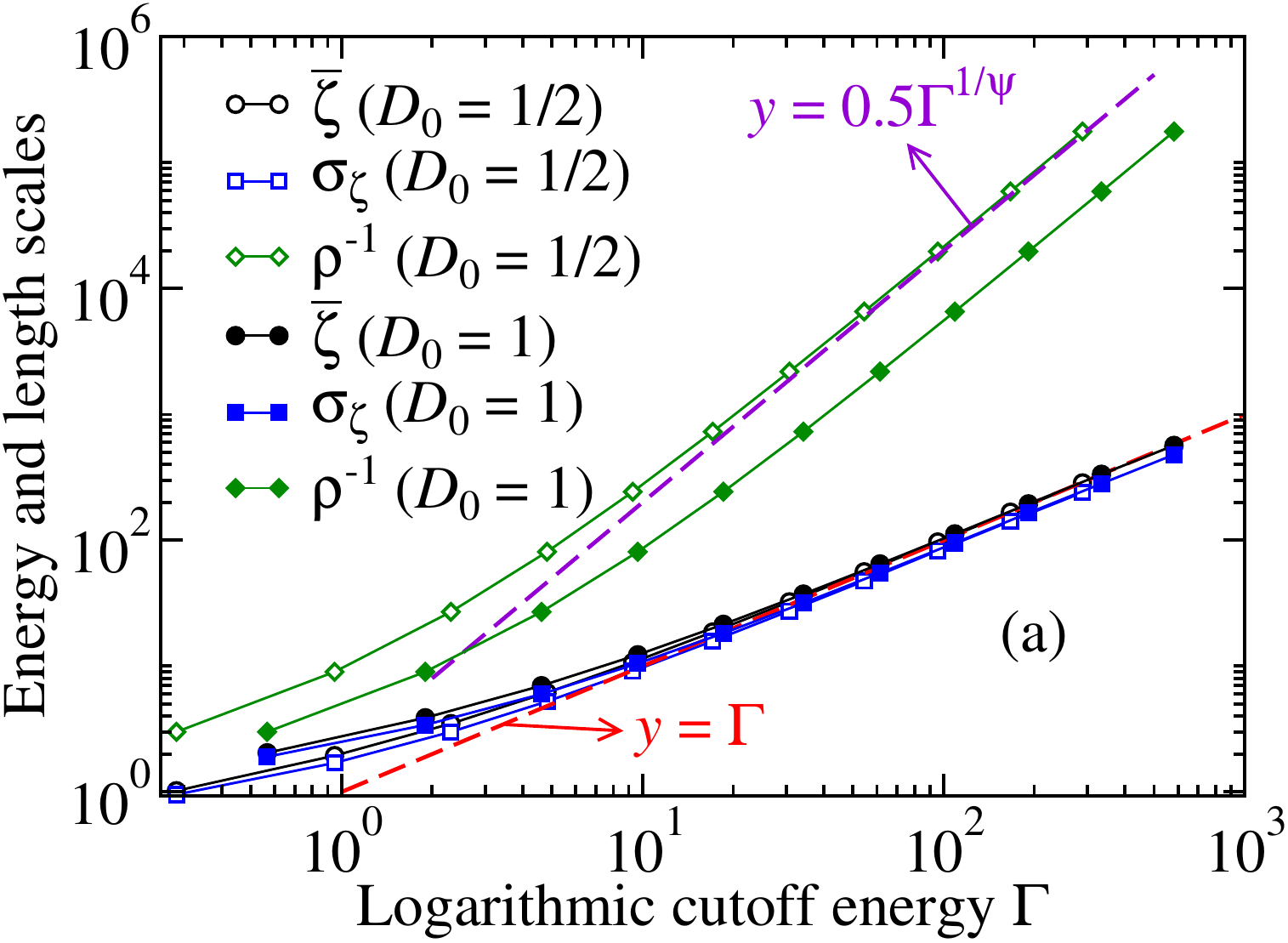}\\
\includegraphics[clip,width=0.88\columnwidth]{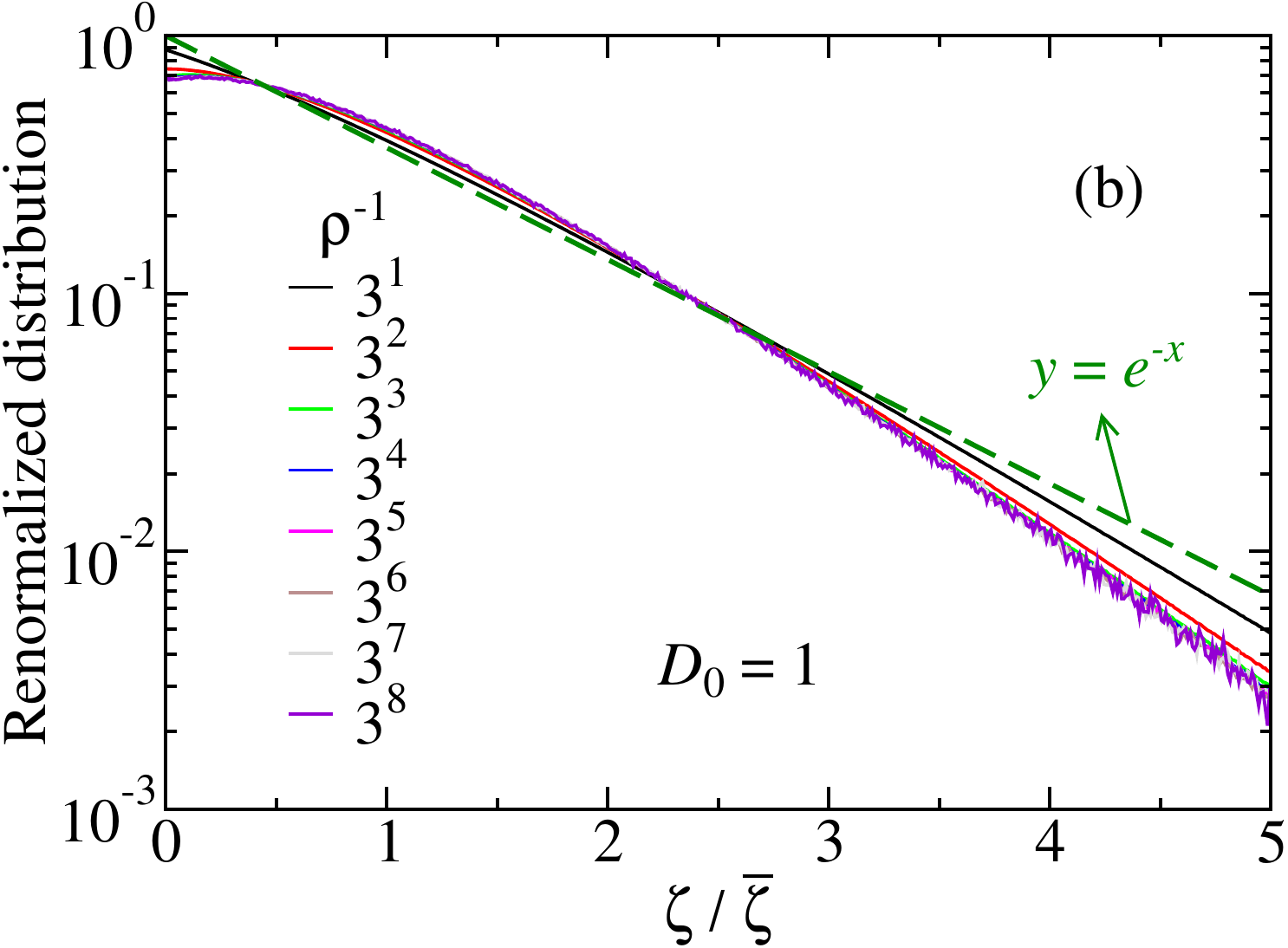}\\
\includegraphics[clip,width=0.88\columnwidth]{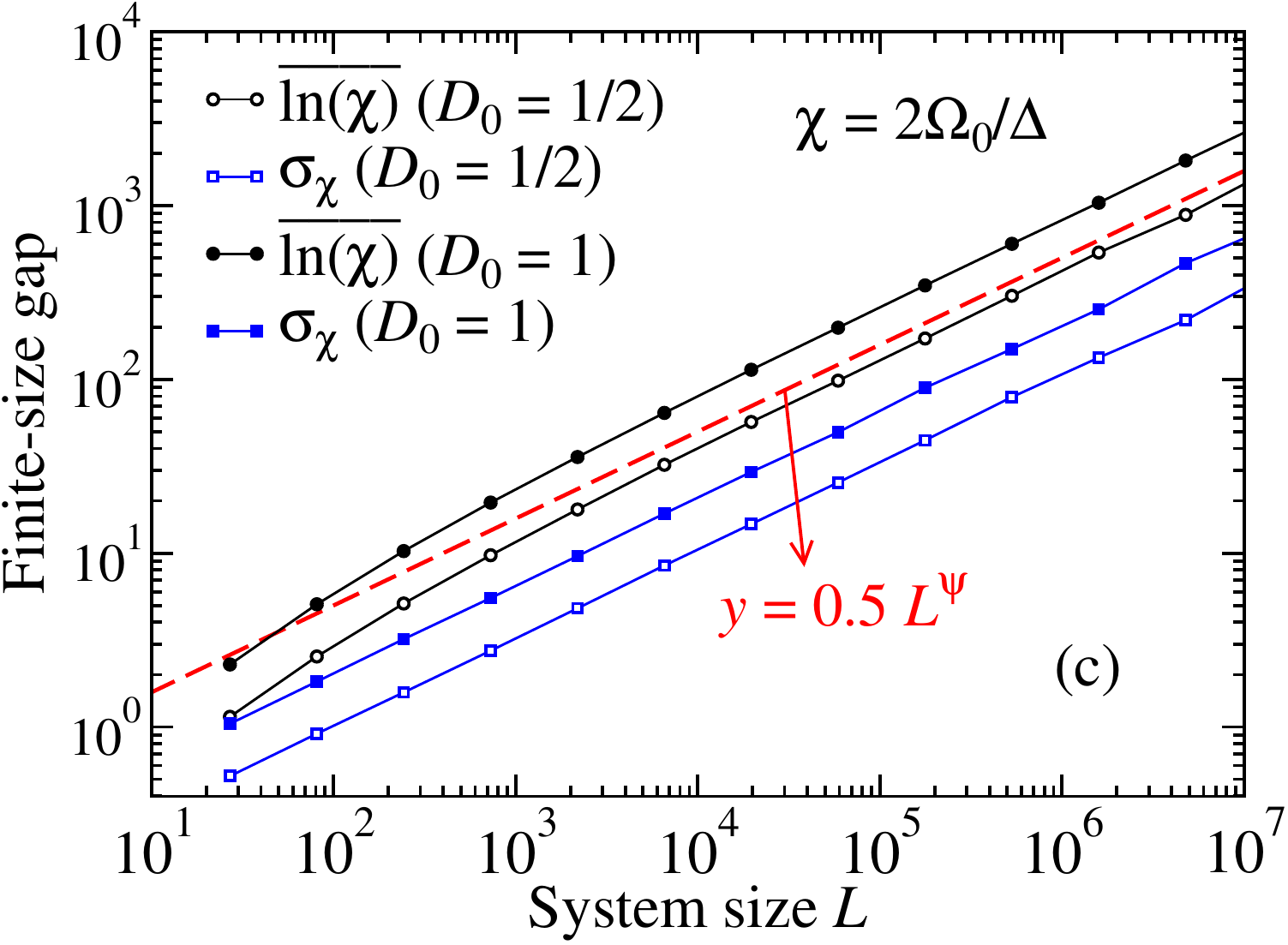}
\par\end{centering}
\caption{(a) The mean value of $\overline{\zeta}$, the standard deviation
$\sigma_{\zeta}$ and the density of active spins $\rho$ as a function
of the logarithmic cutoff energy $\Gamma=\ln\Omega_{0}/\Omega$ ($\Omega=\text{max\ensuremath{\left\{  \lambda_{i}\right\} } }$).
(b) Snapshots of the coupling constant distributions at different
densities $\rho$ along the SDRG flow. (c) The typical value of the
finite-size gap as a function of the system size $L.$ We have considered
chains of up to $L=3^{15}$ spins, with bare disorder strengths $D_{0}=1/2$
and $1$. Here, $\psi=\frac{1}{2}$. The error bars are about the
size of the symbol sizes.\label{fig:z-dz-L-G}}
\end{figure}

We first study the moments of the distribution of the renormalized
couplings along the SDRG flow. As usual in the SDRG approach, it is
convenient to define the logarithmic couplings $\zeta_{i}=\ln\left(\Omega/\lambda_{i}\right)$
and the logarithmic energy cutoff $\Gamma=\ln\left(\Omega_{0}/\Omega\right)$.
We then study the mean value $\overline{\zeta}$ and standard deviation
$\sigma_{\zeta}=\sqrt{\overline{\zeta^{2}}-\overline{\zeta}^{2}}$
as a function of $\Gamma$. Our results for $D_{0}=1/2$ and $D_{0}=1$
and system size of $L=3^{15}$ spins is shown in Fig.~\hyperref[fig:z-dz-L-G]{\ref{fig:z-dz-L-G}(a)}.
Clearly, $\overline{\zeta}\approx\sigma_{\zeta}\approx\Gamma+\text{const}$
in the $\Gamma\rightarrow\infty$ limit. This implies an infinite-disorder
fixed-point critical distribution. Thus, the SDRG method here employed
is justified.

Next, we study the infinite-disorder fixed-point distribution. For
comparison, that distribution for $p=1$ is well known~\citep{fisher94-xxz,igloi-det-z-PRB,hoyosvieiralaflorenciemiranda}.
It is 
\begin{equation}
P_{p=1}^{*}\left(\lambda;\Omega\right)=\frac{1}{D_{\Gamma}\Omega}\left(\frac{\Omega}{\lambda}\right)^{1-1/D_{\Gamma}},\label{eq:P-FP}
\end{equation}
 with $D_{\Gamma}=\Gamma+D_{0}$ {[}equivalent to $\pi^{*}\left(\zeta;\Gamma\right)=e^{-\zeta/D_{\Gamma}}/D_{\Gamma}${]}.
In Fig.~\hyperref[fig:z-dz-L-G]{\ref{fig:z-dz-L-G}(b)}, we show
the distribution of log couplings at different stages of the SDRG
flow for the case $D_{0}=1$. (We find statistically identical result
for $D_{0}=1/2$ which is not shown for clarity.) After the initial
stages of the SDRG flow, the fixed point is reached. Here, the different
stages of the SDRG flow is parametrized by the density $\rho=N_{\Omega}/L$,
where $N_{\Omega}$ is the number of active spins at the cutoff energy
scale $\Omega$. Recall that, after each decimation step of the block
SDRG procedure, $3$ spins are removed. Clearly, the fixed point distribution
differs from the $p=1$ case Eq.~\eqref{eq:P-FP} (dashed line $y=e^{-x}$),
but only slightly. We attribute this small difference to the correlations
arising among renormalized couplings under the flow.

The relation between energy and length scales along the SDRG flow
is shown in Fig.~\hyperref[fig:z-dz-L-G]{\ref{fig:z-dz-L-G}(a)}.
Simply the length scale $\rho^{-1}\sim\Gamma^{1/\psi}$ with universal
tunneling exponent $\psi=1/2$. We also expect the finite-size gap
$\Delta$ to obey the activated dynamical scaling \eqref{eq:etap2}.
This is confirmed in our numerics as shown in Fig.~\hyperref[fig:z-dz-L-G]{\ref{fig:z-dz-L-G}(c)}.
Both the mean value and the width of the distribution of $\ln\left(\Delta\right)$
behave similarly $\sim L^{\psi}$. Here, the finite-size gap is obtained
by decimating the entire chain. The last decimated coupling constant
$\tilde{\lambda}_{\text{final}}$ provides the finite-size gap $\Delta=2\tilde{\lambda}_{\text{final}}$.

\subsubsection{Thermodynamics and correlations}

As the fixed-point distribution is of infinite-randomness type, the
critical thermal entropy and specific heat behave similarly as in
the $p=1$ case discussed in Sec.~\ref{subsec:Thermodynamics}. The
only difference is the residual entropy which modifies Eq.~\eqref{eq:S}
to 
\begin{equation}
S=\frac{2}{3}\left(1+\left(\frac{\sigma_{\ln\lambda}^{2}}{\ln\left(\frac{\Omega_{0}}{T}\right)}\right)^{1/\psi}\right)\ln2,
\end{equation}
 the reasoning being that the ground state have degeneracy $2^{2L/3}$.
The specific heat follows straightforwardly and recovers \eqref{eq:SpecificHeat}.

\begin{figure}
\begin{centering}
\includegraphics[clip,width=0.92\columnwidth]{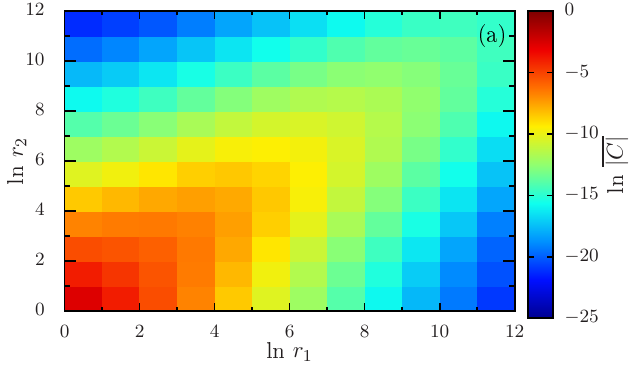}\\
\includegraphics[clip,width=0.92\columnwidth]{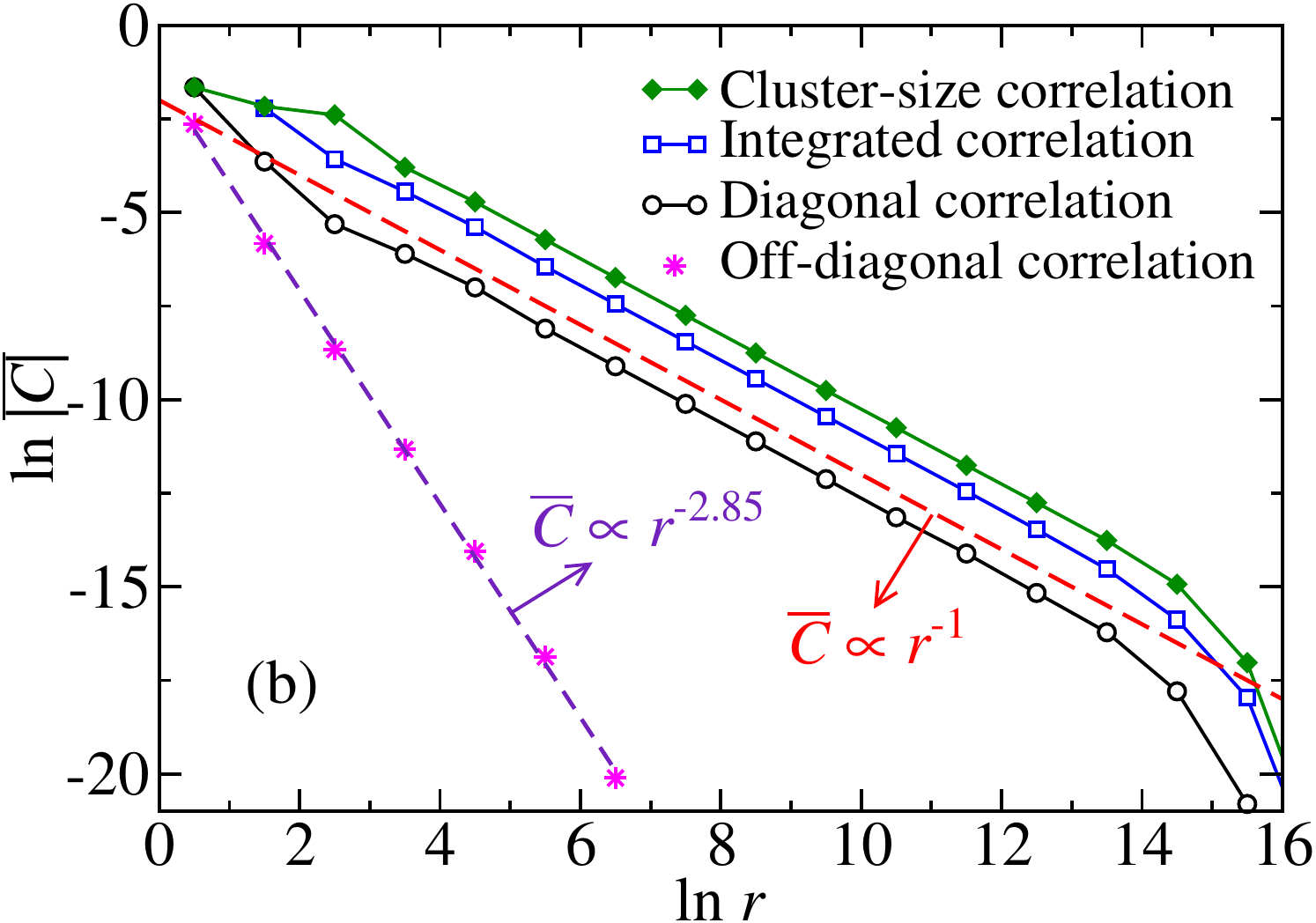}\\
\includegraphics[clip,width=0.92\columnwidth]{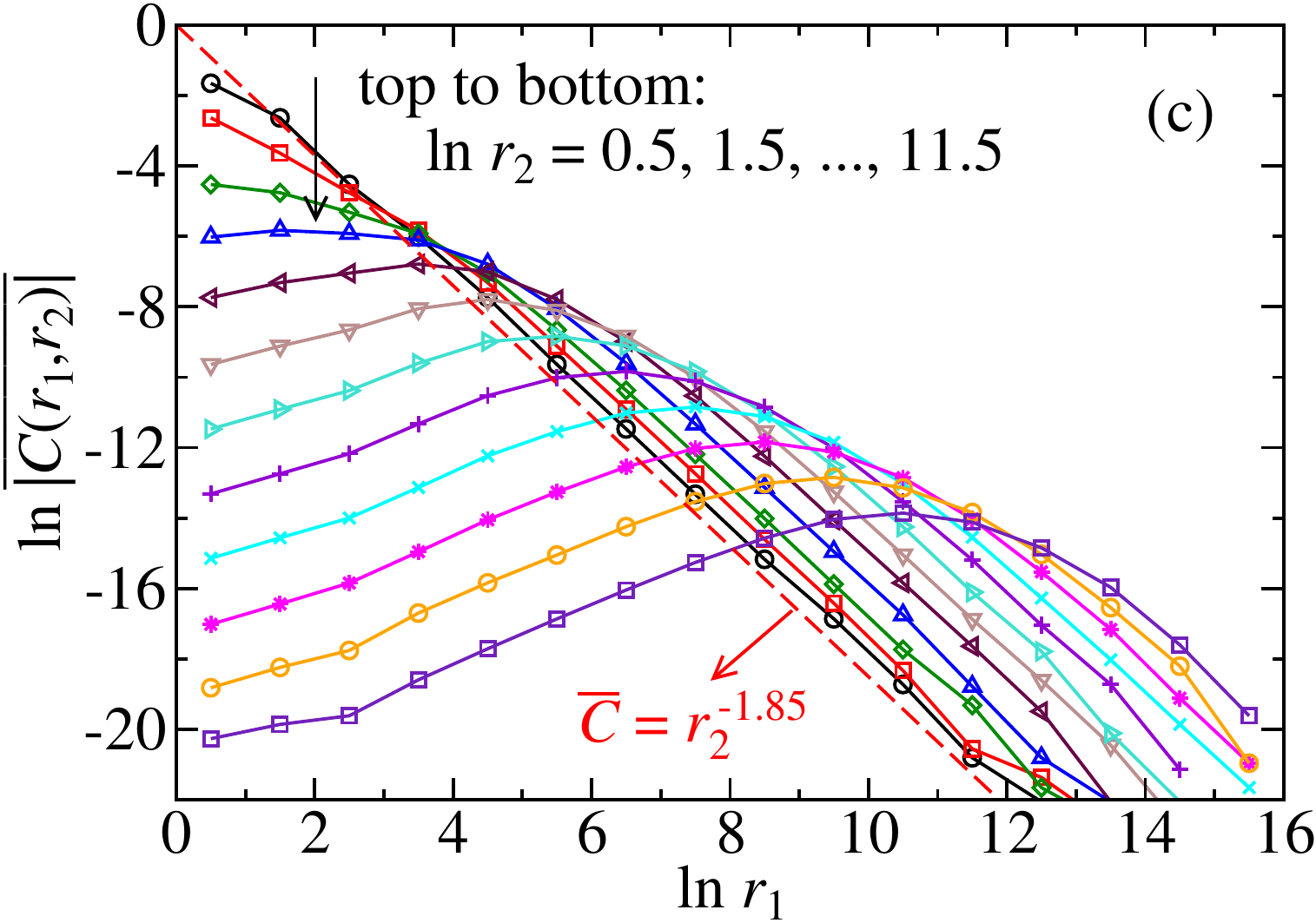}
\par\end{centering}
\caption{Average value of the spin correlations \eqref{eq:Corr}. (a) The average
value of the spin correlation $\overline{\left|C(r_{1},r_{2})\right|}$
{[}see Eq.~\eqref{eq:Corr}{]} as a function of the internal spin-spin
separations $r_{1}$ and $r_{2}$. (b) The average values of the diagonal
$\overline{\left|C(r,r)\right|}$ (open circles), the off-diagonal
$\overline{\left|C(r,2r)\right|}$ (stars), the integrated $\sum_{r_{2}}\overline{\left|C(r,r_{2})\right|}=\sum_{r_{1}}\overline{\left|C(r_{1},r)\right|}$
(open squares), and the cluster-size $\sum_{r_{1},r_{2}}\delta_{r,r_{1}+r_{2}}\overline{\left|C\left(r_{1},r_{2}\right)\right|}$
(closed diamonds) correlations. (c) The average correlation $\overline{\left|C(r_{1},r_{2})\right|}$
as a function of $r_{1}$ for different values of $r_{2}$. The system
size is $L=3^{15}$ averaged over $1\,500$ disorder realizations
which is sufficient to yield error bars of the size of the symbols.
\label{fig:Corr} Solid lines are simple guide to the eyes.}
\end{figure}

In the SDRG framework, the ground-state spin correlation $C_{i,j,k}=\left\langle \sigma_{i}^{x}\sigma_{j}^{z}\sigma_{k}^{z}\right\rangle $
is $\pm1$ if the spins $i<j<k$ are decimated together, and weakly
vanishing (we set to zero) otherwise. To obtain the average value
of $C_{i,j,k}$, we then build a normalized histogram $\overline{\left|C(r_{1},r_{2})\right|}$
in log-scale. After each SDRG decimation (decimation of spins $i$,
$j$, and $k$), we add a unity to the bin ($r_{1},r_{2}$) if $\ln\left(j-i\right)$
is between $\ln r_{1}-\frac{1}{2}$ and $\ln r_{1}+\frac{1}{2}$,
and $\ln\left(k-j\right)$ is between $\ln r_{2}-\frac{1}{2}$ and
$\ln r_{2}+\frac{1}{2}$. This procedure is repeated until the entire
chain is decimated and averaged over many disorder realizations. The
histogram $\overline{\left|C(r_{1},r_{2})\right|}$ is then a proxy
to average value of $C_{i,j,k}$, i.e., 
\begin{equation}
\overline{\left|C(r_{1},r_{2})\right|}=\overline{\left|\left\langle \sigma_{i}^{x}\sigma_{i+r_{1}}^{z}\sigma_{i+r_{1}+r_{2}}^{z}\right\rangle \right|}.\label{eq:Corr}
\end{equation}

We plot in Fig.~\hyperref[fig:Corr]{\ref{fig:Corr}(a)} the mean
value of the spin correlation $\overline{C\left(r_{1},r_{2}\right)}$
as a function of the internal distances $r_{1}$ and $r_{2}$. Clearly,
the correlation is maximum when $r_{1}=r_{2}$, as expected from symmetry. 

In Fig.~\hyperref[fig:Corr]{\ref{fig:Corr}(b)}, we plot the diagonal
correlation $\overline{C\left(r,r\right)}$, one of the possible off-diagonal
correlations $\overline{C\left(r,2r\right)}$, the integrated correlation
$\sum_{r_{2}}\overline{C\left(r,r_{2}\right)}=\sum_{r_{1}}\overline{C\left(r_{1},r\right)}$,
and the cluster-size correlation $\sum_{r_{1}}\overline{\left|C\left(r_{1},r-r_{1}\right)\right|}=\sum_{r_{1},r_{2}}\delta_{r,r_{1}+r_{2}}\overline{\left|C\left(r_{1},r_{2}\right)\right|}$.
They all vanish algebraically $\sim r^{-\phi}$, with universal exponent
$\phi\approx1$ but the off-diagonal correlations which vanishes as
$\overline{C\left(r,2r\right)}\sim r^{-\varphi}$ with $\varphi\approx1.85$. 

In Fig.~\hyperref[fig:Corr]{\ref{fig:Corr}(c)}, we plot the average
correlation $\overline{C\left(r_{1},r_{2}\right)}$ for various values
of $r_{2}$. Interestingly, our numerical data indicates that $\overline{C\left(r_{1},r_{2}\right)}\sim r_{1}^{-\left(\varphi-\phi\right)}$
for fixed $r_{2}$ and $r_{1}\gg r_{2}$.

Analyzing these exponents for other sample sizes, we estimate their
error to be of order $10\%$. Unfortunately, we do not have an analytical
derivation for these exponents. 

How about the typical value of these correlations? Typically, the
triad of spins in \eqref{eq:Corr} are not decimated together and,
thus, develop quite weak correlations. As argued by Fisher~\citep{fisher94-xxz}
and shown in many numerical works~\citep{henelius-girvin,getelina-hoyos-ejpb20,wada-hoyos-prb22}
for the case $p=1$, this weak correlations are of order of the typical
value of the coupling constants involved, $C_{\text{typ}}(r)\sim J_{\text{typ}}(r)$.
Very plausible, this is also the case for any $p$ with the caveat
that more than one coupling constant is involved. Thus, $C_{\text{typ}}\sim J_{\text{typ}}(r_{1})J_{\text{typ}}(r_{2})\dots J_{\text{typ}}(r_{p})$.
Thus, from the dynamical scaling \eqref{eq:dirty-scaling} we then
conclude that the typical value of the correlations decays stretched
exponentially fast with the spin separations. Roughly, we expect 
\begin{equation}
C_{\text{typ}}\left(r_{1},r_{2}\right)\sim e^{-A\left(\ell\left(r_{1},r_{2}\right)\gamma_{D}\right)^{\psi}},\label{eq:Ctyp}
\end{equation}
 where $\ell(r_{1},r_{2})$ is a function which equals $\max\left\{ r_{1},r_{2}\right\} $
for $r_{1}\gg r_{2}$ or $r_{2}\gg r_{1}$ and $\approx cr$ when
$r_{1}\approx r_{2}\approx r$ (with $c$ being disorder-independent
constant of order unity), $\gamma_{D}$ is a disorder-dependent Lyapunov
exponent~\citep{getelina-hoyos-ejpb20,wada-hoyos-prb22} related
to the clean-dirty crossover length~\citep{laflorencie-correlacao-PRB},
and $A$ is a constant of order unity.

\subsubsection{Generalization to $p>2$}

The usual SDRG method of Subsec.~\ref{subsec:usual-SDRG} can be
easily generalized to any $p$. The key point is that new effective
operators $\tilde{h}$ generated under the SDRG flow either commute
or anticommutate and are of two-level type $\tilde{h}^{2}=\mathds{1}$.
In this case, the only type of decimation that can take place is of
second order in perturbation theory, and, thus, the renormalized coupling
constants have the multiplicative structure of Eq.~(\ref{eq:Jtilde-p=00003D2}).
As a consequence, the relation between length and timescales at a
phase transition can only be of activated type Eq.~(\ref{eq:dirty-scaling}),
meaning that, at sufficiently strong disorder where the SDRG is applicable,
the phase transition is governed by a infinite-randomness fixed point. 

The analysis of the block SDRG flow of Subsec.~\ref{subsec:usual-SDRG}
allowed us to conclude that this infinite-randomness critical point
has tunneling exponent $\psi=\frac{1}{2}$ and very different mean
and typical values of the correlation function for the cases $p=1$
and $2$. Although the generalization of the block SDRG method to
other values of $p$ is not straightforward (as it involves many nontrivial
projections, see Appendix~\ref{sec:blockSDRG}), it is plausible
to conjecture that the phase transitions for uncorrelated disorder
are governed by an analogous infinite-randomness critical point for
any $p$. The tunneling exponent is universal and equals $\psi=\frac{1}{2}$.
It not only governs the dynamical scaling but also the typical correlations.
On the other hand, the exponent for the mean correlation function
is universal, i.e., disorder independent, but does depend on the value
of $p$. 

A detailed and quantitative investigation of the critical behavior
and of the associated Griffiths phases is out of the scope of the
present work and is left for future research. 

\section{Finite-size gap statistics\label{sec:FS-gap}}

In this section, we present our numerically exact results on the spectral
gap of the model Hamiltonian \eqref{eq:Hp} for $p=2$. This quantity
is computed from the roots of the polynomial as described in Sec.~\ref{sec:The-model}
and compared with the SDRG predictions described in Sec.~\ref{subsec:SDRGp2}.

\subsection{Technical details\label{subsec:Technical-details}}

We refer the reader to Ref.~\citealp{alcaraz-etal-prb21} for useful
numerical methods for calculating the roots of the polynomial (\ref{eq:polynomial})\textemdash (\ref{eq:recurrence-P}).
We emphasize that our method allow us to evaluate the finite-size
gap in lattice sizes up to $\sim10^{7}$ sites in the neighborhood
of the transition lines with a numerical cost that grows linearly
with the system size competing with the SDRG numerical cost. For systems
that large, typically, the polynomials (\ref{eq:polynomial}) have
coefficients spanning in 500 orders of magnitude. Therefore, the entire
evaluation of these coefficients can only be achieved using high numerical
precision (500 digits). However, as shown in Ref.~\citealp{alcaraz-etal-prb21},
we only need the last $100$ coefficients of the polynomial to obtain
the finite-size gap with quadruple standard numerical precision (32
digits). When applying this method to our Hamiltonian, we found useful
to keep the last coefficient of (\ref{eq:polynomial}) always of order
unity. This is accomplished by factorizing these coefficients when
iterating the recursion relation (\ref{eq:recurrence-P}). With that,
the entire procedure can be accomplished using only standard routines
in FORTRAN with quadruple precision.

\subsection{Off-critical finite-size gap I}

We start analyzing the off-critical Griffiths phases sketched in the
phase diagram of Fig.~\hyperref[fig:PD-dirty]{\ref{fig:PD-dirty}(b)}.
For such, we study three sets of chains: (set I) $\lambda_{A}$ is
uniformly distributed in the interval $\left[0,0.2\right]$; (set
II) $\lambda_{A}$ is uniformly distributed in the interval $\left[0.2,1.2\right]$;
(set III) $\lambda_{A}$ is uniformly distributed in the interval
$\left[0.5,1.5\right]$. In all cases, $\lambda_{B}$ is distributed
uniformly in the interval $\left[0.5,1.5\right]$, and $\lambda_{C}$
is a constant (running from $1.2$ down to $0.9$) which serves as
a tuning parameter. The running $\lambda_{C}$ passes through the
critical point $\lambda^{*}=\lambda_{B,\text{typ}}=e^{\frac{3\ln3-2\ln2}{2}-1}\approx0.95578$
{[}see Eq.~\eqref{eq:delta-BC}{]} but does not include it since
the analysis of the critical system is reported in other sections.
Notice that set III of chains is critical for $\lambda_{C}<\lambda_{C}^{*}$.
For that reason, we consider only $\lambda_{C}>\lambda_{C}^{*}$ in
this section.

\begin{figure}
\begin{centering}
\includegraphics[clip,width=0.8\columnwidth]{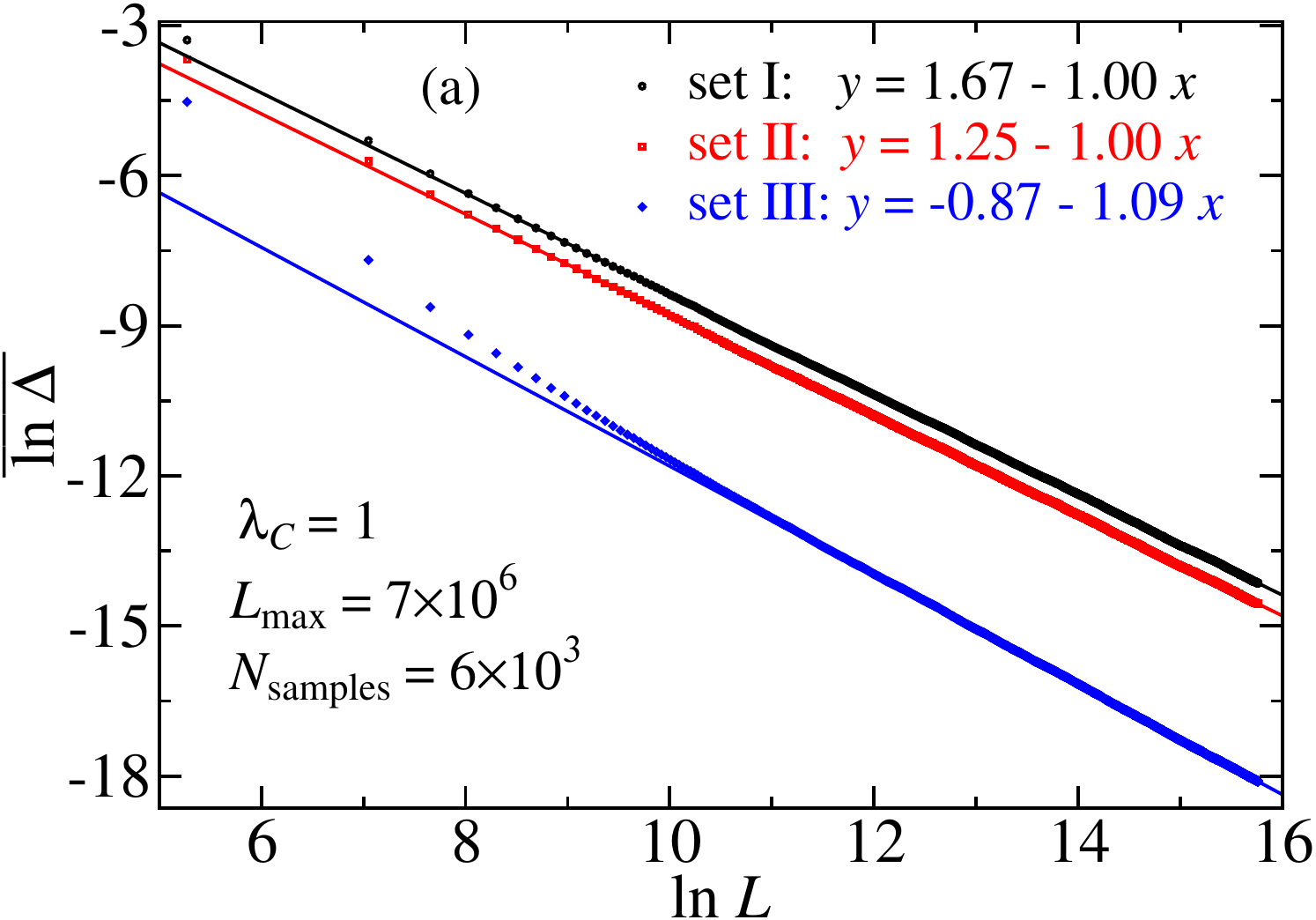}\\
\includegraphics[clip,width=0.8\columnwidth]{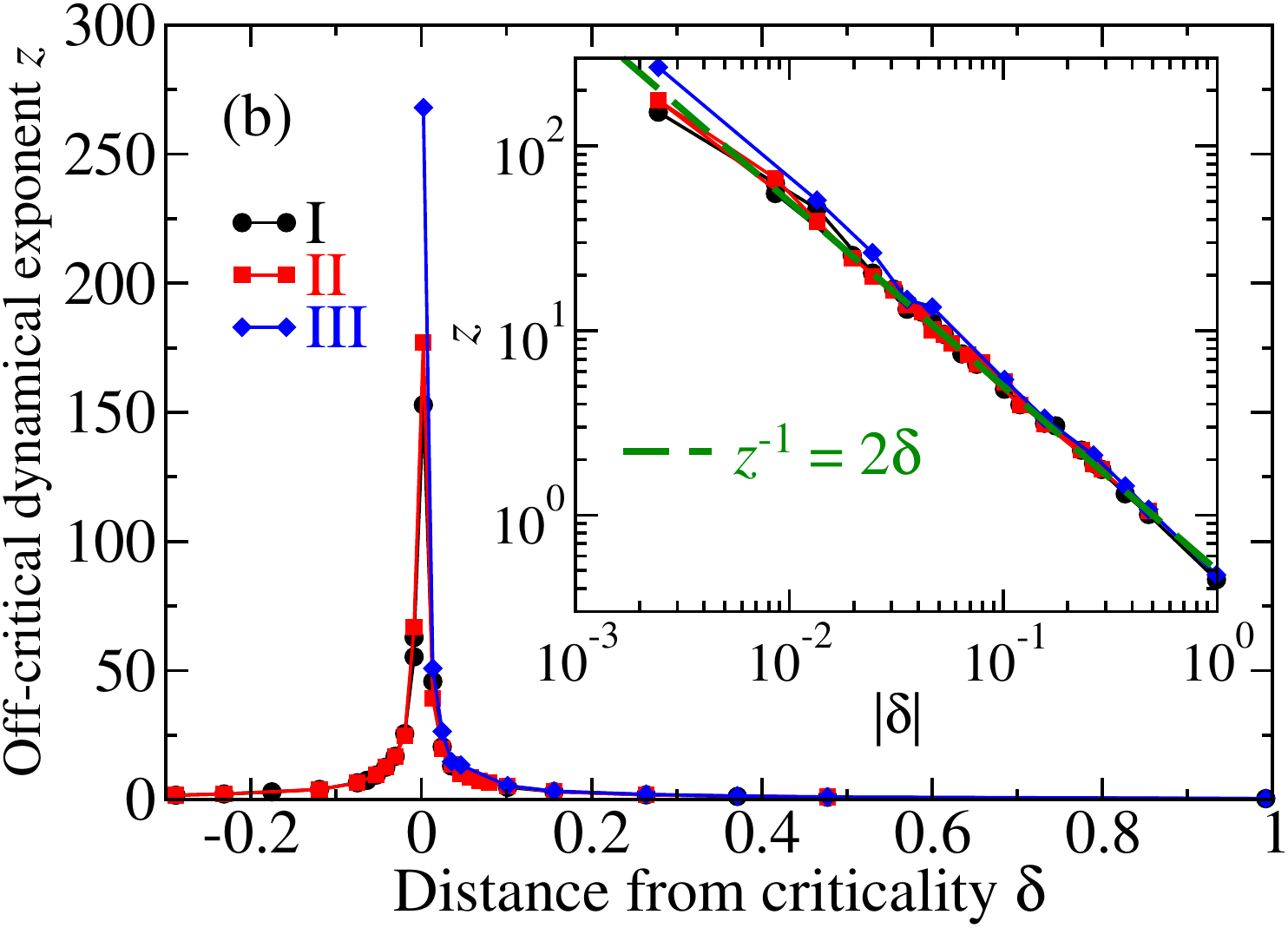}
\par\end{centering}
\caption{\label{fig:FSGap-offcritic-all-random}Finite-size gap analysis of
the Griffiths phases in Fig.~\hyperref[fig:PD-dirty]{\ref{fig:PD-dirty}(b)}.
The coupling constants $\left\{ \lambda_{A,i}\right\} $ are uniformly
distributed in the interval $\left[0,0.2\right]$ (set I), $\left[0.2,1.2\right]$
(set II) and $\left[0.5,1.5\right]$ (set III). In all cases, $\left\{ \lambda_{B,i}\right\} $
are uniformly distributed in the interval $\left[0.5,1.5\right]$,
and $\lambda_{C}$ is uniform and plays the role of a running parameter
throughout the Griffiths phases. (a) The typical finite-size gap $\overline{\ln\Delta}$
as a function of the system size $L$ for $\lambda_{C}=1$. The data
is averaged over $N_{\text{samples}}=6\,000$ disorder realizations
and the error bars are about the size of the data points. Linear fits
for the data are provided in the legends from which the dynamical
exponent $z$ is obtained. The best fits are restricted to system
sizes $L>10^{6}$. (b) The dynamical exponent $z$ as a function of
the distance from criticality Eq.~\eqref{eq:delta-BC}. Lines are
guide to the eyes. Inset: same data in the main panel in log-log scale.
The dashed line is the SDRG asymptotic behavior ($\delta\ll1$) $z=1/(2\delta)$.}
\end{figure}

In case I, we explore the interplay between the two strongest couplings
$\lambda_{B}$ and $\lambda_{C}$ while $\lambda_{A}$ is a much weaker
coupling. The Rare Regions are of $B$-type inside a bulk in the $C$
phase. In case II, the value of $\lambda_{A}$ is increased such that
some Rare Regions of $A$-type also appear. Finally, in case III,
both Rare Regions of $A$- and $B$-type appear equally. At the final
point $\lambda_{C}=\lambda_{C}^{*}$, the multicritical point is reached.

In Fig.~\hyperref[fig:FSGap-offcritic-all-random]{\ref{fig:FSGap-offcritic-all-random}(a)}
we show the relation between $\overline{\ln\Delta}$ (with $\Delta$
being the system finite-size gap) and the system size $L$ for $\lambda_{C}=1$.
From this, we can obtain the off-critical dynamical exponent $z$
by fitting $\Delta\sim L^{-z}$ to the data. Repeating this procedure
for other values of $\lambda_{C}$, we determine how $z$ diverges
as the critical point is approached, see Fig.~\hyperref[fig:FSGap-offcritic-all-random]{\ref{fig:FSGap-offcritic-all-random}(b)}.
It diverges as $z\approx1/\left(2\delta\right)$, with $\delta$ being
the distance from criticality as defined in \eqref{eq:delta-BC}.
We emphasize that this numerically exact result is in agreement with
the SDRG analytical predictions in the $z\rightarrow\infty$ limit.

\subsection{Off-critical finite-size gap II}

We now study the case in which only one of the couplings is disordered,
say, $\lambda_{A}$ uniformly distributed in the interval $\left[0,0.2\right]$.
The corresponding phase diagram is shown in Fig.~\hyperref[fig:PD-dirty]{\ref{fig:PD-dirty}(a)}.
Here, we focus on the off-critical region near the transition between
the $B$- and $C$-phases. Notice that there are no Griffiths phases
surrounding the transition for sufficiently small $\lambda_{A,\text{typ}}$.
Thus, the gap is vanishing only at the transition point $\lambda_{B}=\lambda_{C}$.
Closer to the multicritical point, Griffiths phases appear.

\begin{figure}
\begin{centering}
\includegraphics[clip,width=0.8\columnwidth]{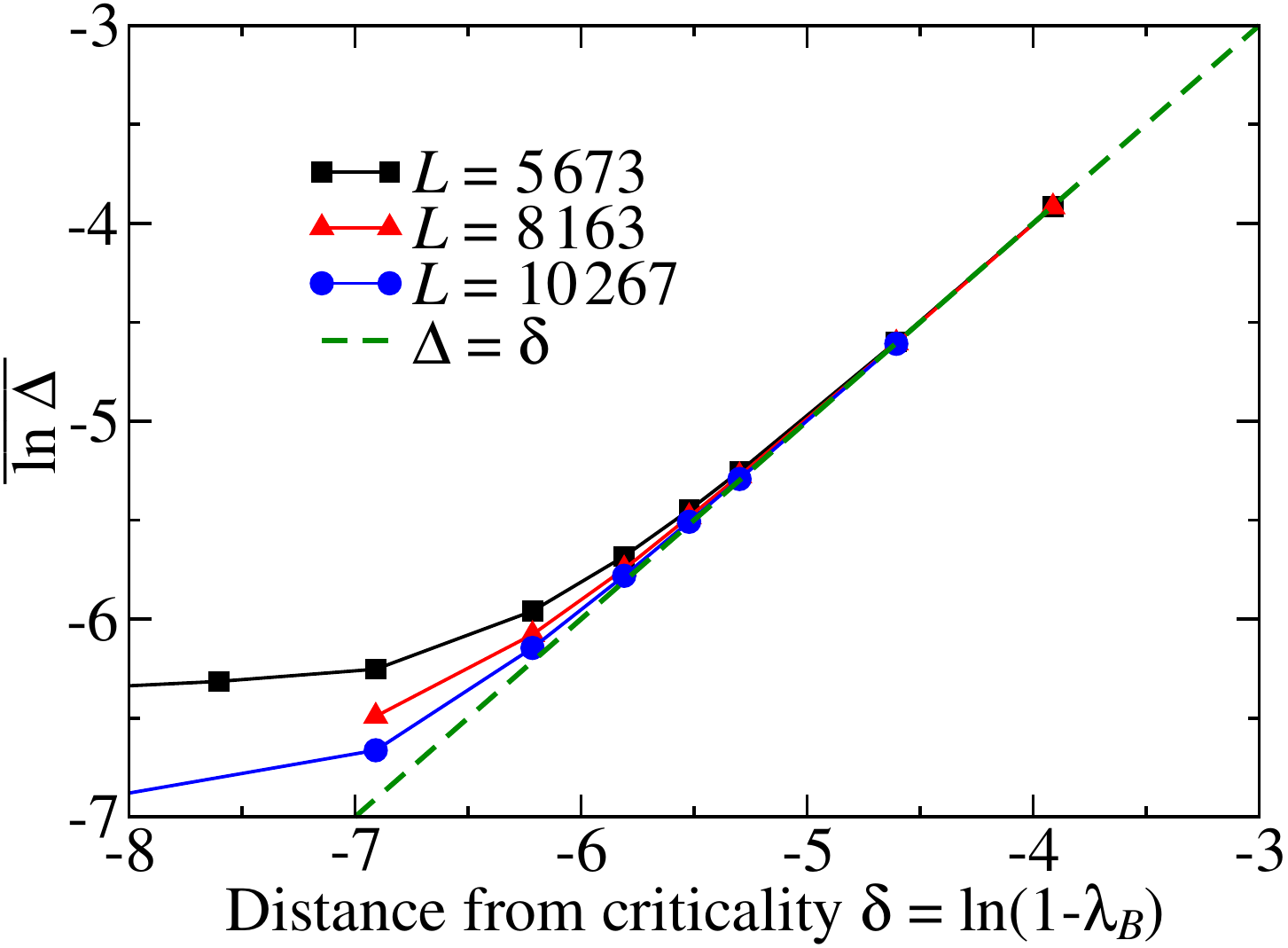}
\par\end{centering}
\caption{\label{fig:Gap-offcritc-noGriffiths}The typical value of the system
gap $\Delta$ as a function of the distance from criticality $\delta=\ln\left(1-\lambda_{B}\right)$
for chains of different sizes $L$. The coupling constants are such
that $\lambda_{A}$ is uniformly distributed in the interval $\left[0,0.2\right]$,
and $\lambda_{B}$ and $\lambda_{C}$ are homogeneous with $\lambda_{C}=1$
and $\lambda_{B}$ being the tuning parameter. The data is average
over $N_{\text{samples}}=1\,500$ disorder realizations. The error
bars are about the symbol sizes. Solid lines are simple guide to the
eyes. The dashed line is the clean behavior in the thermodynamic limit:
$\Delta\sim\delta^{\phi_{\Delta}}$, with $\phi_{\Delta}=1$.}
\end{figure}

We plot in Fig.~\ref{fig:Gap-offcritc-noGriffiths} the typical value
of the system gap $\Delta$ as a function of the distance from criticality
$\delta$. Since $\lambda_{C}$ and $\lambda_{B}$ are homogeneous,
the definition (\ref{eq:delta}) is not useful because of the vanishing
denominator. Here, we simply use $\delta=\ln\left(\lambda_{C}-\lambda_{B}\right)$.
Also, we fix $\lambda_{C}=1>\max\left\{ \lambda_{A}\right\} $ and
use $\lambda_{B}$ as a running parameter. Notice that $\Delta$ diminishes
as in the clean system: $\Delta\sim\delta$. For small $\delta$,
$\Delta$ saturates due to finite-size effects, meaning that the correlation
length is greater than $L$. We recall that our method is not optimal
for gapful systems. This is because larger the gap larger is the number
of coefficients required in the characteristic polynomial~\citep{alcaraz-etal-prb21}.
For that reason, we studied chains of ``small'' sizes up to $10\,267$
sites.

\begin{figure}
\begin{centering}
\includegraphics[clip,width=0.8\columnwidth]{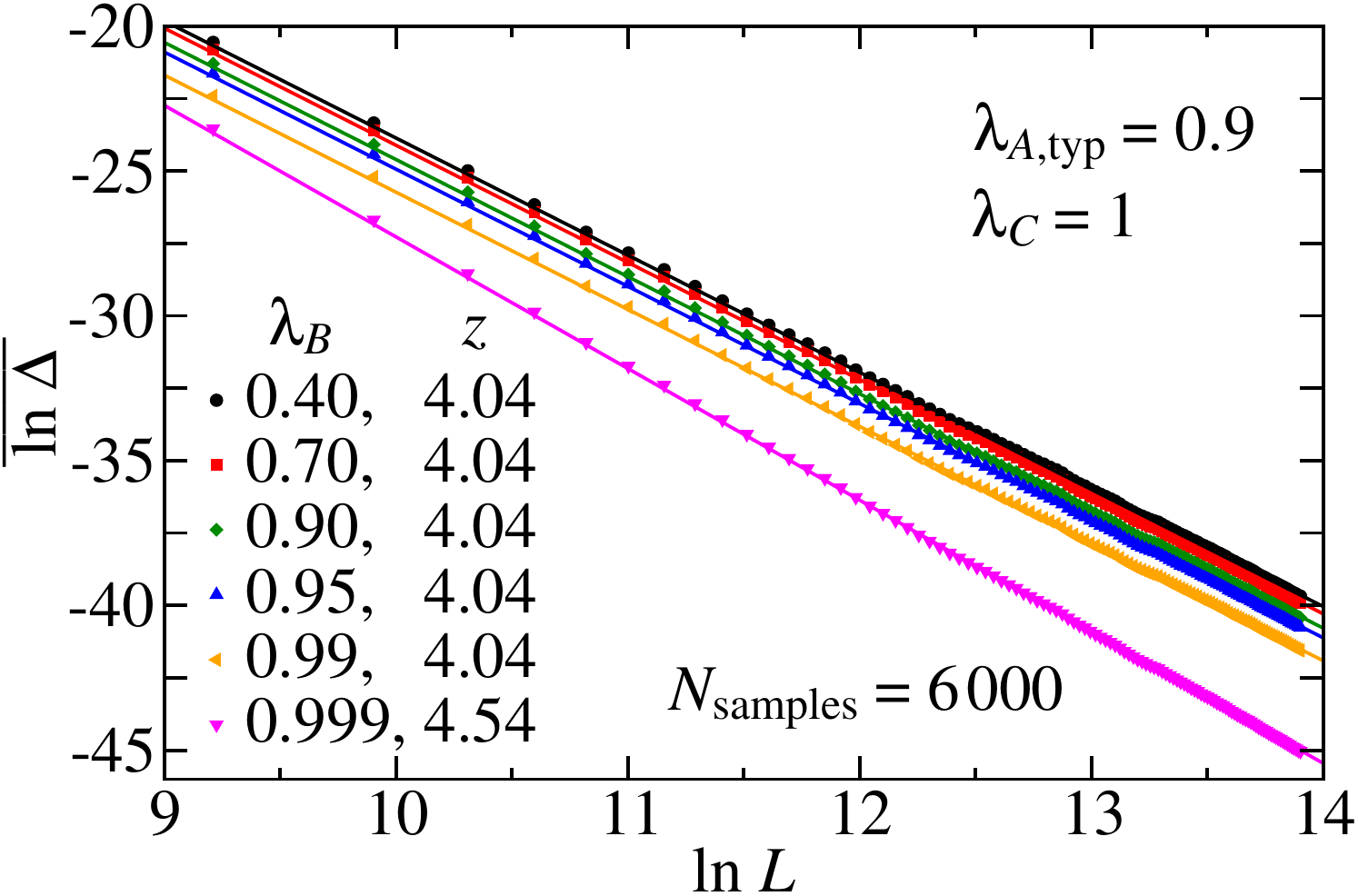}
\par\end{centering}
\caption{\label{fig:Gap-Griffiths-C}The typical value of the finite-size gap
$\Delta$ as a function of the system size $L$. The couplings $\lambda_{B}$
and $\lambda_{C}$ are nonrandom and their values are given in the
legends. The coupling $\lambda_{A}$ is a independent random variable
uniformly distributed in the interval $\left[0,0.9e\right]$ (thus,
typical value $\lambda_{A,\text{typ}}=0.9$). The data is average
over $N_{\text{samples}}=6\,000$ disorder realizations. The error
bars are about the symbol sizes. Solid lines are best fits to Eq.~(\ref{eq:clean-scaling})
in the region $L>e^{12}$. The corresponding Griffiths dynamical exponent
are given in the legends.}
\end{figure}

Having shown the absence of Griffiths phases far from the multicritical
point ($\lambda_{C}>\max\left\{ \lambda_{A}\right\} $), we now show
their existence otherwise. We thus study chains in which $\lambda_{C}=1$,
$\lambda_{A}$ is uniformly distributed in the interval $\left[0,e\lambda_{A,\text{typ}}\right]$,
and $\lambda_{B}$ is the tuning parameter. We report that the finite-size
gap vanishes as $\Delta\sim L^{-z}$ {[}as in Fig.~\hyperref[fig:FSGap-offcritic-all-random]{\ref{fig:FSGap-offcritic-all-random}(a)}{]},
with the value of the off-critical dynamical exponent $z$ increasing
up to a finite value as the transition is approached, see Fig.~\ref{fig:Gap-Griffiths-C}.
Notice that far from the critical point $\lambda_{B}=1$, the off-critical
dynamical exponent is practically insensitive to the distance from
criticality $\delta=1-\lambda_{B}$ as the low-energy behavior is
dominated by Rare Regions which are locally in the $A$ phase.

\subsection{Critical finite-size gap statistics I}

We now address the infinite-randomness critical lines, the dashed
lines in the phase diagram of Fig.~\hyperref[fig:PD-dirty]{\ref{fig:PD-dirty}(a)}
and \hyperref[fig:PD-dirty]{(b)}. To illustrate universality, we
consider two cases: (chain A) $\lambda_{A}$ and $\lambda_{B}$ are,
respectively, uniformly distributed in the intervals $\left[0,0.2\right]$
and $\left[0.5,1.5\right]$, and $\lambda_{C}$ is homogeneous and
equals $\lambda_{B,\text{typ}}=e^{\frac{3\ln3-2\ln2}{2}-1}\approx0.95578$
(ensuring criticality); and (chain B) $\lambda_{A}$, $\lambda_{B}$,
and $\lambda_{C}$ are, respectively, uniformly distributed in the
intervals $\left[0,0.5\right]$, $\left[0,1\right]$, and $\left[0,1\right]$.

\begin{figure}
\begin{centering}
\includegraphics[clip,width=0.85\columnwidth]{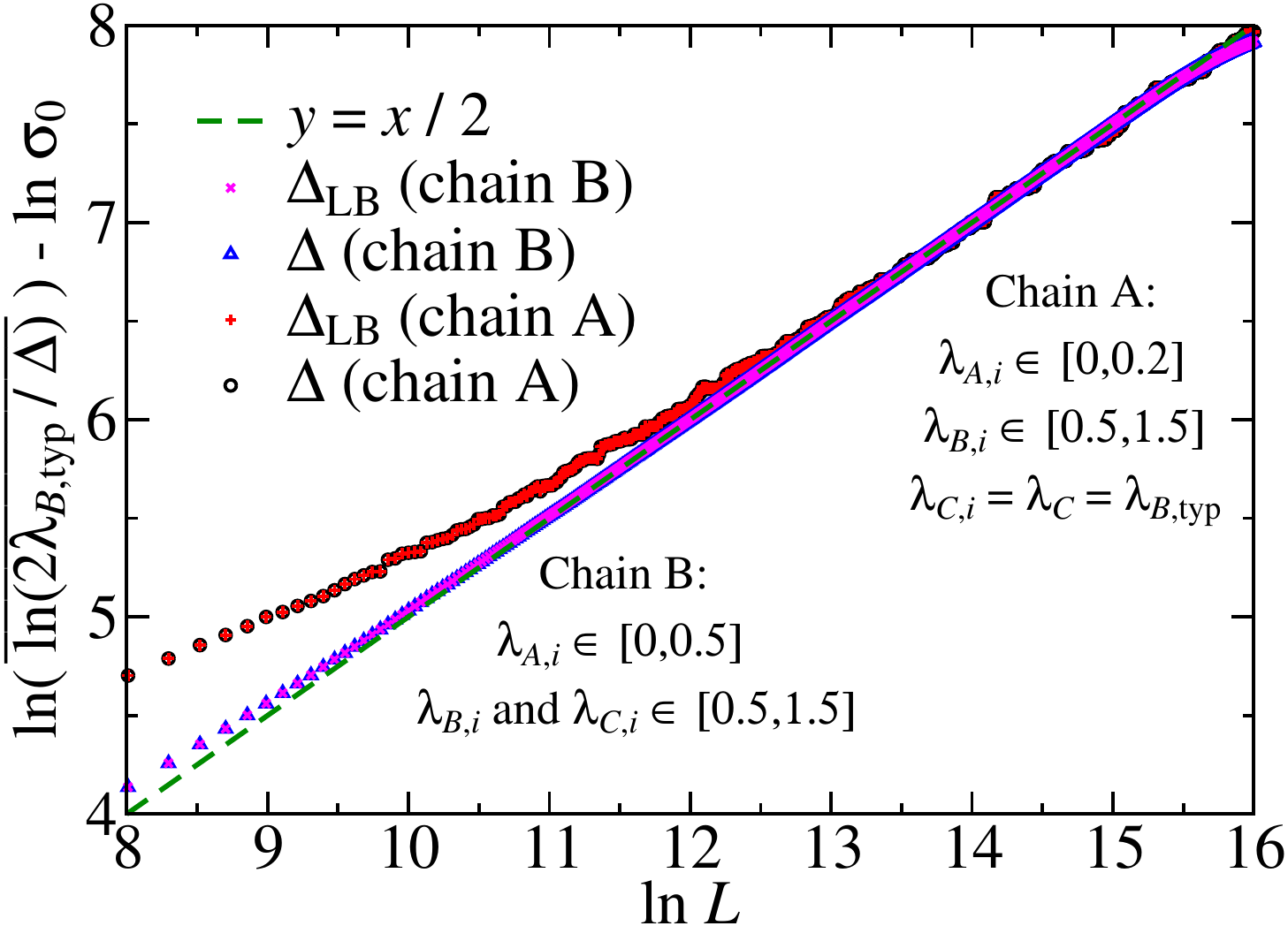}
\par\end{centering}
\caption{\label{fig:FSGap-critic-all-rand}Finite-size gap $\Delta$ (and the
corresponding Laguerre bound $\Delta_{\text{LB}}$) as a function
of the system size for chains A and B (see text). The typical value
is compatible with activated dynamical scaling $\overline{\ln\Delta}\sim-L^{\psi}$,
with universal (disorder-independent) tunneling expoente $\psi=\frac{1}{2}$,
see dashed line and Eq.~\eqref{eq:etap2}. The data is averaged over
$N_{\text{samples}}=10^{3}$ ($4\times10^{4}$) disorder realizations
for chain A (B), and the error bars are about the symbol sizes.}
\end{figure}

In Fig.~\ref{fig:FSGap-critic-all-rand} we plot the typical value
of the finite-size gap {[}and the corresponding Laguerre bound $\Delta_{\text{LB}}$,
see Eqs.~(\ref{eq:LB})\textemdash (\ref{eq:dirty-scaling}){]} for
those critical chains as a function of the system size $L$. As expected
from the activated dynamical scaling \eqref{eq:dirty-scaling}, the
finite-size gap vanishes stretched exponentially fast with universal
(i.e., disorder-independent) tunneling exponent $\psi=1/2$, compatible
with the asymptotic behavior ($L\gg1$) of our data. We call attention
to the fact that $\Delta$ and $\Delta_{\text{LB}}$ become virtual
indistinguishable for $L\gg1$. This feature was already noticed for
the case $p=1$~\citep{alcaraz-etal-prb21}. We conjecture that finite-size
values of $\Delta_{\text{LB}}\rightarrow\Delta$ for $L\rightarrow\infty$
in the case of infinite-randomness criticality. As argued in Ref.~\citealp{alcaraz-etal-prb21},
this is because the largest root of the characteristic polynomial
separates from the other ones as $L$ increases. Thus, only the few
largest coefficients of the polynomial are necessary to accurately
compute it (see Sec.~\ref{subsec:Technical-details}).

\subsection{Critical finite-size gap statistics II}

Here, we investigate the intriguing critical line in which the two
major couplings are uniform, i.e., the horizontal boundary line in
the phase diagram Fig.~\hyperref[fig:PD-dirty]{\ref{fig:PD-dirty}(a)}.
Thus, we take $\lambda_{B}=\lambda_{C}=1$ and $\lambda_{A}$ uniformly
distributed between $0$ and $e\lambda_{A,\text{typ}}$. The typical
value of the $A$ couplings, $\lambda_{A,\text{typ}}$, is used as
a tuning parameter and reaches the multicritical point when $\lambda_{A,\text{typ}}=1$.

\begin{figure}
\begin{centering}
\includegraphics[width=0.9\columnwidth]{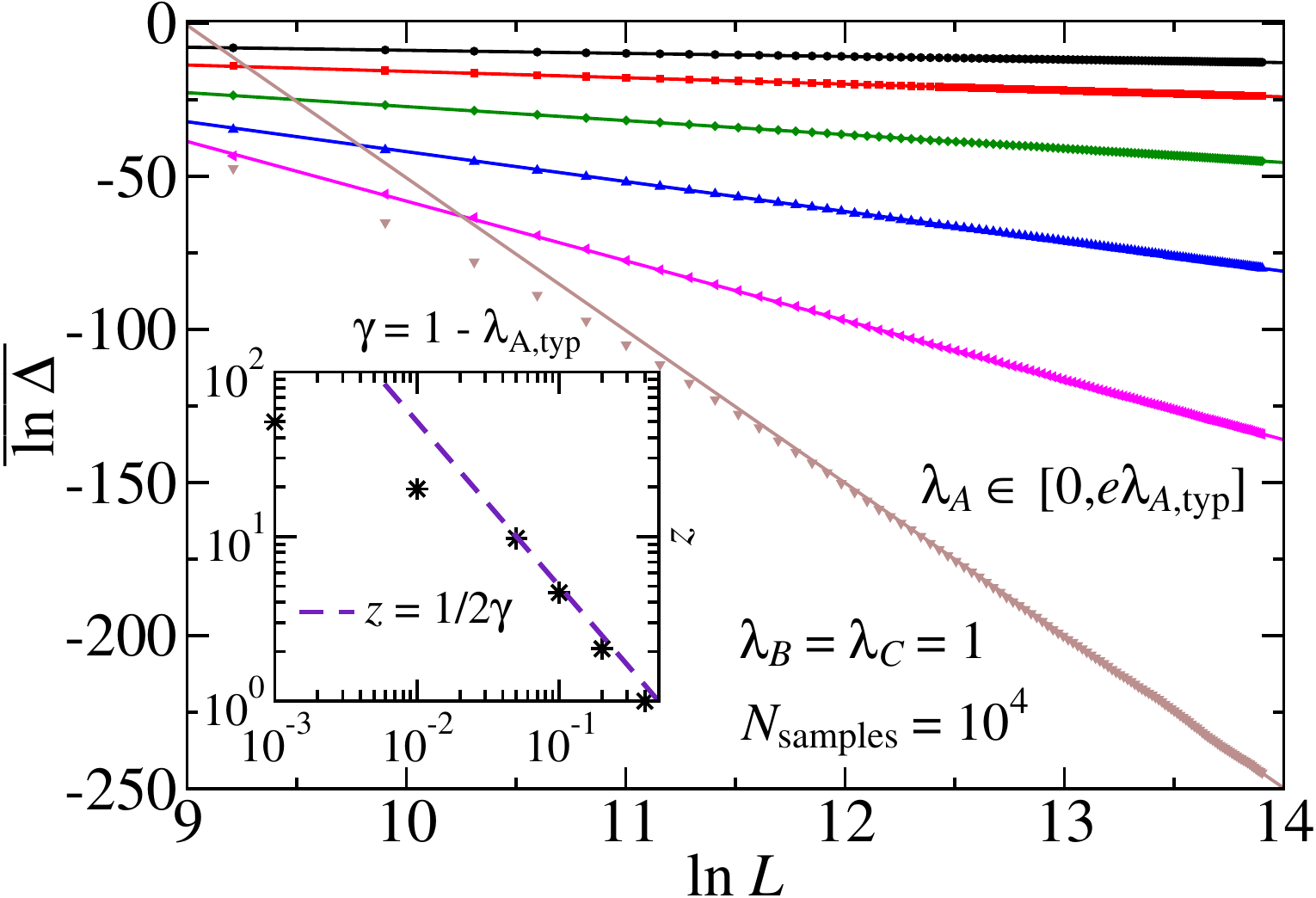}
\par\end{centering}
\caption{\label{fig:FSGap-crit-FRFPs}The typical value of the finite-size
gap $\Delta$ as a function of the system size $L$ for chains with
homogeneous coupling constants $\lambda_{B}=\lambda_{C}=1$ and $\left\{ \lambda_{A}\right\} $
uniformly distributed in the interval $\left[0,e\times\lambda_{A,\text{typ}}\right]$
for various values of the tuning parameter (from top to bottom) $\lambda_{A,\text{typ}}=0.6$,
$0.8$, $0.9$, $0.95$, $0.99$, and $0.999$. The data is averaged
over $N_{\text{samples}}=10^{4}$ disorder realizations and the error
bars are about the symbol sizes. The solid lines are best linear fits
to $\overline{\ln\Delta}\sim-z\ln L$ (constrained to $L>e^{13}$)
from which the critical dynamical exponent is obtained (respectively,
$z=1.00$, $2.08$, $4.56$, $9.76$, $19.48$, and $49.82$) and
shown in the inset as a function of the distance from the multicritical
point $\gamma=1-\lambda_{A,\text{typ}}$. }
\end{figure}

Our results are shown in Fig.~\eqref{fig:FSGap-crit-FRFPs}. Clearly,
sufficiently far from the multicritical point ($\lambda_{A,\text{typ}}\leq\lambda_{A,\text{typ}}^{*}\sim0.6$)
the critical point is in the clean Ising universality class $z=1$
{[}solid line in Fig.~\hyperref[fig:PD-dirty]{\ref{fig:PD-dirty}(a)}{]}.
Approaching the multicritical point ($\lambda_{A,\text{typ}}>\lambda_{A,\text{typ}}^{*}$),
a line of finite-disorder fixed points {[}dotted line in Fig.~\hyperref[fig:PD-dirty]{\ref{fig:PD-dirty}(a)}{]}
is tuned with varying critical dynamical exponent $z$ that diverges
as the multicritical point is approached (see inset). For comparison,
we show the line $z=\left(2\gamma\right)^{-1}$ as predicted by the
SDRG method when $\gamma=1-\lambda_{A,\text{typ}}\rightarrow0$. This
seems to be the trend for $\gamma\apprge0.01$. We cannot rule out
that we are plagued by finite-size effect for smaller $\gamma$. 

\subsection{multicritical finite-size gap statistics}

Here, we finally investigate the multicritical point. We find that
in both phase diagrams of Fig.~\ref{fig:PD-dirty}, the multicritical
point is of infinite-randomness type with universal tunneling exponent
$\psi=1/2$. 

\begin{figure}
\begin{centering}
\includegraphics[clip,width=0.8\columnwidth]{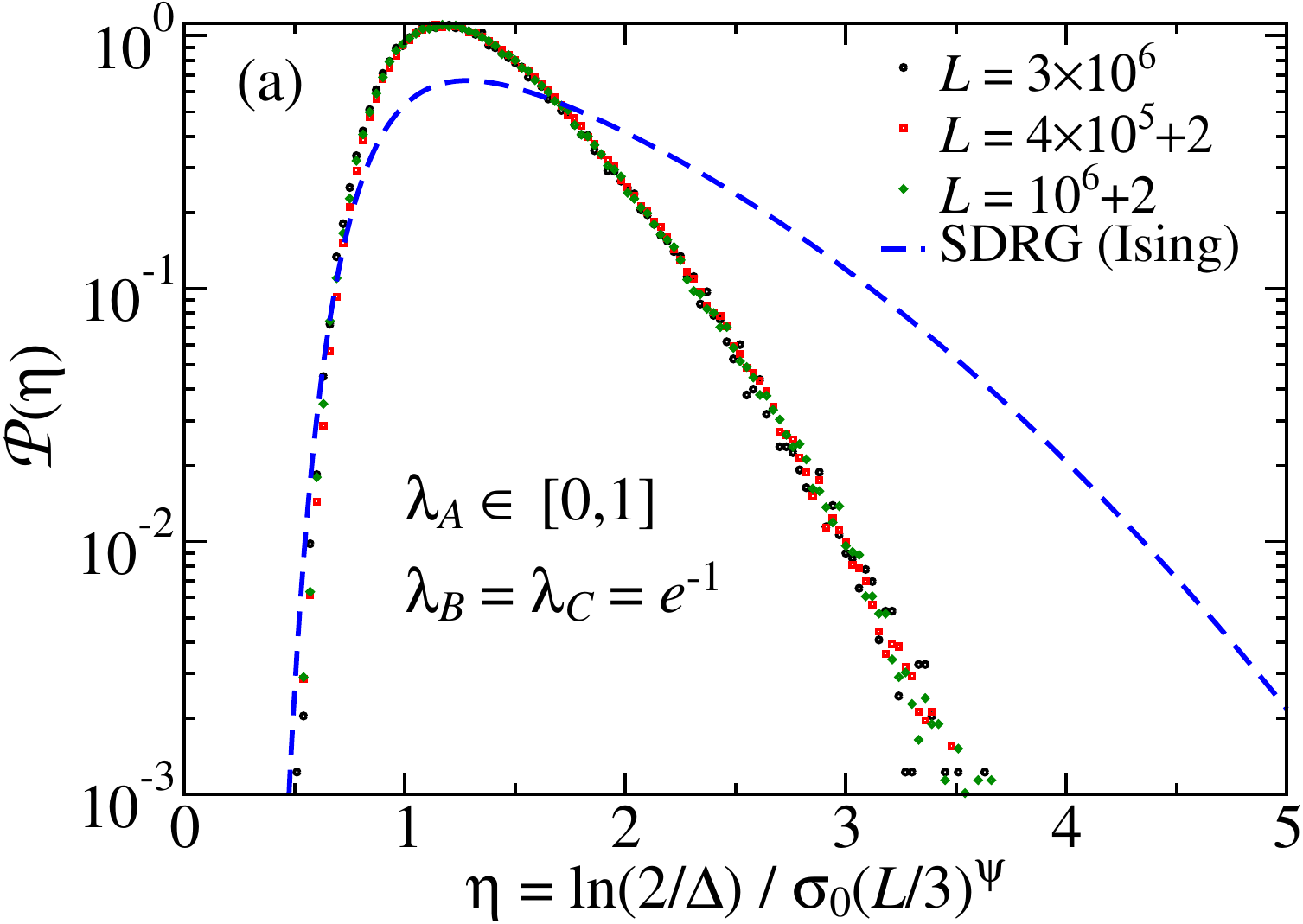}\\
\includegraphics[clip,width=0.8\columnwidth]{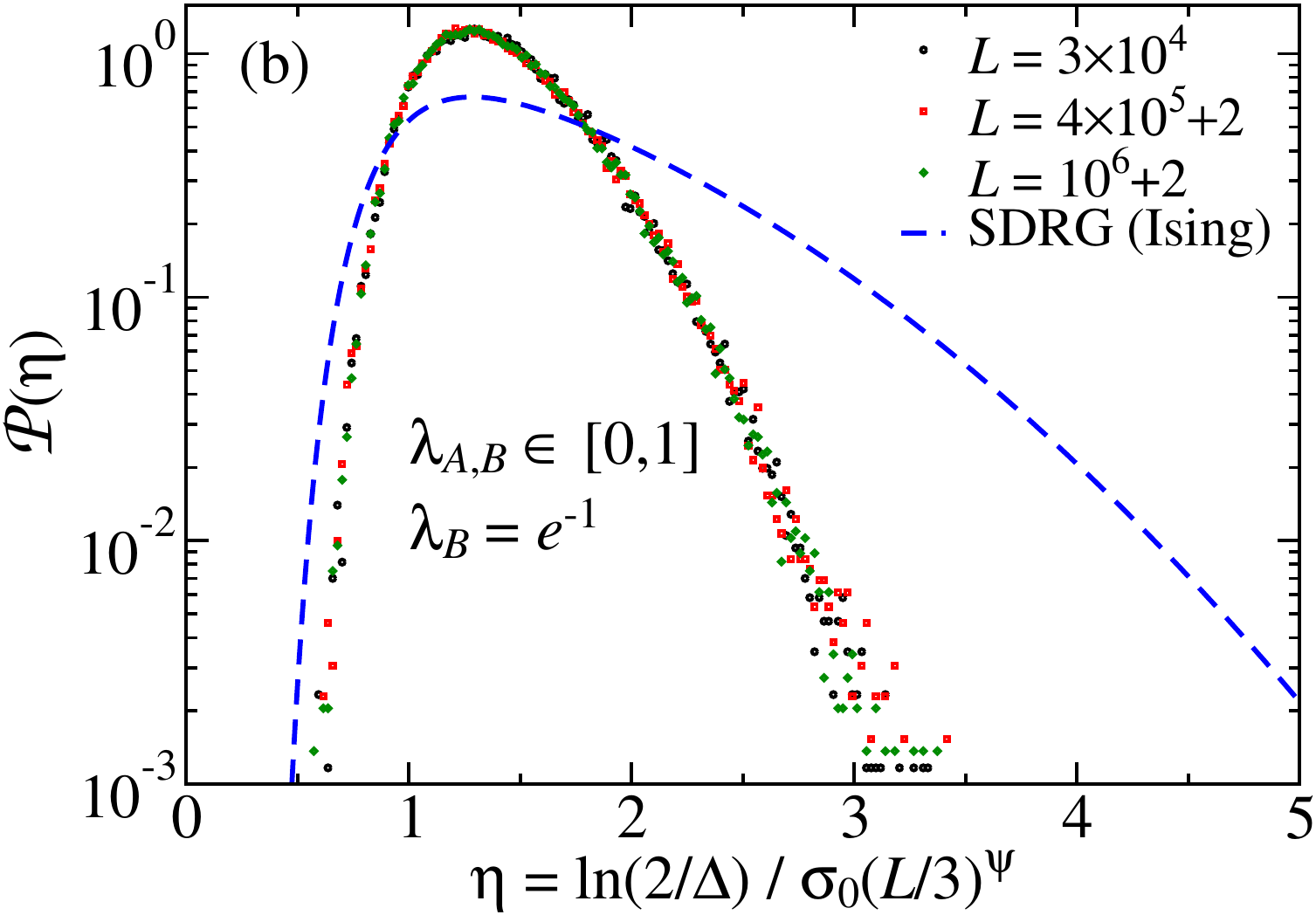}\\
\includegraphics[clip,width=0.8\columnwidth]{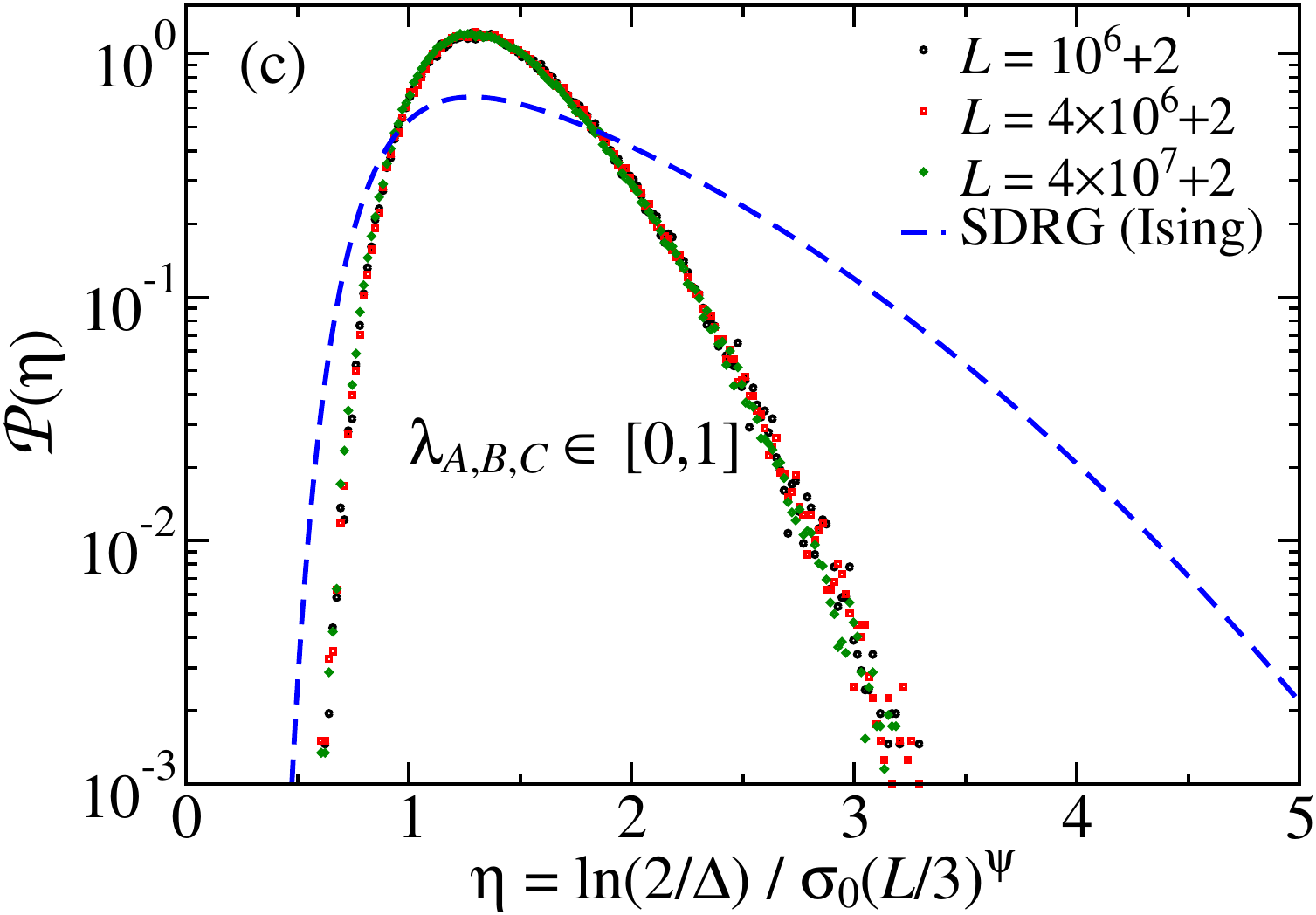}
\par\end{centering}
\caption{\label{fig:P-multi-crit}Finite-size gap distribution ${\cal P}$
for different system sizes $L$. The finite-size gap $\Delta$ is
rescaled according to the scaling variable $\eta$ in Eq.~\eqref{eq:etap2}.
The coupling constants are either homogeneous equal to $e^{-1}$ or
uniformly distributed in the interval $\left[0,1\right]$ as indicated
in the legends. The width $\sigma_{0}$ is, respectively, $\sqrt{1/3}$,
$\sqrt{2/3}$, and $1$ in panels (a), (b), and (c). The normalized
histograms were built using $10^{5}$ to $10^{6}$ disorder realizations.
The dashed line is the SDRG prediction (\ref{eq:PSDRG}) for the transverse-field
Ising chain ($p=1$).}
\end{figure}

In Fig.~\ref{fig:P-multi-crit} we plot the distribution ${\cal P}$
of the finite-size gap $\Delta$ properly rescaled according to Eq.~\eqref{eq:etap2}
for various system lengths and disorder parameters. Clearly, the scaling
variable $\eta$ is sufficient to produce that data collapse for the
system sizes used and this gives us further confidence on the activated
dynamical scaling (\ref{eq:dirty-scaling}) with universal (disorder-independent)
tunneling exponent $\psi=1/2$. We notice, in addition, that the probability
distribution ${\cal P}$ is not universal and is weakly dependent
on the disorder details.\footnote{It is highly nontrivial to understand those details even in the simpler
case of $p=1$ as shown in Ref.~\citealp{getelina-hoyos-ejpb20}.
It involves nonuniversal quantities such as the crossover length between
the clean and infinite-randomness critical points.} Furthermore, ${\cal P}$ is quite different from the SDRG prediction
Eq.~\eqref{eq:PSDRG} for the transverse-field Ising chain ($p=1$),
even though they are governed by the same infinite-disorder fixed
point. Finally, we report (not shown for clarity) that this distribution
depends on the modularity of the lattice size $L$, i.e., ${\cal P}\left(\eta\right)$
depends on $L\mod3$. This is not a surprise since a similar difference
also appears in the clean system. There, the finite-size gap amplitudes
also depend on the modularity, i.e., $\Delta\sim aL^{-z}$ with $a\equiv a\left(L\mod3\right)$.

\section{Conclusions\label{sec:Conclusions}}

We have studied the effects of quenched disorder on a family of free
fermionic models \eqref{eq:Hp} with $(p+1)$-multispin interaction,
paying special attention to the case $p=2$ corresponding to the random
version of the three-spin interacting Fendley model.

When all the coupling constants are generically disordered, the clean
phase transitions (including the multicritical points) are destabilized.
The replacing transition is of infinite-randomness character where
the critical dynamical scaling is activated \eqref{eq:dirty-scaling}
with universal (disorder- and $p$-independent) tunneling exponent
$\psi=1/2$. This exponent governs the low-temperature singular thermodynamic
behavior. The average value of the correlation functions are also
universal and decay algebraically with the spin separations. We do
not have an analytical theory for those exponents and a detailed study
is left for future research. The typical value of the correlations,
on the other hand, behaves quite differently. It decays stretched
exponentially fast with the spin-spin separation. In addition, strong
Griffiths singularities surround the transitions. Although the system
is noncritical with short-range correlations, the spectral gap vanishes.
The singularities are related to the slow dynamics of the domain walls
surrounding the so-called rare regions. This phenomena is very similar
to the domain-wall-induced (or rare-regions-induced) Griffiths singularities
in the dimerized XXZ spin-1/2 chain and in the random transverse-field
Ising model. 

When only one (of few) type of coupling constants are disordered,
the scenario is more involved. When the disordered coupling is weak,
the singular critical behavior of the clean system is stable and there
is no surrounding Griffiths phases. Upon increasing the magnitude
of the disordered coupling, the universality of the transition changes
and is of finite-randomness character. The critical scaling is power-law
conventional \eqref{eq:clean-scaling} but with nonuniversal (disorder-dependent)
dynamical critical exponent $z$. The increasing of the magnitude
of the disordered couplings yields to another effect. It nucleates
rare regions which contributes to off-critical Griffiths singularities.
Interestingly, the finite-randomness criticality extends up to the
multicritical point. 

Although we have explicitly worked out those results for the cases
of $p=1$ and $2$, from symmetry grounds we expect them to be valid
for all $p$ in the family model \eqref{eq:Hp} for sufficiently large
disorder. Evidently, we cannot exclude other scenarios appearing when
$p$ is very large and disorder is weak.

Finally, the agreement between our numerically exact results obtained
for quite large lattice sizes (up to $L\sim10^{7}$ sites) with the
``block'' SDRG is remarkable. We stress that the SDRG methed is not
an exact one. It took more than a decade after its conception to realize
that it can provide asymptotically exact critical exponents and other
universal quantities. This was accomplished comparing the SDRG with
exact diagonalization of a few models such as the transverse-field
Ising chain and the XX spin-1/2 chain and for moderate lattice sizes
($L\sim10^{3}$) and the Heisenberg chain ($L\sim200$). Here, we
have the rare opportunity to show that this is also true for a different
family of models and for quite large chains.

We believe that our method for evaluating the finite-size gaps for
large system sizes with minimal numerical cost ($\sim L$) will be
a useful tool to study the effects of disorder in the phase transition
of other systems.
\begin{acknowledgments}
We thank Jesko Sirker for useful discussions. This work was supported
in part by the Brazilian agencies FAPESP and CNPq. J.A.H. thanks IIT
Madras for a visiting position under the IoE program which facilitated
the completion of this research work. R.A.P. acknowledge support by
the German Research Council (DFG) via the Research Unit FOR 2316.
\end{acknowledgments}

\appendix

\section{The block SDRG method\label{sec:blockSDRG}}

Here, we derive another SDRG decimation procedure. We follow the idea
of Ref.~\onlinecite{hoyos-pre08} where a larger block spin is considered.
It was initially devised to enable the SDRG approach to tackle cases
where the bare disorder in the system is weak. It was important to
correct the SDRG flow from nonphysical features introduced by the
simpler perturbative approach (as in Sec.~\ref{sec:SDRG}). Here,
this approach has a fundamentally different appeal. It consider the
largest spin block in which there are only two energy levels. The
projection is, thus, onto a richer ground-state which may capture
additional features neglected in the simplest approach.

\subsection{Case $p=1$}

Let us start with the simpler case where $p=1$. For convenience we
rewrite the Hamiltonian as 
\begin{equation}
H=\sum_{i}\left\{ -\lambda_{A,i}h_{A,i}-\lambda_{B,i}h_{B,i}\right\} ,
\end{equation}
 where $h_{A,i}=\sigma_{2i-1}^{x}\sigma_{2i}^{z}$ and $h_{B,i}=\sigma_{2i}^{x}\sigma_{2i+1}^{z}$. 

Following the SDRG philosophy, we search for the local Hamiltonian
which exhibits the largest energy gap between its two energy levels.
As the largest spin block still exhibiting only two energy levels
is either $-\lambda_{A,i}h_{A,i}-\lambda_{B,i}h_{B,i}$ or $-\lambda_{B,i}h_{B,i}-\lambda_{A,i+1}h_{A,i+1}$,
we then define the energy cutoff as $\Omega=\max\left\{ \sqrt{\lambda_{A,i}^{2}+\lambda_{B,i}^{2}}\right\} $.\footnote{We could have defined $\Omega=\max\left\{ \sqrt{\lambda_{A,i}^{2}+\lambda_{B,i}^{2}},\sqrt{\lambda_{B,i}^{2}+\lambda_{A,i+1}^{2}}\right\} $.
Since we are interested in the regime where, due to disorder, very
likely neighbor couplings are very distinct from each other, the exact
definition is of no importance.} 

Say that $\Omega=\sqrt{\lambda_{A,2}^{2}+\lambda_{B,2}^{2}}$. In
this case, we treat $H_{0}=-\lambda_{A,2}h_{A,2}-\lambda_{B,2}h_{B,2}$
exactly and project the operators $H_{1}=-\lambda_{B,1}h_{B,1}-\lambda_{A,3}h_{A,3}$
onto its ground-state subspace. More precisely, we are interested
in the projections of the operators $\sigma_{3}^{z}$ and $\sigma_{5}^{x}$.
The ground-state subspace of $H_{0}$ is spanned by $\left|s_{3},s_{5}\right\rangle =\left|s_{3}\right\rangle \otimes\left(\cos\theta\left|\uparrow_{4}\right\rangle +\sin\theta\left|\downarrow_{4}\right\rangle \right)\otimes\left|s_{5}\right\rangle $,
where $\sigma_{3}^{x}\left|s_{3}\right\rangle =s_{3}\left|s_{3}\right\rangle $,
$\sigma_{5}^{z}\left|s_{5}\right\rangle =s_{5}\left|s_{5}\right\rangle $,
$\cos\theta=\lambda_{B,2}s_{5}/\sqrt{2\Omega\left(\Omega-\lambda_{A,2}s_{3}\right)}$,
and $\sin\theta=\left(\Omega-\lambda_{A,2}s_{3}\right)/\sqrt{2\Omega\left(\Omega-\lambda_{A,2}s_{3}\right)}$.
The projected operators are, thus, $\tilde{\sigma}_{3}^{z}=\left|\lambda_{B,2}\right|\tilde{\sigma}^{z}/\Omega$
and $\tilde{\sigma}_{5}^{x}=-\lambda_{A,2}\tilde{\sigma}^{x}\tilde{\tau}^{x}/\Omega$,
where the effective spin-1/2 degrees of freedom $\tilde{\boldsymbol{\sigma}}$
and $\tilde{\boldsymbol{\tau}}$ span the ground-state subspace with
$\tilde{\sigma}^{x}\left|s_{3},s_{5}\right\rangle =s_{3}\left|s_{3},s_{5}\right\rangle $
and $\tilde{\tau}^{z}\left|s_{3},s_{5}\right\rangle =s_{5}\left|s_{3},s_{5}\right\rangle $.
Therefore, the renormalized Hamiltonian is 
\begin{equation}
\tilde{H}_{1}=-\tilde{\lambda}_{B,1}\tilde{h}_{B,1}-\tilde{\lambda}_{A,3}\tilde{h}_{A,3},
\end{equation}
 where $\tilde{h}_{B,1}=\sigma_{2}^{x}\tilde{\sigma}^{z}$, $\tilde{h}_{A,3}=\tilde{\sigma}^{x}\tilde{\tau}^{x}\sigma_{6}^{z}$,
$\tilde{\lambda}_{B,1}=\lambda_{B,1}\left|\lambda_{B,2}\right|/\Omega$,
and $\tilde{\lambda}_{A,3}=-\lambda_{A,2}\lambda_{A,3}/\Omega$. Notice
that the algebra \eqref{eq:algebrap} is preserved in the renormalized
system. In addition, the extra degree of freedom $\tilde{\boldsymbol{\tau}}$
appears only in $\tilde{h}_{A,3}$, and, thus, we interpret the renormalized
chain as being two new chains where the renormalized operators are
$\tilde{h}_{B,1}=\sigma_{2}^{x}\tilde{\sigma}^{z}$, $\tilde{h}_{A,3}=\tilde{\sigma}^{x}\sigma_{6}^{z}$
in both chains. The corresponding renormalized couplings are $\tilde{\lambda}_{B,1}=\lambda_{B,1}\left|\lambda_{B,2}\right|/\Omega$
and $\tilde{\lambda}_{A,3}=\mp\lambda_{A,2}\lambda_{A,3}/\Omega$. 

In the regime of interest where $\lambda_{A,2}$ and $\lambda_{B,2}$
are very distinct from each other, say, $\lambda_{B,2}\gg\lambda_{A,2}$,
then $\Omega\rightarrow\left|\lambda_{B,2}\right|$, the renormalized
operator $\tilde{\boldsymbol{\sigma}}\rightarrow\boldsymbol{\sigma}_{3}$,
and the renormalized couplings become $\tilde{\lambda}_{B,1}=\lambda_{B,1}$
and $\tilde{\lambda}_{A,3}=\mp\lambda_{A,2}\lambda_{A,3}/\left|\lambda_{B,2}\right|$,
which recovers the usual SDRG result \eqref{eq:H1tilde}. If, on the
other hand, $\lambda_{B,2}\ll\lambda_{A,2}$, then $\Omega\rightarrow\left|\lambda_{A,2}\right|$
and the renormalized couplings become $\tilde{\lambda}_{B,1}=\lambda_{B,1}\left|\lambda_{B,2}\right|/\left|\lambda_{A,2}\right|$
and $\tilde{\lambda}_{A,3}=\mp\lambda_{A,3}$, which, at first glance,
does not seem to recover the usual SDRG result \eqref{eq:H1tilde}.
This is not the case. One can correct the signs of the couplings by
appropriately redefining the effective spin-1/2 degrees of freedom
$\tilde{\boldsymbol{\sigma}}$ and $\tilde{\boldsymbol{\tau}}$.

In sum, the block SDRG approach here derived is similar to the usual
SDRG derived in Sec.~\ref{subsec:p1}. Qualitatively, they are the
same in the sense that 2 original operators are removed from the system
in each decimation step and that the renormalized couplings are smaller
than the original ones (both approaches are self-consistent). The
only difference is quantitative: the effective couplings are renormalized
differently. However, this difference vanishes in the strong-disorder
limit.

\subsection{Case $p=2$}

The Hamiltonian of interest is 
\begin{equation}
H=\sum_{i}\left\{ -\lambda_{A,i}h_{A,i}-\lambda_{B,i}h_{B,i}-\lambda_{C,i}h_{C,i}\right\} ,
\end{equation}
 where $h_{A,i}=\sigma_{3i-2}^{x}\sigma_{3i-1}^{z}\sigma_{3i}^{z}$,
$h_{B,i}=\sigma_{3i-1}^{x}\sigma_{3i}^{z}\sigma_{3i+1}^{z}$, and
$h_{C,i}=\sigma_{3i}^{x}\sigma_{3i+1}^{z}\sigma_{3i+2}^{z}$. 

Following the block SDRG philosophy, we search for the largest local
coupling $\Omega=\max\left\{ \left|\lambda_{A,i}\right|,\left|\lambda_{B,i}\right|,\left|\lambda_{C,i}\right|\right\} $,
say, $\Omega=\left|\lambda_{B,2}\right|$, and consider the largest
block of operators which still exhibits only two energy levels. Thus,
we consider as unperturbed Hamiltonian $H_{0}=-\lambda_{A,2}h_{A,2}-\lambda_{B,2}h_{B,2}-\lambda_{C,2}h_{C,2}$.
The renormalized system is obtained by projecting $h_{B,1}$, $h_{C,1}$,
$h_{A,3}$, and $h_{B,3}$ onto the ground-state subspace of $H_{0}$.

The construction of the eigenstates of $H_{0}$ is straightforward
but tedious. There are only two energy levels with energies $\pm E$,
where $E=\sqrt{\lambda_{A,2}^{2}+\lambda_{B,2}^{2}+\lambda_{C,2}^{2}}$.
Each level is $2^{4}$ degenerate as $H_{0}$ involves $5$ spins-1/2
(spins $\boldsymbol{\sigma}_{4,5,\dots,8}$). The ground-state manifold
is the set $\left\{ \left|g_{s_{4},s,s_{7},s_{8}}\right\rangle \right\} $
where $\left|g_{s_{4},s,s_{7},s_{8}}\right\rangle =\left|s_{4},s_{7},s_{8}\right\rangle \otimes\left|s_{s_{4},s_{7},s_{8}}\right\rangle $
since $H_{0}$ is diagonal in the operators $\sigma_{4}^{x}$, $\sigma_{7}^{z}$,
and $\sigma_{8}^{z}$. Here, $\sigma_{4}^{x}\left|g_{s_{4},s,s_{7},s_{8}}\right\rangle =s_{4}\left|g_{s_{4},s,s_{7},s_{8}}\right\rangle $,
$\sigma_{7}^{z}\left|g_{s_{4},s,s_{7},s_{8}}\right\rangle =s_{7}\left|g_{s_{4},s,s_{7},s_{8}}\right\rangle $,
$\sigma_{8}^{z}\left|g_{s_{4},s,s_{7},s_{8}}\right\rangle =s_{8}\left|g_{s_{4},s,s_{7},s_{8}}\right\rangle $,
and the two states $\left|s_{s_{4},s_{7},s_{8}}\right\rangle $ ($s=\pm1$)
encodes the states of spins $5$ and $6$. They are 
\[
\left|\pm_{s_{4},s_{7},s_{8}}\right\rangle =\frac{\left|a_{\vec{s}}\right\rangle +\left|b_{\vec{s}}\right\rangle }{2\sqrt{1+\left\langle b_{\vec{s}}|a_{\vec{s}}\right\rangle }}\pm\frac{\left|a_{\vec{s}}\right\rangle -\left|b_{\vec{s}}\right\rangle }{2\sqrt{1-\left\langle b_{\vec{s}}|a_{\vec{s}}\right\rangle }},
\]
 where 
\begin{eqnarray*}
\left|a_{\vec{s}}\right\rangle  & = & \left|a_{s_{4},s_{7},s_{8}}\right\rangle =N_{-}\left[\lambda_{C,2}s_{7}s_{8}\left|\rightarrow_{5},\uparrow_{6}\right\rangle \right.\\
 &  & \left.+\left(E-\lambda_{B,2}s_{7}\right)\left|\rightarrow_{5},\downarrow_{6}\right\rangle -\lambda_{A,2}s_{4}\left|\leftarrow_{5},\downarrow_{6}\right\rangle \right],\\
\left|b_{\vec{s}}\right\rangle  & = & \left|b_{s_{4},s_{7},s_{8}}\right\rangle =N_{+}\left[\left(E+\lambda_{B,2}s_{5}\right)\left|\rightarrow_{5},\uparrow_{6}\right\rangle \right.\\
 &  & \left.+\lambda_{C,2}s_{7}s_{8}\left|\rightarrow_{5},\downarrow_{6}\right\rangle +\lambda_{A2}s_{4}\left|\leftarrow_{5},\uparrow_{6}\right\rangle \right],
\end{eqnarray*}
 $N_{\pm}=1/\sqrt{2E\left(E\pm\lambda_{B,2}s_{7}\right)}$ are normalization
constants, and $\left\langle b_{\vec{s}}|a_{\vec{s}}\right\rangle =s_{7}s_{8}\lambda_{C,2}/\sqrt{\lambda_{A2}^{2}+\lambda_{C2}^{2}}$.

We define the four spin-1/2 operators $\tilde{\boldsymbol{\sigma}}_{4}$,
$\tilde{\boldsymbol{\sigma}}$, $\tilde{\boldsymbol{\sigma}}_{7}$,
and $\tilde{\boldsymbol{\sigma}}_{8}$ which span the ground-state
subspace in the following manner: $\tilde{\sigma}_{4}^{x}\left|g_{s_{4},s,s_{7},s_{8}}\right\rangle =s_{4}\left|g_{s_{4},s,s_{7},s_{8}}\right\rangle $,
$\tilde{\sigma}^{z}\left|g_{s_{4},s,s_{7},s_{8}}\right\rangle =s\left|g_{s_{4},s,s_{7},s_{8}}\right\rangle $,
$\tilde{\sigma}_{7}^{z}\left|g_{s_{4},s,s_{7},s_{8}}\right\rangle =s_{7}\left|g_{s_{4},s,s_{7},s_{8}}\right\rangle $,
and $\tilde{\sigma}_{8}^{z}\left|g_{s_{4},s,s_{7},s_{8}}\right\rangle =s_{8}\left|g_{s_{4},s,s_{7},s_{8}}\right\rangle $. 

We are now ready to project the $H_{1}$ onto the ground states of
$H_{0}$. Let us start by projecting the operator $\sigma_{4}^{z}$.
The matrix element is 
\begin{align*}
\left\langle g_{t_{4},t,t_{7},t_{8}}\left|\sigma_{4}^{z}\right|g_{s_{4},s,s_{7},s_{8}}\right\rangle  & =\delta_{s_{4},-t_{4}}\delta_{s_{7},t_{7}}\delta_{s_{8},t_{8}}\left\langle t_{-s_{4},s_{7},s_{8}}|s_{s_{4},s_{7},s_{8}}\right\rangle \\
=\frac{\delta_{s_{4},-t_{4}}\delta_{s_{7},t_{7}}\delta_{s_{8},t_{8}}}{\sqrt{\lambda_{A,2}^{2}+\lambda_{C,2}^{2}}} & \left(-\frac{\left|\lambda_{A,2}\right|\lambda_{B,2}s_{7}s\delta_{s,t}}{E}+s_{7}s_{8}\lambda_{C,2}\delta_{s,-t}\right).
\end{align*}
 Thus, the projected operator is 
\[
\tilde{\sigma}_{4}^{z}\left(-\frac{\left|\lambda_{A,2}\right|\lambda_{B,2}}{E\sqrt{\lambda_{A,2}^{2}+\lambda_{C,2}^{2}}}\tilde{\sigma}^{z}\tilde{\sigma}_{7}^{z}+\frac{\lambda_{C,2}}{\sqrt{\lambda_{A,2}^{2}+\lambda_{C,2}^{2}}}\tilde{\sigma}^{x}\tilde{\sigma}_{7}^{z}\tilde{\sigma}_{8}^{z}\right).
\]

Similarly, projecting $\sigma_{8}^{x}$, $\sigma_{7}^{x}\sigma_{8}^{z}$,
and $\sigma_{4}^{z}\sigma_{5}^{z}$ we obtain, respectively, 
\[
\frac{\left|\lambda_{A,2}\right|}{\sqrt{\lambda_{A,2}^{2}+\lambda_{C,2}^{2}}}\tilde{\sigma}_{8}^{x}-\frac{\lambda_{B,2}\lambda_{C,2}}{E\sqrt{\lambda_{A,2}^{2}+\lambda_{C,2}^{2}}}\tilde{\sigma}^{y}\tilde{\sigma}_{8}^{y},
\]
\[
\frac{\left|\lambda_{A,2}\right|}{E}\tilde{\sigma}_{7}^{x}\tilde{\sigma}_{8}^{z},\mbox{ and }-\text{sign}\left(\lambda_{A,2}\right)\frac{\lambda_{C,2}}{E}\tilde{\sigma}_{4}^{y}\tilde{\sigma}^{y}\tilde{\sigma}_{7}^{z}\tilde{\sigma}_{8}^{z}.
\]

Collecting all those terms, we find that $\tilde{H}_{1}=-\tilde{\lambda}_{B,1}\tilde{h}_{B,1}-\tilde{\lambda}_{C,1}\tilde{h}_{C,1}-\tilde{\lambda}_{A,3}\tilde{h}_{A,3}-\tilde{\lambda}_{B,3}\tilde{h}_{B,3},$
where 
\[
\tilde{\lambda}_{C,1}=-\text{sign}\left(\lambda_{A,2}\right)\frac{\lambda_{C,1}\lambda_{C,2}}{E},\ \tilde{h}_{C,1}=\sigma_{3}^{x}\tilde{\sigma}_{4}^{y}\tilde{\sigma}^{y}\tilde{\sigma}_{7}^{z}\tilde{\sigma}_{8}^{z},
\]

\[
\tilde{\lambda}_{A,3}=\frac{\left|\lambda_{A,2}\right|\lambda_{A,3}}{E},\ \tilde{h}_{A,3}=\tilde{\sigma}_{7}^{x}\tilde{\sigma}_{8}^{z}\sigma_{9}^{z},
\]
 
\[
\tilde{\lambda}_{B,1}\tilde{h}_{B,1}=\lambda_{B,1}\sigma_{2}^{x}\sigma_{3}^{z}\tilde{\sigma}_{4}^{z}\left(\frac{\lambda_{C,2}E\tilde{\sigma}^{x}\tilde{\sigma}_{8}^{z}-\left|\lambda_{A,2}\right|\lambda_{B,2}\tilde{\sigma}^{z}}{E\sqrt{\lambda_{A,2}^{2}+\lambda_{C,2}^{2}}}\right)\tilde{\sigma}_{7}^{z},
\]
 
\[
\tilde{\lambda}_{B,3}\tilde{h}_{B,3}=\lambda_{B,3}\left(\frac{\left|\lambda_{A,2}\right|E\tilde{\sigma}_{8}^{x}-\lambda_{B,2}\lambda_{C,2}\tilde{\sigma}^{y}\tilde{\sigma}_{8}^{y}}{E\sqrt{\lambda_{A,2}^{2}+\lambda_{C,2}^{2}}}\right)\sigma_{9}^{z}\sigma_{10}^{z}.
\]

\subsubsection{The algebra of the renormalized operators}

As in the simpler SDRG approach (Sec.~\ref{subsec:SDRGp2}) the algebra
\eqref{eq:algebrap} is not preserved. However, it is almost preserved.
The only operators changing the algebra structure are the operators
$\tilde{h}_{B,1}$ and $\tilde{h}_{B,3}$. They anticommutate with
the nearest- and next-nearest-neighbor operators and commute with
the farther-neighbor operators, preserving the algebra \eqref{eq:algebrap}.
However, they do not commute with each other. Instead, $\left[\tilde{\lambda}_{B,1}\tilde{h}_{B,1},\tilde{\lambda}_{B,3}\tilde{h}_{B,3}\right]\propto\lambda_{B,1}\lambda_{A,2}\lambda_{C,2}\lambda_{B,3}/E^{2}.$
Interestingly, in the regime $\left|\lambda_{B,2}\right|\gg\left|\lambda_{A,2}\right|,\left|\lambda_{C,2}\right|$
(which is the one we are interested in), they can be considered commuting
operators since 
\begin{equation}
\frac{\left[\tilde{\lambda}_{B,1}\tilde{h}_{B,1},\tilde{\lambda}_{B,3}\tilde{h}_{B,3}\right]}{E_{B,1}E_{B,3}}\mbox{ is of order }{\cal O}\left(\frac{\lambda_{A,2}\lambda_{C,2}}{\lambda_{B,2}^{2}}\right)\rightarrow0,\label{eq:commutator}
\end{equation}
 where $\pm E_{B,1}$ ($\pm E_{B,3}$) are the energy levels of $\tilde{h}_{B,1}$
($\tilde{h}_{B,3}$). Precisely, 
\[
E_{B,1}=\left|\lambda_{B,1}\right|\sqrt{\frac{\lambda_{C,2}^{2}}{\lambda_{A,2}^{2}+\lambda_{C,2}^{2}}+\frac{\lambda_{A,2}^{2}}{\lambda_{A,2}^{2}+\lambda_{C,2}^{2}}\left(\frac{\lambda_{B,2}}{E}\right)^{2}}.
\]
 (The value of $E_{B,3}$ is that of $E_{B,1}$ with $\lambda_{B,1}$
replaced by $\lambda_{B,3}$ and $\lambda_{A,2}$ interchanged with
$\lambda_{C,2}$.) In the regime $\left|\lambda_{B,2}\right|\gg\left|\lambda_{A,2}\right|,\left|\lambda_{C,2}\right|$,
then $E_{B,1}\approx\left|\lambda_{B,1}\right|$ and $E_{B,3}\approx\left|\lambda_{B,3}\right|$.
In addition, the renormalized operators simplify to 

\begin{equation}
\tilde{h}_{B,1}=\sigma_{2}^{x}\sigma_{3}^{z}\tilde{\sigma}_{4}^{z}\left(\frac{\lambda_{C,2}\tilde{\sigma}^{x}\tilde{\sigma}_{8}^{z}-\text{sign}\left(\lambda_{B,2}\right)\left|\lambda_{A,2}\right|\tilde{\sigma}^{z}}{\sqrt{\lambda_{A,2}^{2}+\lambda_{C,2}^{2}}}\right)\tilde{\sigma}_{7}^{z},\label{eq:B1}
\end{equation}
 
\begin{equation}
\tilde{h}_{B,3}=\left(\frac{\left|\lambda_{A,2}\right|\tilde{\sigma}_{8}^{x}-\text{sign}\left(\lambda_{B,2}\right)\lambda_{C,2}\tilde{\sigma}^{y}\tilde{\sigma}_{8}^{y}}{\sqrt{\lambda_{A,2}^{2}+\lambda_{C,2}^{2}}}\right)\sigma_{9}^{z}\sigma_{10}^{z}.\label{eq:B3}
\end{equation}
 and the corresponding renormalized couplings are simply $\tilde{\lambda}_{B,1}=\lambda_{B,1}$
and $\tilde{\lambda}_{B,3}=\lambda_{B,3}$. 

It is important to notice that the commutator \eqref{eq:commutator}
is vanishing only when $\left|\lambda_{B,2}\right|\gg\left|\lambda_{A,2}\right|,\left|\lambda_{C,2}\right|$.
In any other regime, it is of order unity. This justify the choice
of the block Halmiltonian $H_{0}$. Having localized the largest coupling
in the system, the block must take into account the nearest-neighbors.

Having devised a decimation procedure that preserves the algebra \eqref{eq:algebrap},
we now show further simplifications which appear at and near the critical
points. At or near the transition lines of Fig.~\hyperref[fig:PD-dirty]{\ref{fig:PD-dirty}(b)},
one of the couplings are much smaller than the competing ones. Without
loss of generality, say that $\left|\lambda_{C,2}\right|\ll\left|\lambda_{A,2}\right|$.
Near the multicritical point, on the other hand, all couplings are
of the same order of magnitude. However, under renormalization, the
effective disorder is large. Thus, very likely either $\left|\lambda_{A,2}\right|\gg\left|\lambda_{C,2}\right|$
or $\left|\lambda_{A,2}\right|\ll\left|\lambda_{C,2}\right|$. Without
loss of generality, let us consider that $\left|\lambda_{C,2}\right|\ll\left|\lambda_{A,2}\right|$. 

Thus, the regime $\left|\lambda_{B,2}\right|\gg\left|\lambda_{A,2}\right|\gg\left|\lambda_{C,2}\right|$
is quite general near and at the transitions. In that case the $B$-type
operators and renormalized coupling constants simplify to 
\[
\tilde{\lambda}_{B,1}=-\text{sign}\left(\lambda_{B,2}\right)\lambda_{B,1},\ \tilde{h}_{B,1}=\sigma_{2}^{x}\sigma_{3}^{z}\tilde{\sigma}_{4}^{z}\tilde{\sigma}^{z}\tilde{\sigma}_{7}^{z},
\]
 
\[
\tilde{\lambda}_{B,3}=\lambda_{B,3},\mbox{ and }\tilde{h}_{B,3}=\tilde{\sigma}_{8}^{x}\sigma_{9}^{z}\sigma_{10}^{z}.
\]
 (An analogous simplification is obtained in the case $\left|\lambda_{A,2}\right|\ll\left|\lambda_{C,2}\right|$
after a convenient redefinition of $\tilde{\boldsymbol{\sigma}}$.)

As a final simplification, notice that the new effective degrees of
freedom $\tilde{\boldsymbol{\sigma}}_{4}$ and $\tilde{\boldsymbol{\sigma}}$
appear only in $\tilde{h}_{B,1}$ and $\tilde{h}_{C,1}$ with the
combination $\tilde{\sigma}_{4}^{z}\tilde{\sigma}^{z}$ and $\tilde{\sigma}_{4}^{y}\tilde{\sigma}^{y}$
which commute with each other. Thus, we can diagonalize the system
in those degrees of freedom. The eigenstates are $(\left|\uparrow_{4},\uparrow_{\sim}\right\rangle \pm\left|\downarrow_{4},\downarrow_{\sim}\right\rangle )/\sqrt{2}$
and $(\left|\uparrow_{4},\downarrow_{\sim}\right\rangle \pm\left|\uparrow_{4},\downarrow_{\sim}\right\rangle )/\sqrt{2}$.
This means that we can fix the degrees of freedom of spins $4$ and
$5$ in one of these states and obtain four different effective Halmiltonians,
namely, 
\begin{equation}
\tilde{H}_{1}=\pm\lambda_{B,1}\tilde{h}_{B,1}\pm\tilde{\lambda}_{C,1}\tilde{h}_{C,1}-\tilde{\lambda}_{A,3}\tilde{h}_{A,3}-\lambda_{B,3}\tilde{h}_{B,3},\label{eq:H1-new}
\end{equation}
 where the renormalized operators are $\tilde{h}_{B,1}=\sigma_{2}^{x}\sigma_{3}^{z}\tilde{\sigma}_{7}^{z}$,
$\tilde{h}_{C,1}=\sigma_{3}^{x}\tilde{\sigma}_{7}^{z}\tilde{\sigma}_{8}^{z}$,
$\tilde{h}_{A,3}=\tilde{\sigma}_{7}^{x}\tilde{\sigma}_{8}^{z}\sigma_{9}^{z}$,
and $\tilde{h}_{B,3}=\tilde{\sigma}_{8}^{x}\sigma_{9}^{z}\sigma_{10}^{z}$,
and the renormalized couplings are 
\begin{equation}
\tilde{\lambda}_{C,1}=\frac{\lambda_{C,1}\lambda_{C,2}}{\Omega}\mbox{ and }\tilde{\lambda}_{A,3}=\frac{\left|\lambda_{A,2}\right|\lambda_{A,3}}{\Omega}.\label{eq:lambda-new}
\end{equation}
 The decimation procedure \eqref{eq:H1-new} and \eqref{eq:lambda-new}
is depicted in Fig.~\hyperref[fig:Decimation-p2]{\ref{fig:Decimation-p2}(a)}. 

\subsubsection{On the difference between the usual and block SDRG approaches}

At first glance, the block SDRG approach here devised is not qualitatively
different from usual SDRG described in Sec.~\ref{subsec:SDRGp2}.
Clearly, it has the advantage, however, of providing a clear reason
for neglecting the new operator $\tilde{h}_{AC}$ in Eq.~\eqref{eq:hab}
near and at the phase transitions. Analyzing a bit further the differences
between these approaches, we notice that the usual SDRG method generates
a hybrid operator $\tilde{h}_{AC}$ originated from $A$- and $C$-type
original operators. No such operator is generated in the block SDRG
method. Instead, the $B$-type operators \eqref{eq:B1} and \eqref{eq:B3}
are actually $B$-type operators plus $AB$- and $CB$-type operators
as well. These hybrid operators, however, in the regime of different
local energy scales ($\left|\lambda_{B,2}\right|\gg\left|\lambda_{A,2}\right|\gg\left|\lambda_{C,2}\right|$)
can be neglected. The reason is the following. Consider for instance
the terms inside parenthesis in Eq.~\eqref{eq:B1}, clearly, the
term originating the $AB$-type operators (the first term) can be
viewed as a small tilt to the molecular field of the $B$-type operator
(the second term). In the regime $\left|\lambda_{A,2}\right|\gg\left|\lambda_{C,2}\right|$,
this small transverse molecular field can be neglected.

This possibility of neglecting a hybrid operator in detrimental of
a ``pure'' one does not appear in the usual SDRG. Maybe because the
local Hilbert space (that of $H_{0}$) is not large enough to accommodate
more than one possibility of renormalization.

\section{Simplified SDRG flow\label{sec:SDRG-flow-Eq}}

In this section, we consider the simplified version of the SDRG decimation
procedure for the $p=2$ case. As stated in Sec.~\ref{subsec:SDRGp2},
the first decimation is such that five operators are removed $h_{1,2,\dots,5}$
and three new ones are inserted $\tilde{h}_{1,2,3}$ in the system
(see Fig.~\ref{fig:Decimation-p2}). If, for some reason, one could
neglect $\tilde{h}_{2}$, the algebra structure would not change after
decimation. This allows for a simplification of the problem. The new
operator $\tilde{h}_{1}$ ($\tilde{h}_{3}$) corresponds to a renormalization
of the couplings on the sites $3i-2$ ($3i-1$). We then can write
an equation for the transformation of the coupling constant distributions.
The transformation for the distribution of the couplings $3i-2$ when
the cutoff $\Omega$ diminished to $\Omega-d\Omega$ is 
\begin{equation}
{\cal P}_{A}\left(\lambda,\Omega-d\Omega\right){\cal N}={\cal P}_{A}\left(\lambda,\Omega\right)+R_{B}\left[{\cal P}_{A}\right]+R_{C}\left[{\cal P}_{A}\right],\label{eq:Pa}
\end{equation}
 where ${\cal N}$ is a normalization constant (which we define later)
and
\begin{align}
R_{X}\left[{\cal P}_{Y}\right] & ={\cal P}_{X}\left(\Omega,\Omega\right)d\Omega\int d\lambda_{1}d\lambda_{4}{\cal P}_{Y}\left(\lambda_{1},\Omega\right){\cal P}_{Y}\left(\lambda_{4},\Omega\right)\nonumber \\
\times & \left[-\delta\left(\lambda-\lambda_{1}\right)-\delta\left(\lambda-\lambda_{4}\right)+\delta\left(\lambda-\frac{\lambda_{1}\lambda_{4}}{\Omega}\right)\right],
\end{align}
 is a functional quantifying the change in ${\cal P}_{Y}$ when a
$X$-type coupling constant is decimated. Here, ${\cal P}_{X}\left(\Omega,\Omega\right){\cal P}_{Y}\left(\lambda_{1}\right){\cal P}_{Y}\left(\lambda_{4}\right)d\Omega d\lambda_{1}d\lambda_{4}$
is the probability of having the decimation of an $X$-type coupling
which involves the neighboring $Y$-type couplings $\lambda_{1}$
and $\lambda_{4}$. We need to sum over all possibilities for the
values of these neighboring couplings. The first two deltas correspond
to the removal of these $Y$-type couplings. The last one corresponds
to the addition of the renormalized coupling $\tilde{\lambda}=\frac{\lambda_{1}\lambda_{4}}{\Omega}$.

The normalization constant is important to keep the distribution ${\cal P}_{A}$
normalized after the cutoff is reduced. Integrating both sides of
Eq.~\eqref{eq:Pa} from $\lambda=0$ to $\Omega-d\Omega$, the l.h.s.
is simply ${\cal N}$. Up to linear order in $d\Omega$, the r.h.s
is $1-\left({\cal P}_{A}\left(\Omega,\Omega\right)+{\cal P}_{B}\left(\Omega,\Omega\right)+{\cal P}_{C}\left(\Omega,\Omega\right)\right)d\Omega$.
This is the expected result if one counts that a decimation of type
$B$ removes a net fraction of ${\cal P}_{B}\left(\Omega,\Omega\right)d\Omega$
couplings of type $A$. In addition, a decimation of $A$ type removes
a fraction of ${\cal P}_{A}\left(\Omega,\Omega\right)d\Omega$ couplings
of type $A$. The beta function for the distribution $P_{A}$ simplifies
to 
\begin{equation}
-\frac{\partial{\cal P}_{A}}{\partial\Omega}={\cal P}_{A}\left(\Omega\right){\cal P}_{A}-{\cal P}_{B\cup C}\left(\Omega\right)\left({\cal P}_{A}-{\cal P}_{A}\otimes{\cal P}_{A}\right),
\end{equation}
 where ${\cal P}_{X}\left(\Omega\right)={\cal P}_{X}\left(\Omega,\Omega\right)$,
${\cal P}_{X}={\cal P}_{X}\left(\lambda,\Omega\right)$, ${\cal P}_{B\cup C}\left(\Omega\right)={\cal P}_{B}\left(\Omega\right)+{\cal P}_{C}\left(\Omega\right)$,
and 
\begin{equation}
{\cal P}_{A}\otimes{\cal P}_{A}=\int d\lambda_{1}d\lambda_{4}{\cal P}_{A}\left(\lambda_{1},\Omega\right){\cal P}_{A}\left(\lambda_{4},\Omega\right)\delta\left(\lambda-\frac{\lambda_{1}\lambda_{4}}{\Omega}\right).
\end{equation}
 The equivalent equations for ${\cal P}_{B,C}$ are obtained by exchanging
$A\rightleftharpoons B$ and $A\rightleftharpoons C$.

At criticality, ${\cal P}_{A}={\cal P}_{B}={\cal P}_{C}$ and, thus,
\[
\frac{\partial{\cal P}_{A}}{\partial\Omega}={\cal P}_{A}\left(\Omega\right)\left({\cal P}_{A}-2{\cal P}_{A}\otimes{\cal P}_{A}\right).
\]
 Using the ansatz 
\begin{equation}
{\cal P}_{A}=\frac{1}{z\left(\Omega\right)\Omega}\left(\frac{\Omega}{\lambda}\right)^{1-1/z\left(\Omega\right)},\label{eq:Pa*}
\end{equation}
 then 
\[
\frac{1}{{\cal P}_{A}}\frac{\partial{\cal P}_{A}}{\partial\Omega}=-\frac{\dot{z}}{z}-\frac{1}{z\Omega}+\frac{\dot{z}}{z^{2}}\ln\frac{\Omega}{\lambda},
\]

\[
{\cal P}_{A}\otimes{\cal P}_{A}=\Omega\int_{\lambda}^{\Omega}\frac{dx}{x}{\cal P}_{A}\left(x\right){\cal P}_{A}\left(\frac{\lambda\Omega}{x}\right)=\frac{{\cal P}_{A}}{z}\ln\frac{\Omega}{x},
\]
\[
\frac{\dot{z}}{z}+\frac{1}{z\Omega}-\frac{\dot{z}}{z^{2}}\ln\frac{\Omega}{\lambda}=-\frac{1}{z\Omega}+2\frac{1}{z^{2}\Omega}\ln\frac{\Omega}{x}.
\]
 Thus, 
\[
\frac{\dot{z}}{z}=-\frac{1}{z\Omega}-\frac{1}{z\Omega},\mbox{ and }\dot{z}=-\frac{2}{\Omega},
\]
 which are the same. So the ansatz is acceptable. Then, 
\[
z\left(\Omega\right)=D+2\Gamma,
\]
 where $\Gamma=\ln\left(\Omega_{0}/\Omega\right)$ and $D=z\left(\Omega_{0}\right)$. 

The relation between the number of coupling constants and the cutoff
energy scale is 
\[
N\left(\Omega-d\Omega\right)=N\left(\Omega\right)-3N\left(\Omega\right){\cal P}_{A}\left(\Omega\right)d\Omega,
\]
 which simplifies to 
\[
\frac{d\ln N}{d\Omega}=3{\cal P}_{A}\left(\Omega\right)=\frac{3}{z\Omega},
\]
 from which we obtain 
\[
\ln\frac{N\left(\Omega\right)}{L}=-\frac{3}{2}\ln\frac{D+2\Gamma}{D}.
\]
 Thus, 
\[
\ell\equiv\frac{L}{N}=\left(1+\frac{2\Gamma}{D}\right)^{\frac{1}{\psi}},
\]
 with tunneling exponent $\psi=\frac{2}{3}>\frac{1}{2}$.

\bibliography{/home/hoyos/Documents/referencias/referencias}

\end{document}